\documentclass[preprint2]{aastex1}

\usepackage{natbib}
\bibpunct{(}{)}{;}{a}{}{,}
\usepackage{graphicx}
\usepackage{amsmath}

\shorttitle{Neutral I/S helium parameters from IBEX-Lo}
\shortauthors{Bzowski, Kubiak, M\"obius et al. }

\begin{document}

\title{Neutral interstellar helium parameters based on IBEX-Lo observations and test particle calculations}

\author{M. Bzowski\altaffilmark{1}, M.A. Kubiak\altaffilmark{1}, E. M\"obius\altaffilmark{2}, P. Bochsler\altaffilmark{2,3}, T. Leonard\altaffilmark{2}, D. Heirtzler\altaffilmark{2},\\ H. Kucharek\altaffilmark{2}, J.M. Sok\'o\l\altaffilmark{1}, M. H{\l}ond\altaffilmark{1}, G.B. Crew\altaffilmark{4}, N.A. Schwadron\altaffilmark{2}, S.A. Fuselier\altaffilmark{5}, \\D.J. McComas\altaffilmark{6,7}}

\altaffiltext{1}{Space Research Centre PAS, Warsaw, Poland}
\altaffiltext{2}{Space Science Center \& Department of Physics, University of New Hampshire, Durham NH, USA}
\altaffiltext{3}{Physikalisches Institut, University of Bern, Bern, Switzerland}
\altaffiltext{4}{Haystack Observatory, Massachusetts Institute of Technology, Westford MA, USA}
\altaffiltext{5}{Lockheed Martin, Space Physics Lab, 3251 Hanover Street, Palo Alto, CA 94304, USA;  stephen.a.fuselier@linco.com}
\altaffiltext{6}{Southwest Research Institute, San Antonio TX, USA}
\altaffiltext{7}{University of Texas in San Antonio, San Antonio TX, USA}

\begin{abstract}
Because of its high ionization potential and weak interaction with hydrogen, Neutral Interstellar Helium is almost unaffected at the heliospheric interface with the interstellar medium and freely enters the solar system. This second most abundant species provides some of the best information on the characteristics of the interstellar gas in the Local Interstellar Cloud. The Interstellar Boundary Explorer (IBEX) is the second mission to directly detect NISHe. We present a comparison between recent IBEX NISHe observations and simulations carried out using a well-tested quantitative simulation code. Simulation and observation results compare well for times when measured fluxes are dominated by NISHe (and contributions from other species are small). Differences between simulations and observations indicate a previously undetected secondary population of neutral helium, likely produced by interaction of interstellar helium with plasma in the outer heliosheath. Interstellar neutral parameters are statistically different from previous in situ results obtained mostly from the GAS/Ulysses experiment, but they do agree with the local interstellar flow vector obtained from studies of interstellar absorption: the newly-established flow direction is ecliptic longitude ${79.2}\degr$, latitude ${-5.1}\degr$, the velocity is $\sim 22.8 \, \mathrm{km s}^{-1}$, and the temperature is 6200~K. These new results imply a markedly lower absolute velocity of the gas and thus significantly lower dynamic pressure on the boundaries of the heliosphere and different orientation of the Hydrogen Deflection Plane compared to prior results from Ulysses. A different orientation of this plane also suggests a new geometry of the interstellar magnetic field and the lower dynamic pressure calls for a compensation by other components of the pressure balance, most likely a higher density of interstellar plasma and strength of interstellar magnetic field.
\end{abstract}


	\section{Introduction}
The Sun is moving through a surrounding warm, partially ionized interstellar cloud \citep{fahr:68a, blum_fahr:70b, bertaux_blamont:71, holzer_axford:71, axford:72} called the Local Interstellar Cloud (LIC). Because the Sun emits a supersonic stream of solar wind plasma (primarily protons and electrons with an embedded magnetic field), it inflates a bubble, called the heliosphere, which effectively shields out the LIC plasma from a region $\sim 100$~AU around the Sun. In contrast, neutral interstellar helium (NISHe) atoms penetrate freely through the heliospheric interface and since He has a high ionization potential and low cross section for charge exchange with solar wind protons, almost all of these atoms are able to reach Earth's orbit. Thus, NISHe is an important source of information on the physical state of the LIC. 

Experimental studies of NISHe began with sounding rockets \citep{paresce_etal:73a, paresce_etal:74a, paresce_etal:74b} and advanced to spacecraft \citep{weller_meier:74}. These early studies focused on the characteristic pattern of UV emissions from neutral interstellar helium and hydrogen and yielded the first estimates of the density, inflow direction, bulk velocity, and temperature of the neutral interstellar gas. The discovery by \citet{mobius_etal:85a} of the He$^{+}$ pickup ions in the solar wind (i.e., ions that result from ionization of neutral interstellar gas in the inner heliosphere) created a new method for analyzing the neutral component of interstellar gas by in-situ pick-up ion measurements in the solar wind. A third analysis method -- direct in situ measurements of the incoming NISHe atoms with a neutral particle detector -- was successfully implemented by \citet{witte_etal:92a} in the GAS experiment on board the Ulysses spacecraft. Analysis of GAS/Ulysses measurements by \citet{witte_etal:93}, capped by \citet{witte:04, witte_etal:04a}, created a benchmark set of NISHe gas parameters. The density was determined to be $0.015 \pm 0.0028 \, \mathrm{cm}^{-3}$, flow (downwind) direction (in J2000 coordinates) ecliptic longitude ${75.2}\degr \pm {0.5}\degr$ and latitude ${-5.2}\degr \pm {0.2}\degr$, velocity $26.3 \pm 0.4 \, \textrm{km s}^{-1}$, temperature $6300 \pm 340 \, \mathrm{K}$. A resume of measurements of the NISHe gas with the use of various techniques can be found in \citet{rucinski_etal:03} and \citet{mobius_etal:04a}.

The most recent attempt at reaching consensus values of the NISHe flow parameters (prior to the launch of the IBEX mission) was performed by a team organized by the International Space Science Institute (ISSI) in Bern, Switzerland \citep{mobius_etal:04a}. This consensus development involved parallel analysis of direct observations of NISHe flow by GAS/Ulysses \citep{witte:04}, observations of the He$^{+}$ pickup ions by SWICS/Ulysses and SWICS/ACE and NOZOMI \citep{gloeckler_etal:04b}, and measurements of the backscattered heliospheric {He~\footnotesize{I}} glow from EUVE \citep{vallerga_etal:04a} and UVCS/SOHO \citep{lallement_etal:04b}. The consensus set of parameters that emerged from this study was: density $n=0.0148 \pm 0.0020 \, \mathrm{cm}^{-3}$, flow direction in the J2000 ecliptic coordinates (longitude, latitude) $\lambda = {75.38}\degr \pm {0.56}\degr$, $\beta = {-5.31}\degr \pm {0.28}\degr $\footnote{Which corresponds to the \emph{inflow} (upwind) direction $\lambda = {255.4}\degr$, $\beta = {5.31}\degr$.}, flow velocity $v=26.24 \pm 0.45 \, \textrm{km s}^{-1}$, and temperature $T=6306 \pm 390 \, \mathrm{K}$. 

The IBEX mission was launched in 2008 to discover the global interaction between the solar wind and the interstellar medium \citep{mccomas_etal:09a, mccomas_etal:09c}. Part of this discovery is based on ground-breaking new measurements of interstellar neutral gas. The main goal of interstellar neutral gas studies with IBEX is to discover and analyze neutral interstellar oxygen and its expected secondary population coming from the outer heliosheath. Initial results on this topic were reported by \citet{mobius_etal:09b} and are expanded by \citet{bochsler_etal:12a}. However, interstellar oxygen is highly processed (``filtered'') at the heliospheric boundary. Therefore, drawing meaningful conclusions about this interstellar species is possible only after critical evaluation of the flow of NISHe gas, which is a topic of this paper as well as some other papers in this special issue \citep{mobius_etal:12a, lee_etal:12a}.

The Science Team of the Interstellar Boundary Explorer (IBEX) mission \citep{mccomas_etal:09a} present a series of articles by \citet{mobius_etal:12a, bochsler_etal:12a, lee_etal:12a, saul_etal:12a, hlond_etal:12a}, including also this paper, on results from measurements of NISHe gas and other neutral interstellar species. These neutral species were measured in 2009 and 2010 by the IBEX-Lo sensor \citep{fuselier_etal:09b} on board the IBEX spacecraft. The other papers in the series focus on analytic modeling of helium parameters \citep{lee_etal:12a}, measurements of oxygen and neon \citep{bochsler_etal:12a}, and hydrogen \citep{saul_etal:12a}, and determining the accurate spacecraft pointing critical for all interstellar neutral studies \citep{hlond_etal:12a}. This paper and its companion paper \citep{mobius_etal:12a} focus on NISHe measurements. \citet{mobius_etal:12a} provide a detailed description of the geometry and other details of observations and discuss the data selection for analysis. In particular, they define the select ISM flow observation times as used throughout this series of papers. They further discuss the flow parameters of NISHe based on comparison of the data with the approximate analytical model by \citet{lee_etal:12a}. We provide a detailed description of NISHe simulations performed using the Warsaw Test Particle Model and compare them with NISHe measurements. Both papers demonstrate evidence that the flow parameters of the NISHe gas are significantly different than previously thought and that, surprisingly, a secondary population of the helium gas seems to be present at Earth's orbit.

We begin  the paper with a detailed description of the model used to understand and analyze the results. We discuss experimental and observational aspects of the modeling pipeline (``things that must be taken into account''), the Warsaw Test Particle Model of the flow of NISHe gas in the heliosphere, relevant heliospheric conditions during observations and how these conditions are accounted for in the modeling. Then we discuss the data selection we did specifically for this study: we show which IBEX orbits were used in the analysis, how we identify the component of the primary population of NISHe gas in the observed signal processed by the IBEX-Lo collimator, what role various miscellaneous observational effects play, and what bias they introduce into results if unaccounted for. After these simulation details, data preparation for fitting of the NISHe gas flow parameters is discussed. We finish these preparatory sections by presentation of the fitting method used and demonstrate the results of the analysis. We close with an extended discussion of notable consequences for the physics of the heliosphere that result from the finding that the NISHe parameters are different than previously reported. Finally, we show evidence on the existence of a significant secondary helium population.

	\section{Model of the gas flow}
The goal of the numerical model used in this study is to simulate measurements of the flux of NISHe gas by the IBEX-Lo instrument in such a way that these simulation results are directly comparable with the measurements. Hence, the model simulates the NISHe flux for each of the IBEX orbits for which measurements were available (during the 2009 and 2010 measurement campaigns). In the following we discuss from the top down the specifics of how the geometrical, instrumental, and orbital conditions are introduced into the model; the simulation process used to obtain flux values as they would be observed in each orbit; the core of the simulation pipeline that calculates angular distribution of the flux of the NISHe gas flow in the inner heliosphere; and the heliospheric conditions adopted for the modeling. 

	\subsection{Specifics taken into account}
To achieve the highest possible realism and fidelity of the simulations, the simulation pipeline accurately addressed all relevant geometry, timing and instrumental aspects, including the following.

\begin{figure}[t]
		\centering
  	\plotone{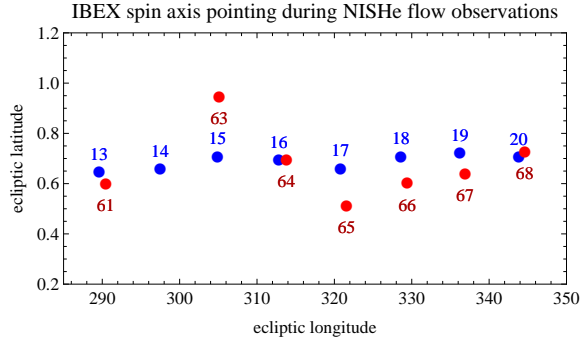}
  	\caption{Ecliptic J2000 coordinates of the IBEX spin axis during the two NISHe measurements campaigns: 2009 (blue) and 2010 (red). The orbit numbers are shown at the corresponding points.}
 		\label{figAxisPointing}
		\end{figure}

\begin{figure}[t]
			\centering
			\includegraphics[scale=0.8]{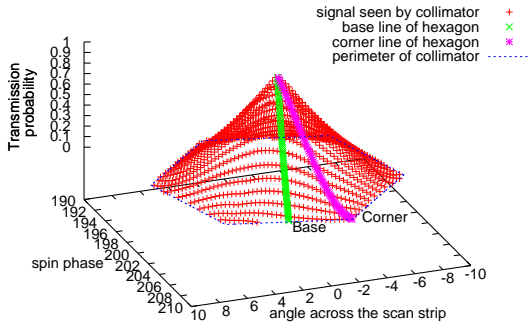}
			\caption{Transmission function of the IBEX-Lo collimator as used in the simulation program. The transmission function is the probability of transmission for an atom that goes through the collimator at an angle $\rho$ off the boresight axis, at an azimuth angle $\theta$. The base of the field of view is hexagonal and the transmission function is calculated as a linear interpolation between the transmission at one of the corner lines (magenta in the plot) and the adjacent base line (green). The angle $\rho$ goes along the radial lines, examples of which are the magenta and green lines, $\theta$ goes counterclockwise from the polar line.}
			\label{figColimTransFun}
			\end{figure}

\begin{figure}[t]
			\centering
			\includegraphics[scale=0.8]{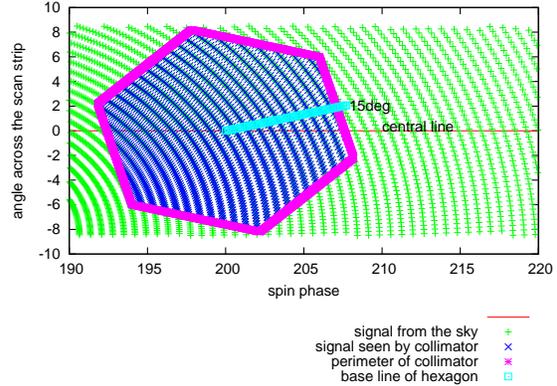}
			\caption{Geometry of the collimator relative to the Visible Strip of the sky. The limits of the instantaneous field of view are drawn in the thick magenta line forming a hexagon. Green crosses mark the centers of the sky pixels at which the NISHe flux is calculated. Blue symbols mark the sky pixels within the collimator field of view at a given instant. The collimator scans the Visible Strip along the center line, constantly changing its spin angle. The collimator polar angle $\rho$ is counted from the boresight along polar line (e.g., the cyan line shown in the figure) and the azimuthal angle $\theta$ goes counterclockwise from the polar line.}
			\label{figColimFOV}
			\end{figure}

		\begin{itemize}
		\item[--] The Visible sky Strip. IBEX-Lo observes a strip on the sky almost exactly perpendicular to the IBEX rotation axis. The field of view (FOV) of IBEX-Lo was defined in the simulation program according to the FOV specification for IBEX-Lo \citep{fuselier_etal:09b} and, in a separate study \citep{hlond_etal:12a}, it was verified that the pointing of IBEX-Lo indeed agrees with its specified pointing in the spacecraft system to better than $0.15\degr$. The spin axis of IBEX is close to the ecliptic during science operations and  pointed $< 1\degr$ above (Fig. \ref{figAxisPointing}). For each orbit the Visible Strip of sky viewed by the sensor was calculated based on the exact pointing of the IBEX spin axis determined by the IBEX Science Operations Center (ISOC) \citep{schwadron_etal:09a} and illustrated in Fig. \ref{figAxisPointing}. 
		\item[--] Collimator shape and transmission function. Transmission function $T(\rho, \theta)$ of the collimator was adopted from pre-launch calibration \citep{fuselier_etal:09b}; its shape is shown in Fig. \ref{figColimTransFun}. The value of the transmission function at a given location within the FOV of the sensor, described by the polar coordinates $(\rho, \theta)$ relative to the boresight axis of the collimator, corresponds to the percentage of the flux that is able to enter the sensor at a given area element $\sin \rho \,\mathrm{d} \rho \; \mathrm{d} \theta$. The field of view of the low-angular resolution portion of the collimator of IBEX-Lo is hexagonal in shape \citep{fuselier_etal:09b} and its arrangement relative to the sky strip scanned during spacecraft rotation is shown in Fig. \ref{figColimFOV}. The profiles of the transmission function from the boresight to the corner and to the side of the hexagonal base were fitted with the third and second order polynomials, respectively. Both the shape of the collimator transmission function and its arrangement in the spacecraft reference system were exactly simulated in the program.
		\item[--] Positions and velocity of Earth relative to the Sun. We used the ephemeris obtained from the SPICE-based program developed and operated by the (ISOC) \citep{acton:96a, schwadron_etal:09a}. These include actual solar distances and ecliptic longitudes of Earth as well as Earth velocity vectors relative to the Sun for all dates for which simulations were performed.
		\item[--] Velocity vectors of the IBEX satellite relative to Earth. They were taken from the same SPICE-based program source as for the Earth orbit; together with the Earth information these spacecraft velocity vectors were used to calculate the state vectors of IBEX relative to the Sun for the simulations.
		\item[--] Selection of observations. We use the observations selected from the IBEX-Lo data set with data drop-outs, spacecraft pointing knowledge problems, and other spacecraft and sensor conditions that affect overall flux and direction removed (see \citet{mobius_etal:12a} for a detailed description of the select ISM flow observation times).
	\end{itemize}

The simulation pipeline accepts as input: parameters of the NISHe gas in the LIC, energy limits of the incoming atoms in the spacecraft inertial frame to be adopted as flux integration boundaries, parameters that describe heliospheric conditions (time series of the photoionization rate and solar wind density and velocity averaged over Carrington rotations), the number of the orbit for which the simulation is to be performed (i.e., dates and times of the simulation), the spin axis pointing for the orbit, the list of select ISM flow observation times for the orbit from \citet{mobius_etal:12a}, and the state vectors of the IBEX spacecraft relative to the Sun for the observations. It returns collimator-averaged fluxes of the NISHe gas as function of IBEX spin angle averaged by the selected times and the collimator transmission function. The simulation pipeline product can be directly compared with the observed count rates for the given orbit after linear scaling. In the simulations carried out for this study, the integration boundaries were adopted from zero to infinity, so effectively the integration was over the full energy range of the incoming NISHe atoms. As discussed by \citet{mobius_etal:12a}, such an approach is valid because IBEX-Lo actually does not measure incoming He atoms directly, rather it detects H, O, and C atoms sputtered off the conversion surface, so He atoms of all relevant energies contribute significantly to the sputtered H signal collected by the energy steps 1, 2, and 3 (energy passbands between 0.01 and $\sim 0.075$~keV) of IBEX-Lo. Details of calibration of the IBEX-Lo instrument for detection of H, He, and O atoms are provided by \citet{saul_etal:12a} and \citet{bochsler_etal:12a}.

		\subsection{Simulation of NISHe flux for a single orbit}
The core of the simulation program calculates the flux of NISHe relative to the IBEX spacecraft located at a point $\vec{r}$ relative to the Sun, traveling at a velocity $\vec{v}$ at a time $t$ for a line of sight determined by ecliptic coordinates $(\lambda_{\mathrm{LOS}},\beta_{\mathrm{LOS}})$. This part of the simulation set is described in the following section. Here we discuss simulations of the NISHe flux averaged over the IBEX-Lo collimator FOV and select ISM flow observation times in a given orbit.

The simulation pipeline is organized as follows. With the select ISM flow observation times transformed into Julian days, a series of dates at halves of full Julian days that straddle and fill in the selected intervals is determined. Subsequently, the Visible Strip is determined based on the spin axis pointing for the given orbit and coordinates of its boundaries in the ecliptic reference system are calculated. The Visible Strip is then transformed into heliographic reference system (HGI, \citet{franz_harper:02a}) and mapped on a grid of equal-area, equi-distant pixels whose boundaries and centers in the heliographic coordinates are adopted following the HealPix scheme \citep{gorski_etal:05a} with the resolution parameter $N = 64$, which corresponds to 49152 pixels for the whole sky. Thus the angular resolution of the Visible Strip coverage is better than 1~deg$^2$. Centers of these pixels make the simulations mesh~$(\lambda_{\mathrm{LOS}},\beta_{\mathrm{LOS}})$.

The Visible Strip does not change during one orbit, so during all select ISM flow observation times in a given orbit the instrument is looking at the same portion of the sky. The simulations are carried out for all pixel centers within the Visible Strip for all select ISM flow observation times in a given orbit in the inertial frame of the IBEX spacecraft. The inertial frame is determined by the IBEX velocity vector $\vec{v}$ relative to the Sun, which is obtained from the ISOC. 

With the detailed map of the NISHe flux within the Visible Strip for a given day, we calculate the flux transmitted through the collimator. We do so by sliding the collimator boresight along the spin angle $\psi$ in 1-degree steps (see Fig. \ref{figColimFOV}), integrating the flux as convolution of the transmission function $T(\rho , \theta)$ with the flux $F_{\mathrm{He}}(\rho (\psi), \theta(\psi))$:
		\begin{eqnarray}
		F_{\mathrm{He,coll}}(\psi)={\int\limits_{\theta=0}^{2 \pi}}{\int\limits_{\rho=0}^{\rho_{1}(\theta)}}T(\rho , \theta) \nonumber \\
	F_{\mathrm{He}}(\rho (\psi), \theta(\psi)) \sin \rho \,\mathrm{d} \rho \,\mathrm{d} \theta
		\label{eqFluxFmColl1}
		\end{eqnarray}
where $F_{\mathrm{He}}\left(\rho \left(\psi\right), \theta \left(\psi\right)\right)$ is the flux calculated at $\left(\rho , \theta \right)$ for a given spin phase angle $\psi$ of the collimator boresight and $\rho_{1}\left(\theta \right)$ describes the hexagonal boundary of the field of view. For a different boresight $\psi$ the same flux element will be located differently relative to the boresight direction and consequently will contribute to the collimator-averaged flux with a different weight. In practice, the collimator FOV was divided into regions of approximately equal areas distributed symmetrically around the boresight at a series of $\left(\rho_{i},\theta_{j}\right)$. The integration over the FOV was in fact a summation of the flux with appropriate weights:
 	 \begin{eqnarray}							
		F_{\mathrm{He,coll}}\left( \psi \right)={\sum_{i=1}^{N_i}}{\sum_{j=1}^{N_j}}{\sum_{k=1}^{N_{ij}}}T\left( \rho_i , \theta_j \right) \nonumber \\F_{\mathrm{He}}\left(\rho_{ik}, \theta_{jk}\right)s_{ij}/N_{ij}
		\label{eqFluxFmColl2}
		\end{eqnarray}		
where $i$ marks the radial and $j$ the azimuthal index of the mesh, $s_{ij}$ is the unity-normalized area of the $i, j$ field and $k$ counts from 1 to $N_{ij}$ the pixels at $\left(\rho_{ik},\theta_{jk}\right)$ in the $\left(i,j\right)$-th field, in which the field-averaged $F_{\mathrm{He}}$ flux is calculated. Since the $s_{ij}$ fields are equal-area, the number of sky pixels per integration field is approximately constant, which adds to the numerical stability of the calculation scheme. 

Following the procedure described in the preceding paragraphs, we obtain a series of collimator-integrated fluxes for given days, which subsequently are time-averaged over the select ISM flow observation times. The result of this averaging is taken as the simulation result for a given set of parameters of NISHe gas for a given orbit. The procedure of calculating the collimator- and orbit-averaged flux was repeated for all orbits within the 2009 and 2010 observation seasons. 

	\subsection{Model of NISHe flux in the inner heliosphere}
In the inertial frame of IBEX, the flux of NISHe gas $F_{\mathrm{He}}\left(\lambda, \beta, \vec{r},t\right)$ that goes into the ecliptic-coordinates direction $\left(\lambda, \beta \right)$ at the location described by the heliocentric vector $\vec{r}$ at a time $t$ is calculated by
		\begin{eqnarray}							
		F_{\mathrm{He}}\left(\lambda, \beta, \vec{r}, t\right)=\int\limits_{0}^{\infty}v_{\mathrm{He,sc}}f_{\mathrm{He}}
		\left(\vec{v}_\mathrm{{He,ecl}}, \vec{r}, t\right)\nonumber \\\hat{\vec{e}}\left(\lambda, \beta\right) v_{\mathrm{He,sc}}^2 \mathrm{d}v_{\mathrm{He,sc}},
		\label{eqHeFluxSC}
		\end{eqnarray}	
where $v_{\mathrm{He,sc}}$ is the magnitude of the He atom velocity vector $\vec{v}_{\mathrm{He,sc}}$ in the inertial frame of the spacecraft, $\hat{\vec{e}}\left(\lambda, \beta \right)$ is the unity vector pointing toward $\left(\lambda, \beta\right)$, and  $f_{\mathrm{He}}\left(\vec{v}_{\mathrm{He,ecl}},\vec{r},t\right)$ is the distribution function of the NISHe gas for time $t$ and solar frame-velocity $\vec{v}_{\mathrm{He,ecl}}$ at the location specified by the solar-frame radius vector $\vec{r}$. Assuming that the flow of the NISHe gas in the Local Interstellar Cloud is constant, the distribution function $f_{\mathrm{He}}$ at $\vec{r}$ is time dependent only because of variations in the helium ionization rate. 

The transition from the spacecraft inertial frame to the solar inertial frame is done by a simple vector subtraction: with the IBEX solar-inertial velocity $\vec{v}_{\mathrm{IBEX}}\left(t\right)$ the relation between the IBEX-inertial $\vec{v}_{\mathrm{He,sc}}$ and solar-inertial $\vec{v}_\mathrm{He,ecl}$ velocities is:
		\begin{equation}							
		\vec{v}_{\mathrm{He,ecl}}\left(t\right)=\vec{v}_{\mathrm{He,sc}}-\vec{v}_{\mathrm{IBEX}}\left(t\right)
		\label{eqVelChange}
		\end{equation}
The conversion to the solar-inertial frame during the integration specified in Eq.~(\ref{eqHeFluxSC}) is done separately for each value of $v_{\mathrm{He,sc}}$ and the calculation of the local distribution function $f_{\mathrm{He}}$ is performed in the solar-inertial frame. 

The model of neutral interstellar gas flow in the inner heliosphere, used to calculate the maps of NISHe flux at Earth orbit, is a derivative of the Warsaw Test Particle Model developed since the mid 1990s \citep{rucinski_bzowski:95b}. Previous versions, as well as its development history, are found in \citet{tarnopolski_bzowski:09}. Recent applications of this code in interpreting measurements of neutral interstellar hydrogen in the inner heliosphere are discussed by \citet{bzowski_etal:08a, bzowski_etal:09a} and its use in interpreting interstellar helium measurements by \citet{gloeckler_etal:04b}. Predictions of neutral interstellar deuterium flux at IBEX, obtained using the model, can be found in \citet{tarnopolski_bzowski:08a}. Details of test-particle calculations of NISHe in the inner heliosphere are in \citet{rucinski_etal:03}.The model was used by \citet{mobius_etal:09b} to verify the detection by IBEX of the NISHe atoms.

In order to be used in the determination of the flow parameters of the NISHe gas from IBEX-Lo observations, the model had to be modified. Modification to the model was done in three main areas: (1) atom dynamics, (2) inertial frame, and (3) ionization rate as function of time and location in the heliosphere. 

The first modification was the most straightforward: since the resonance radiation pressure force acting on the neutral He atoms in the heliosphere is practically negligible, the radiation pressure module in the code could be switched off. The atoms now move solely under the $1/r^2$ solar gravity force. This change greatly simplified the requirements for the atom tracking module. Nevertheless, this module still had to maintain its ability to accurately link the time on orbit with the locus on orbit and the current sophisticated Runge-Kutta tracking scheme was not replaced to save on the development time and maintain sufficient homogeneity of the code in view of planned future applications of the model to interstellar hydrogen analysis. A version of the code with the full radiation pressure module installed was used to calculate the predictions of the neutral interstellar H signal discussed later on in the paper.

Since the calculation of the NISHe flux needs to be done in the IBEX spacecraft inertial frame, the input direction in space and speed of the atom are formulated in the moving frame and transformed to the solar frame. Therefore, initial values of the atom velocity are taken relative to the Sun, not to the spacecraft. The integration over speed, which returns the flux relative to the spacecraft from a given direction in space, is performed in the spacecraft reference frame, but parameters of the integrand function are converted to the solar inertial frame in the heliographic reference system. This change to the solar HGI frame is needed because the ionization rate model, which is used to calculate the survival probability of the atom, uses the solar equator as the natural reference plane. 

The transformation from the spacecraft-inertial frame to the solar-inertial frame requires only specifying the velocity vector of the spacecraft at the desired moment of time. No further assumptions need to be made, which facilitates adoption of various spacecraft velocity vectors in the calculation scheme. 

The ionization rate, which is discussed in greater detail in the following section, is time-dependent. We determine all the quantities relevant for the calculation of the net ionization rate as a function of time by interpolating between Carrington-period averaged quantities. Thus the model is fully time-dependent and uses current best parameters obtained directly from measurements, which adds to the accuracy of the results. 

	\subsection{Heliospheric conditions: ionization of NISHe gas}
Helium has the highest first ionization potential of all elements (27.587 eV) and hence the ionization losses of the NISHe gas in the heliosphere are relatively low. Where IBEX makes its measurements (at 1 AU), as much as $~70\%$ of the atoms from the original population are able to survive \citep{rucinski_etal:03}. Nevertheless, ionization has to be taken into account in the analysis because it modifies the shape of the observed helium beam. Ionization changes the apparent velocity distribution of the NISHe beam because it more readily removes slower atoms from the ensemble than faster ones and thus the mean velocity vector of the remaining distribution differs from the conditions when no ionization is operating (this effect is much more pronounced for hydrogen and was discussed in this context by \citet{lallement_etal:85b} and \citet{bzowski_etal:97}). The selective ionization results in a change in the ecliptic longitude at which the maximum of the NISHe beam is observed by a few tenth of a degree and, if unaccounted for, biases the derived speed and longitude of the flow direction. Similarly, this effect reduces the width of the beam somewhat, which if neglected, leads to an underestimation of the temperature. 

\begin{figure}[t]
		\centering
		\plotone{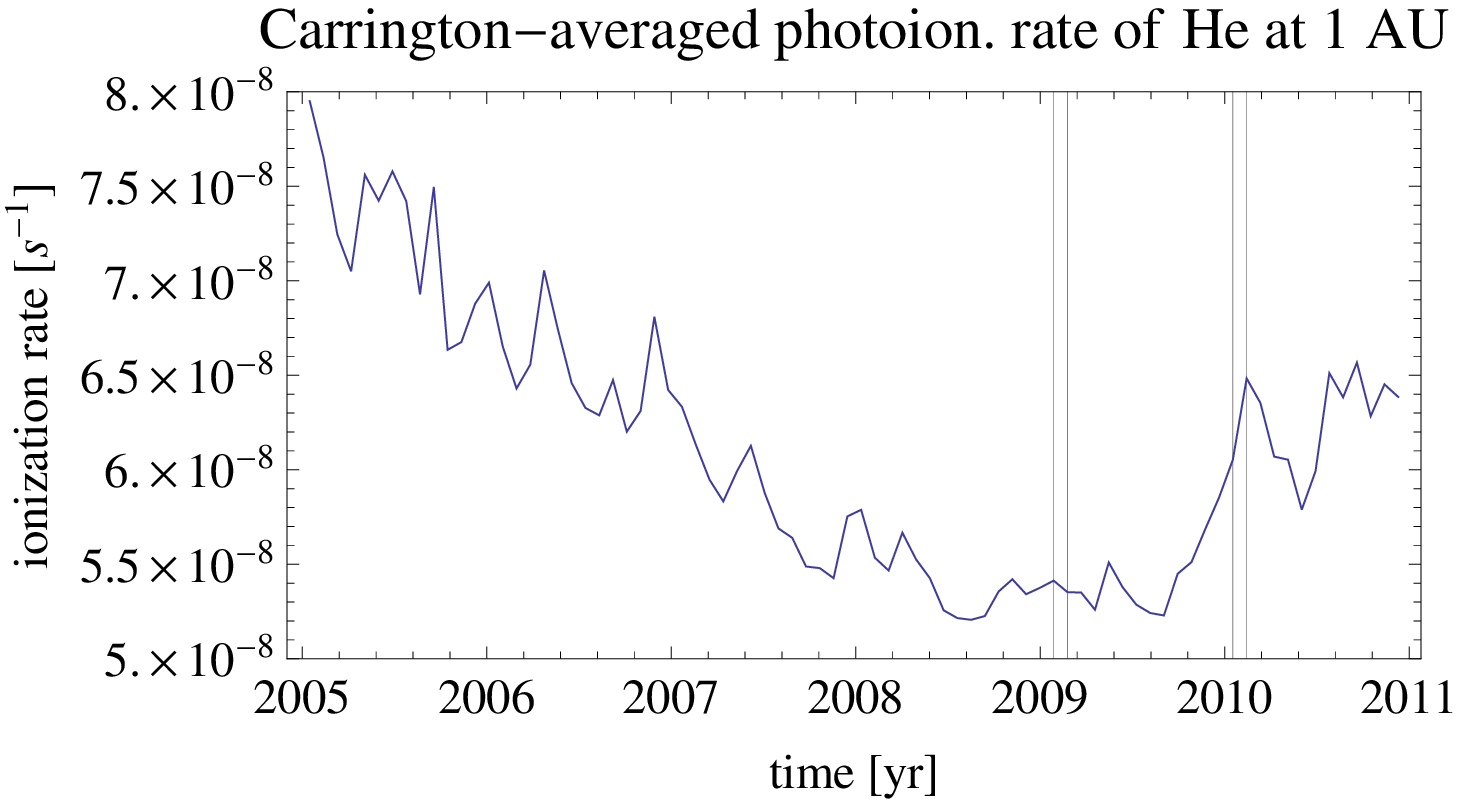}
		\caption{Time series of Carrington period-averages of the photoionization rate of neutral helium at a distance of 1 AU from the Sun. They are calculated \citep[in preparation]{bochsler_etal:12b} based on direct integration of the solar spectrum as measured by TIMED/SEE experiment \citep{woods_etal:05a} and calibrated with the CELIAS/SEM observations \citep{judge_etal:98}, using the photoionization cross section from \citet{verner_etal:96}. Two pairs of vertical lines mark the time intervals of the NISHe flow observations by IBEX-Lo in 2009 and 2010.}
		\label{figPhotoIonRate}
		\end{figure}

Heliospheric conditions that affect the flow of the NISHe gas in the inner heliosphere were extensively discussed by \citet{mcmullin_etal:04a}. The dominant ionization process is solar photoionization, which varies throughout the solar cycle from about $5.5 \times 10^{-8} \, \mathrm{s}^{-1}$ at minimum to $\sim 1.5 \times 10^{-7} \, \mathrm{s}^{-1}$ at maximum. In the present study, following \citet{bochsler_etal:12b} (in preparation), we adopted the cross section after \citet{samson_etal:94a, verner_etal:96} and we directly integrated the spectra obtained from TIMED/SEE \citep{woods_etal:05a}.  We verified the agreement of the results with the measurements from CELIAS/SEM \citep{judge_etal:98}. As seen in Fig. \ref{figPhotoIonRate}, measurements of the NISHe flow in the 2009 season followed a prolonged period of very low solar activity and very stable photoionization rate. In contrast, measurements in the 2010 season occurred during a period of increasing activity, with the photoionization rate higher by $~15\%$ than during the preceding measurement season.

		\begin{figure}[t]
		\centering
		\plotone{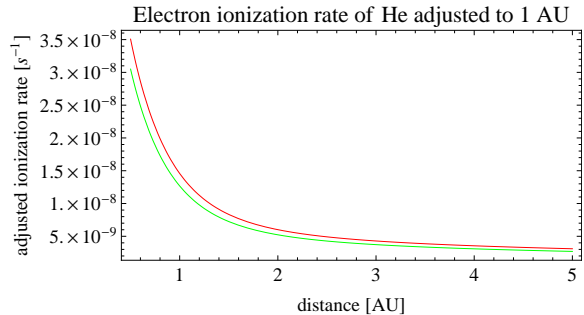}
		\caption{Electron-impact ionization rate of helium for the 2009 season (red) and 2010 season (green), adjusted to 1 AU by $r^2$.}
		\label{figElecIonRate}
		\end{figure}
	
As pointed out by \citet{auchere_etal:05a, auchere_etal:05c}, the photoionization rate of helium appears to vary weakly with heliolatitude, with the polar rate probably being about 80\% -- 85$\%$ of the equatorial rate. This latitudinal variation was accounted for by implementing the following relation:
		\begin{equation}							
		\beta_{\mathrm{ph}}\left(\phi\right)=\beta_{\mathrm{ph}}\left(0\right)\sqrt{a_{\beta_{\mathrm{ph}}}\sin^{2}\phi+\cos^{2}\phi}
		\label{eqBetPhFlat}
		\end{equation}
where $a_{\beta_{\mathrm{ph}}}$ is the  latitudinal ``flattening'' parameter adopted to be 0.8. In test simulations we verified, however, that this flattening has a small effect on the expected NISHe flux in the heliospheric tail region, and practically no effect at the interval of ecliptic longitudes where IBEX measurements were taken. The weakness of this effect can be easily explained by the fact that the trajectories of NISHe atoms detected by IBEX-Lo remain close to the ecliptic throughout their travel from the LIC to Earth's orbit and therefore never experience the ionization rates relevant for higher latitudes.

Another ionization process of neutral helium is ionization by impact of solar wind electrons. The importance of this ionization process for NISHe in the heliosphere was first pointed out by \citet{rucinski_fahr:89, rucinski_fahr:91}. As discussed by \citet{mcmullin_etal:04a}, who used more recent measurements of solar wind electrons, this rate close to the ecliptic plane at 1 AU from the Sun is equal to about $2\times 10^{-8}\, \mathrm{s}^{-1}$, i.e., it is at an appreciable level of $\sim 30\%$ of the photoionization rate, but due to the rapid cooling of the solar wind electrons it falls off with solar distance much faster than $1/r^{2}$, i.e., faster than the drop-off of the photoionization rate. 

We expanded the electron-ionization model used by \citet{mcmullin_etal:04a} assuming the thermal behavior of solar wind electrons as conforming to the core + halo model \citep{pilipp_etal:87c}. Following the approach adopted by \citet{bzowski:08a} to develop an electron ionization rate model for hydrogen, we used the solar wind electron temperature and density measurements by \citet{scime_etal:94, issautier_etal:98, maksimovic_etal:00a} and implemented the model by \citet{rucinski_fahr:89, rucinski_fahr:91}, where the cross section for electron impact ionization by \citet{lotz:67} is convolved with the Maxwellian distribution function separately for the core and halo temperatures, assuming the radial dependence of the temperatures and the proportions between the core and halo population densities as compiled by \citet{bzowski:08a}. The total electron density was implemented as tied to the density of solar wind protons (enhanced by the doubled average alpha particle abundance). The radial behavior of thereby obtained ionization rates for the 2009 and 2010 seasons is presented in Fig. \ref{figElecIonRate}. The electron-impact ionization is important only in the final phase of a NISHe atom flight before detection by IBEX, when its distance from the Sun is close to 1~AU. One has to note, however, that because IBEX measures only atoms near their perihelia, i.e., those which travel nearly tangentially to the 1~AU circle around the Sun, the influence of electron impact ionization is stronger than when they are observed at ecliptic longitudes in the upwind hemisphere.

The least significant ionization process for neutral helium in the inner heliosphere is charge exchange with solar wind particles: protons and alphas \citep{rucinski_etal:96a, rucinski_etal:98, mcmullin_etal:04a}. While significantly less intense, we include this process for completeness. The instantaneous charge exchange rate is defined by \citet{mcmullin_etal:04a} in their Equations 2, 3, and 4, from the formula:
		\begin{eqnarray}		\beta_{\mathrm{He,cx}}\left(t\right)=n_{\mathrm{p}}\left(t\right)\left|\vec{v}_{\mathrm{HeENA}}-\vec{v}_{\mathrm{SW}}\right|\nonumber \\	\left[2\alpha_{\mathrm{\alpha}}\sigma_{\mathrm{He,\alpha}}\left(\left|\vec{v}_{\mathrm{HeENA}}-\vec{v}_{SW}\right|\right) \right. \nonumber \\ \left. + \sigma_{\mathrm{He,p}}\left(\left|\vec{v}_{\mathrm{HeENA}}-\vec{v}_{\mathrm{SW}}\right|\right)\right]	
		\label{eqBetHeCX}
		\end{eqnarray}
where $\left|\vec{v}_{\mathrm{HeENA}}-\vec{v}_{\mathrm{SW}}\right|$ is the relative speed between a He atom at $\vec{v}_{\mathrm{HeENA}}$ and the radially expanding solar wind at $\vec{v}_{SW}(t)$, $\alpha_{\mathrm{alpha}}\approx 0.04$ is a typical abundance of solar wind alphas relative to protons, $n_p(t)$) is the local proton density taken from the OMNI-2 compilation of solar wind observations \citep{king_papitashvili:05}, and $\sigma_{\mathrm{He,p}}$, $\sigma_{\mathrm{He,\alpha}}$ are the charge exchange cross sections for the reaction given by Eq.~(2) and a sum of reactions given by Eq.~(3) and (4) by \citet{mcmullin_etal:04a}). The  net rate from the three charge exchange processes that were taken into account is $\sim2.6 \times 10^{-9} \, \mathrm{s}^{-1}$ regardless of the activity phase, which is of the order of 4\% of the typical photoionization rate. Thus typically the rate of charge exchange losses is less than the uncertainty in the photoionization rate. We implemented it only to make sure that we do not miss a sudden increase in total ionization rate due to possible high flux events in the solar wind (like Coronal Mass Ejections, CMEs), when the solar wind density may increase by an order of magnitude.

	\section{Initial insights from modeling of NISHe flow}
Before deciding which of the many effects should be taken into account in the simulation pipeline we carried out a study of the expected behavior of the NISHe signal and its dependence on various aspects of the measurement process.

	\subsection{Orbit selection for the analysis}
Since IBEX-Lo is able to observe helium only indirectly, via sputtering products from the conversion surface, which include H atoms \citep{mobius_etal:09a, mobius_etal:09b, mobius_etal:12a, saul_etal:12a}, we determined from the simulation in which orbits the flux expected from the NISHe flow should exceed the flux expected from neutral interstellar hydrogen. We compared collimator-averaged total NISHe fluxes expected assuming the prior consensus NISHe flow parameters \citep{mobius_etal:04a}, which are very close to the parameters obtained by \citet{witte:04} from Ulysses, with the neutral interstellar hydrogen flux in Energy Step 2 (center energy 27~eV) of the IBEX-Lo detector \citep{fuselier_etal:09b}. For this comparison, we assumed that the population of interstellar hydrogen at IBEX is a mixture of the primary population of interstellar hydrogen and a secondary component due to charge exchange with the heated and compressed plasma in front of the heliopause \citep{malama_etal:06}. We used the parameters of the two populations as determined by \citet{bzowski_etal:08a} based on pickup ion measurements on Ulysses \citep{gloeckler_etal:09a}. 

	\begin{figure*}[t]
	\centering
	\begin{tabular}{ccc}
	\includegraphics[scale=.7]{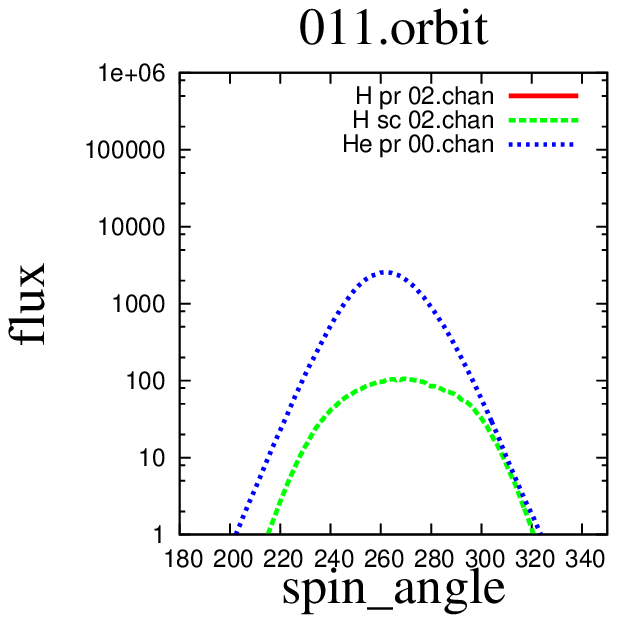}&\includegraphics[scale=.7]{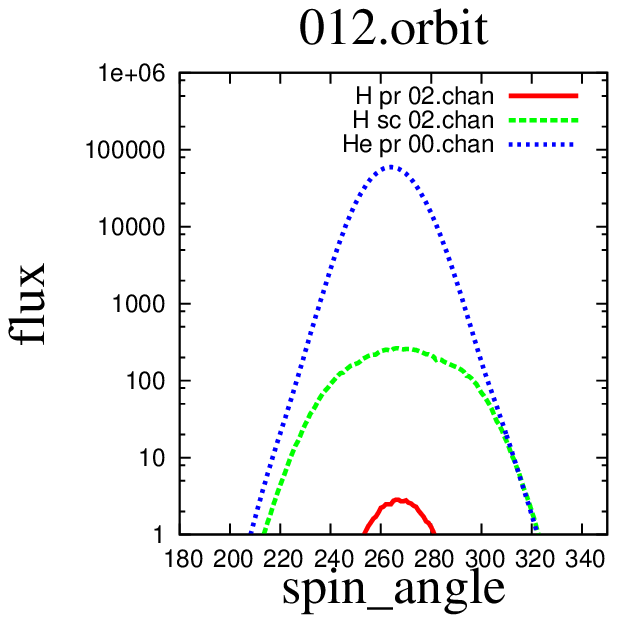}&\includegraphics[scale=.7]{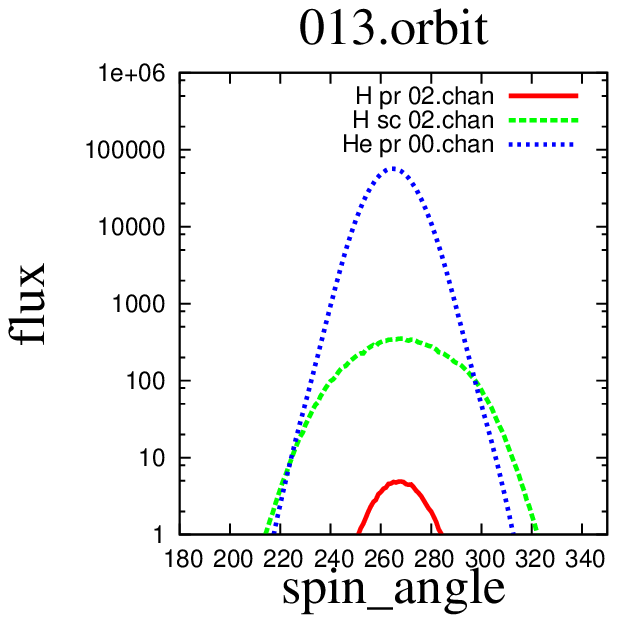}\\
	\includegraphics[scale=.7]{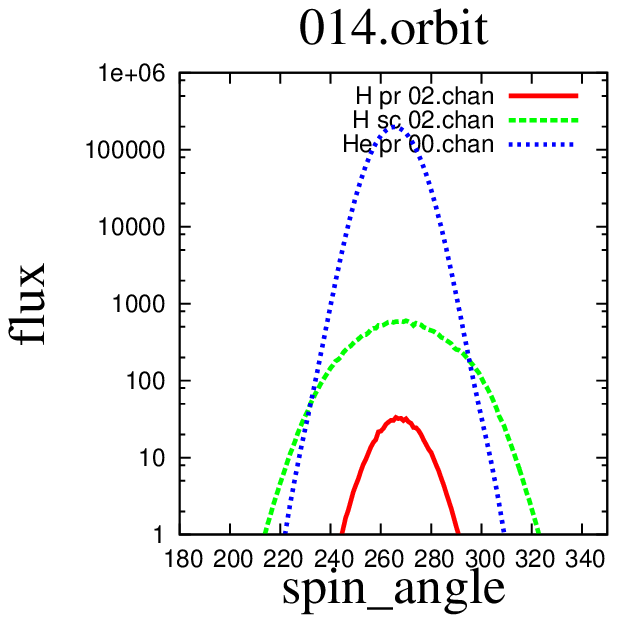}&\includegraphics[scale=.7]{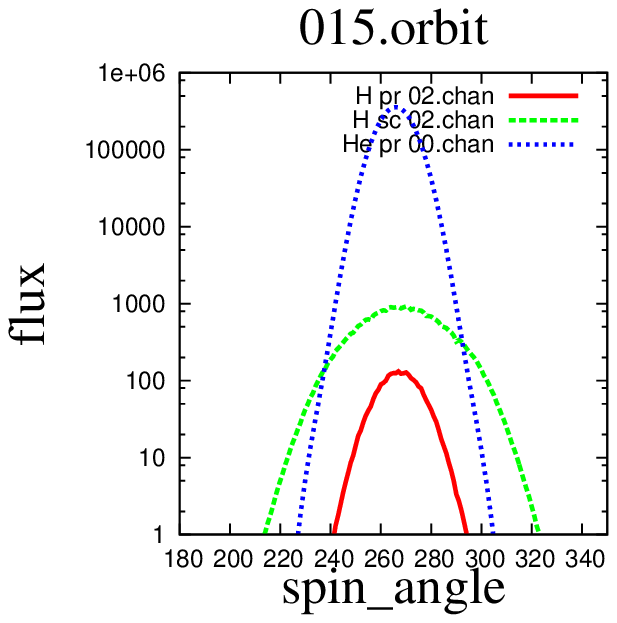}&\includegraphics[scale=.7]{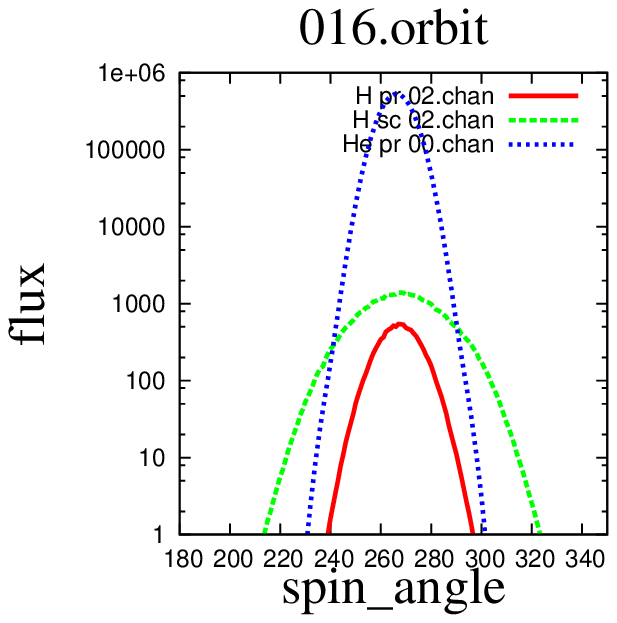}\\
	\includegraphics[scale=.7]{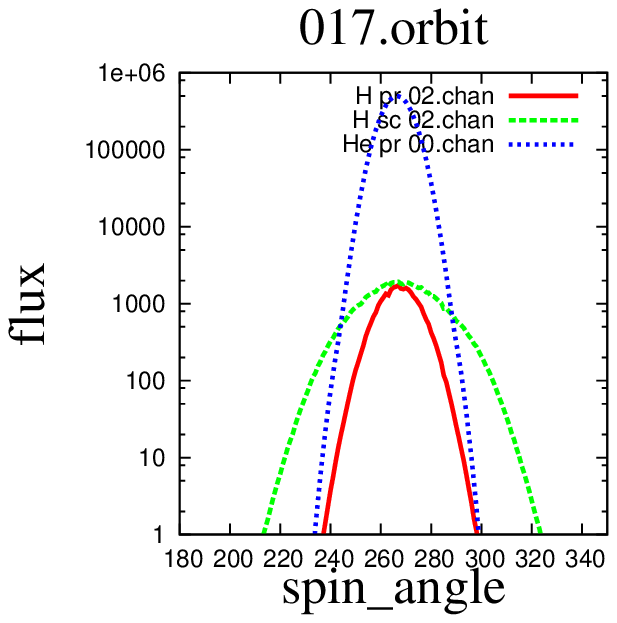}&\includegraphics[scale=.7]{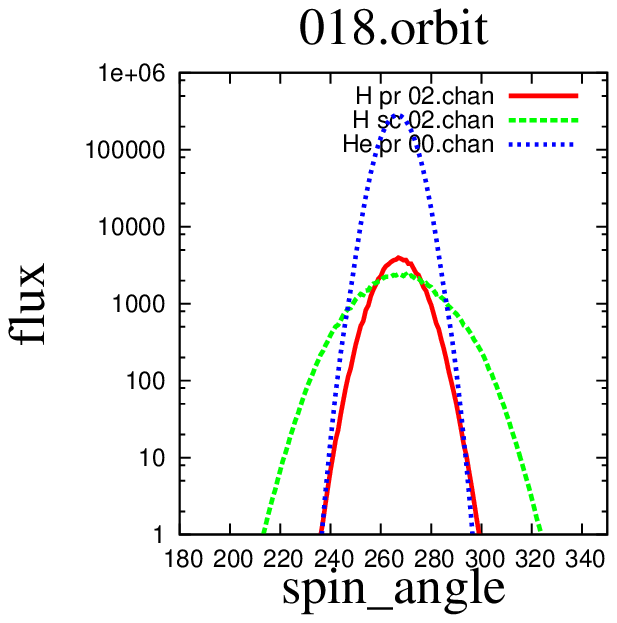}&\includegraphics[scale=.7]{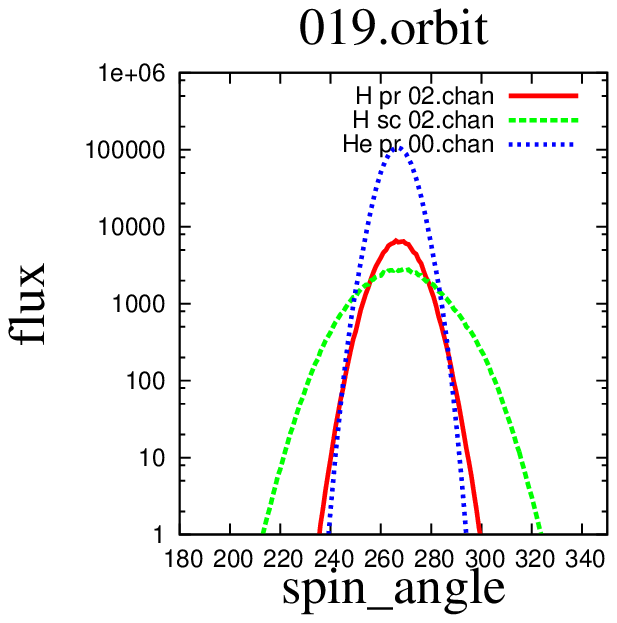}\\
	\includegraphics[scale=.7]{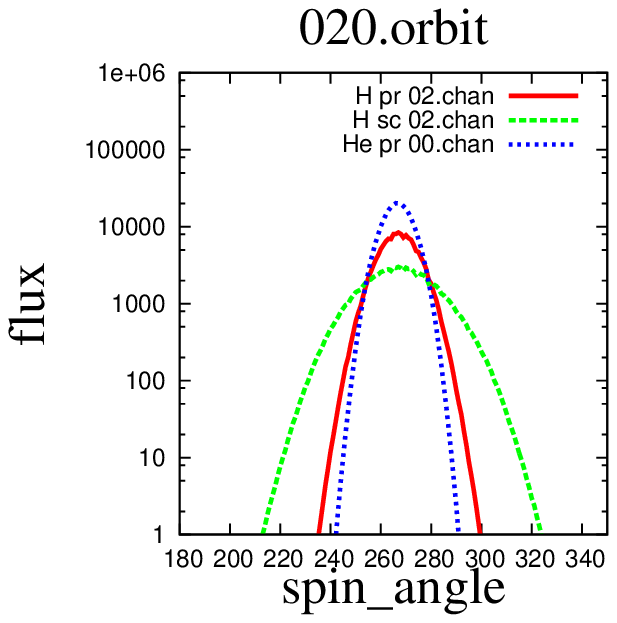}&\includegraphics[scale=.7]{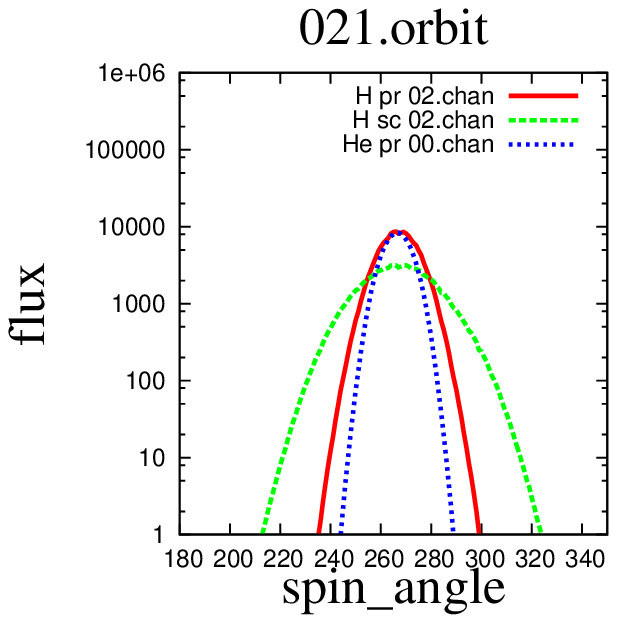}&\includegraphics[scale=.7]{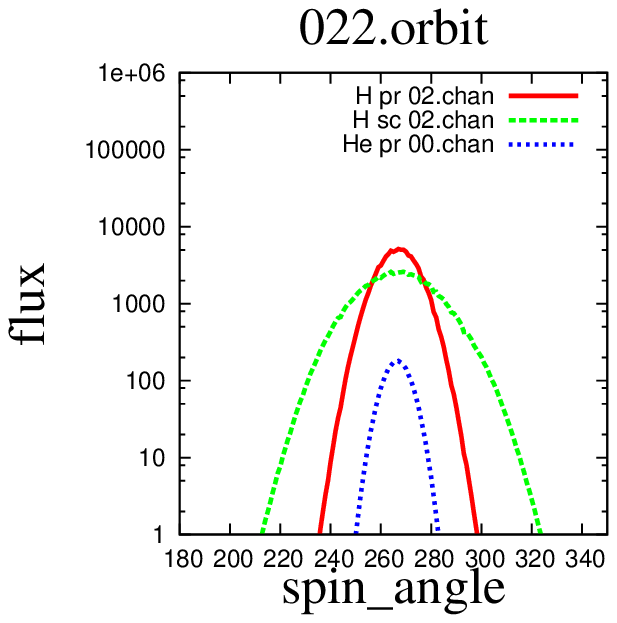}\\
	\end{tabular}
	\caption{Simulated collimator-averaged flux of neutral interstellar helium (blue) integrated over all energies, compared with the primary (red) and secondary (green) populations of neutral interstellar hydrogen at IBEX orbits 11 through 22, integrated over the energy range corresponding to the IBEX-Lo energy step 2.}
	\label{figHvsHeFlux}
	\end{figure*}
	\clearpage

As shown in Fig.~\ref{figHvsHeFlux}, in orbit 11 the helium signal dominates and the only appreciable NISH flux (2 orders of magnitude lower than the He flux) is from the secondary population. The dominance of He over H increases from orbit 11 to 17, but the intensity of the primary H population gradually increases and in orbit 17 it exceeds the peak intensity of the secondary hydrogen. However, He fluxes are still significantly higher than the combined hydrogen fluxes. Starting in Orbit 14, the wings of the H signal become wider than the wings from He, but these wings are more than 3 orders of magnitude lower than the peak of the He flux. The situation changes in Orbit 20, when the hydrogen primary population is only a few times weaker than He and thus might appear as an extra component in the total signal. The secondary hydrogen wings should be at a level of $\sim 1\%$ of the He peak. In orbit 21 (when IBEX is viewing the nose of the heliosphere), H exceeds He and in Orbit 22 H becomes dominant. The observation of interstellar H is discussed by \citet{saul_etal:12a}.

Even though a change in the solar wind or interstellar parameters may change details, the basic conclusion is that the best orbits to study the NISHe flow are orbits 13 through 19 -- 20 and their equivalent during the second ISN season for IBEX (see Fig.~\ref{figAxisPointing}; further justification is provided in the data selection section). Since the NISHe population is highly peaked and at the peaks it exceeds the H populations by more than 3 orders of magnitude, it is appropriate to analyze the Gaussian cores of the signal as due solely to the NISHe flow. Since the NISH flow should be mostly visible at the wings and since it is expected to consist of at least 2 populations, making the signal fairly complex, we decided to remove these non-Gaussian wings from the NISHe analysis.

	\subsection{Collimator-averaged signal as function of spin phase}
To differentiate the signal from background, secondary populations, and other potential biasing, we investigated in greater detail how the collimator-averaged signal would appear if we assume no background or secondaries and further assume that the NISHe gas distribution function in the LIC is the purely Maxwellian function: 
		\begin{eqnarray}							
		f_{\mathrm{He,Maxw}}\left(\vec{v}\right)=n_{0}\left(\frac{m_{\mathrm{He}}}{2\pi kT}\right)^{3/2} \nonumber \\
		\mathrm{exp}\left[-\frac{m_{\mathrm{He}}}{2kT}\left(\vec{v}-\vec{v}_\mathrm{B}\right)^2\right]
		\label{eqMaxwDiFu}
		\end{eqnarray}
with the density $n_0$, temperature $T$ and a shift in phase space by the bulk velocity $\vec{v}_\mathrm{B}$. We further assume that instantaneous observations with high spin-phase resolution are performed during various orbits in one observation season at the moments when the ecliptic longitude of the spin axis of the IBEX satellite is precisely equal to the ecliptic longitude of the Sun. We will refer to such conditions as the Exact Sun-Pointing (ES) conditions.

\begin{figure*}[t]
		\centering
		\begin{tabular}{ccc}	\includegraphics[scale=.55]{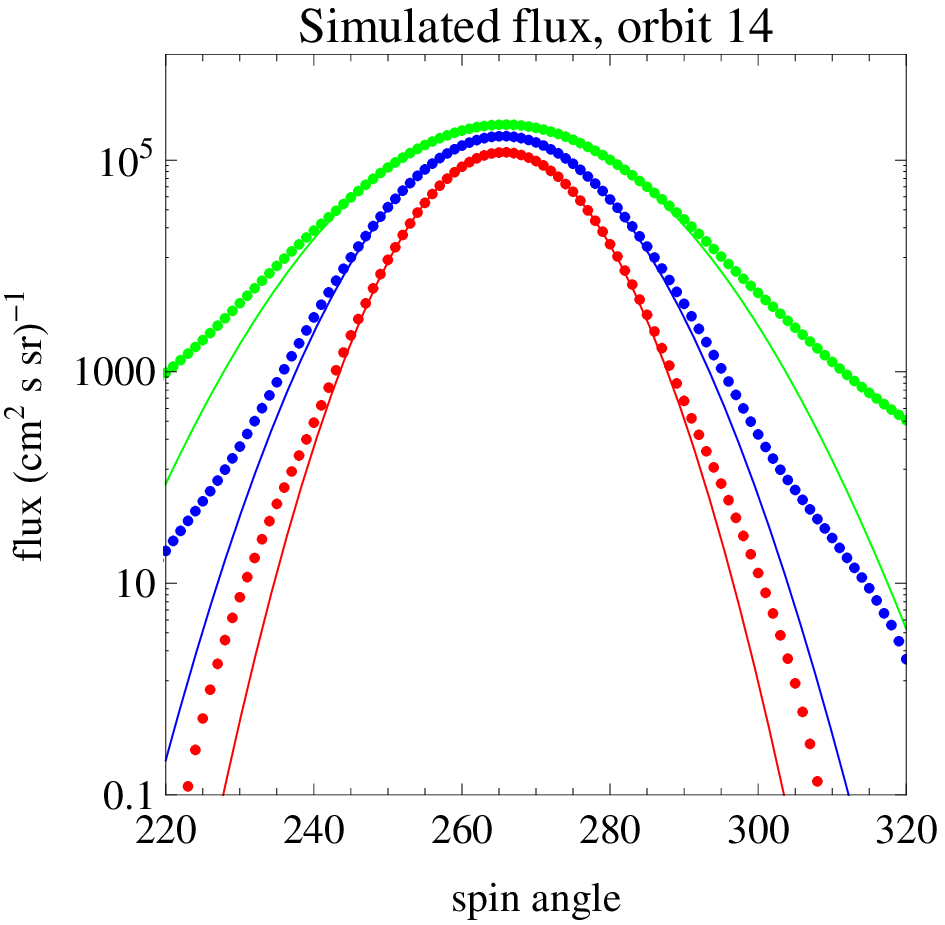}&\includegraphics[scale=.55]{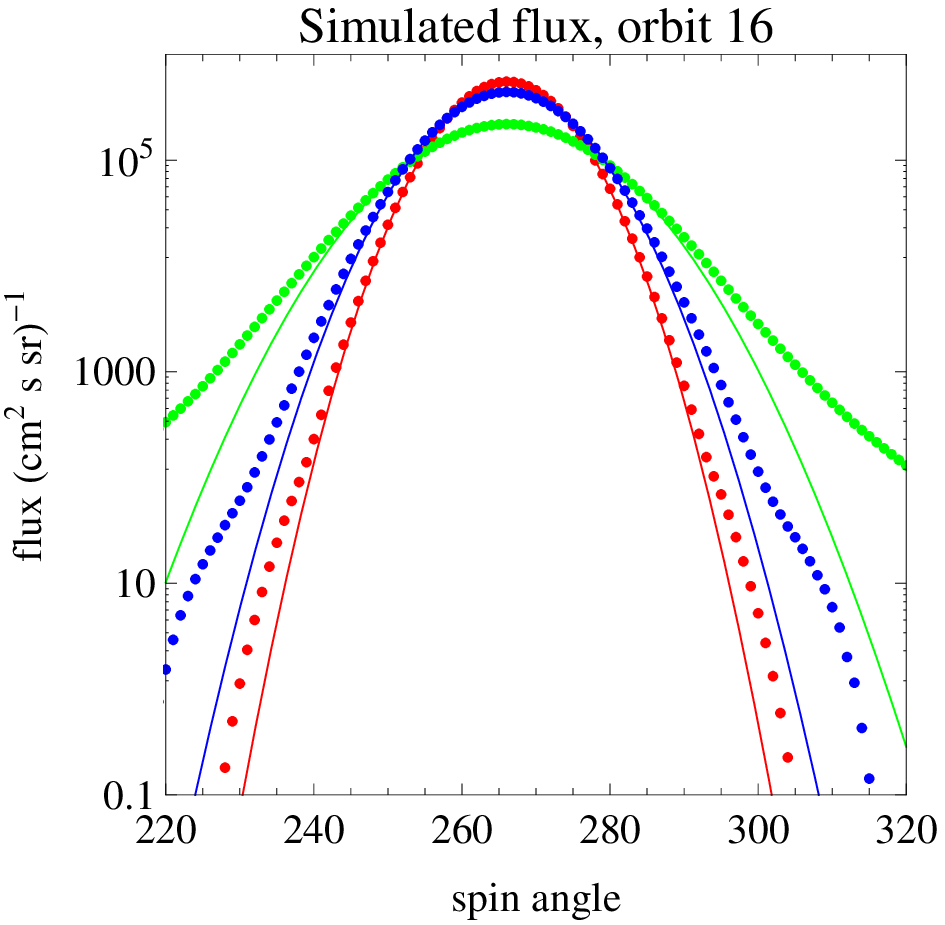}&\includegraphics[scale=.55]{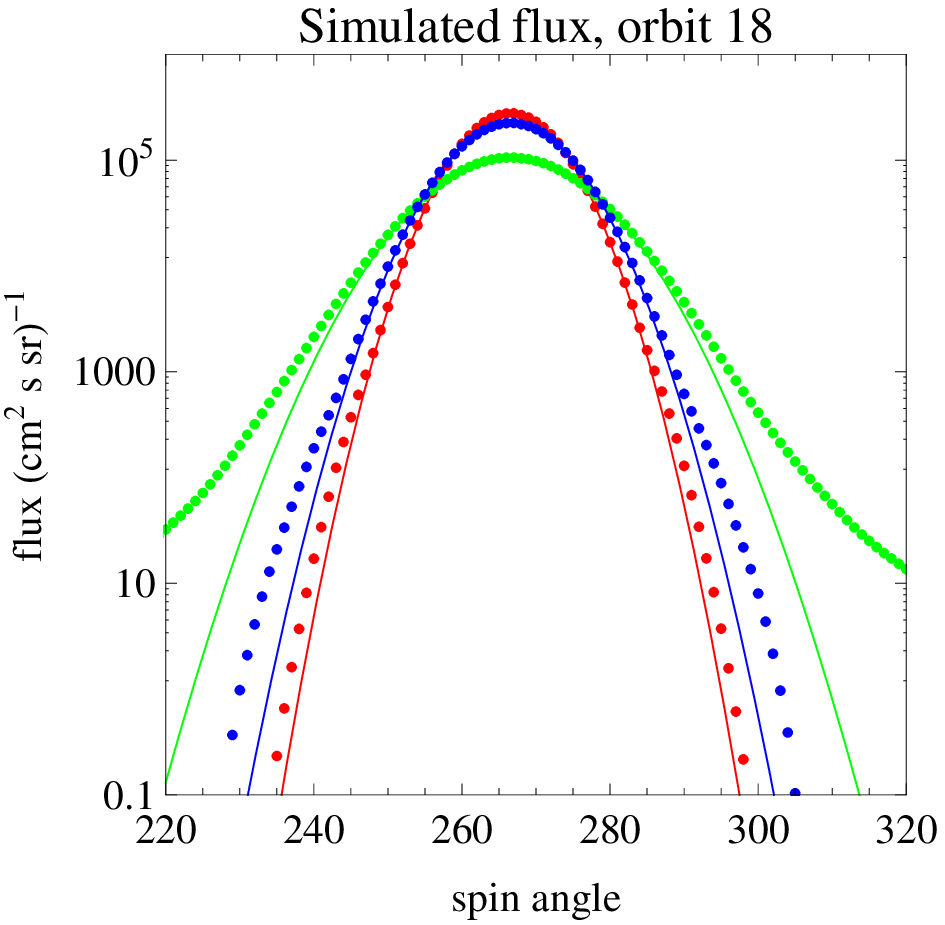}\\
		\end{tabular}
		\caption{Examples of simulated flux of the NISHe flow for IBEX-Lo during orbits 14 (before the passage of the flux maximum -- left-hand panel), 16 (at the orbit when the maximum of flux appears, middle panel), and 18 (after the passage through the flux maximum, right-hand panel). The dotted lines represent simulations results, solid lines represent fits of the simulations to the Gausian formula in Eq.~(\ref{eqGaussBeam}). A wide range of parameters for the NISHe gas were used for the simulations to demonstrate that, regardless of the parameter choice, the simulated NISHe beam observed by IBEX-Lo is composed of a Gaussian core and non-Gaussian wings. Specifically, the parameter sets shown are the following: $\lambda = {75.4}\degr$, $\beta = {-5.31}\degr$, $v = 26.4 \, \textrm{km s}^{-1}$, $T=6318$~K (red), $\lambda = {75.4}\degr$, $\beta = {-5.31}\degr$, $v = 18.744 \, \textrm{km s}^{-1}$, $T=10 000$~K (green), $\lambda = {79.0}\degr$, $\beta = {-5.20}\degr$, $v = 22.0 \, \textrm{km s}^{-1}$, $T=6318$~K (blue).}
		\label{figSimuGauss}
		\end{figure*}

Simulations performed for a number of parameter sets that covered the expected range of the parameters of the NISHe gas in the LIC suggest that at the orbits where the helium signal is expected to be the strongest (i.e., from orbit 13 to 20 and the equivalent ones during the later seasons) the observed count rate as function of spin angle $\psi$ can be approximated by a Gaussian core:
		\begin{equation}							
		F_{\mathrm{obs}}\left(\psi \right)=f_{0}\,\exp\left[-\left(\frac{\psi-\psi_{0}}{\sigma}\right)^2\right]
		\label{eqGaussBeam}
		\end{equation}
with elevated wings. This is illustrated in Fig. \ref{figSimuGauss} for 3 selected orbits and 3 different parameter sets. The parameters of the Gaussians (peak height $f_0$, peak width $\sigma$ and spin angle of the peak $\psi_0$) depend on the choice of parameters of the NISHe gas in the LIC, but the feature of a Gaussian core and elevated non-Gaussian wings is always present. The Gaussian core is a result of convolution of the true Gaussian signal with the near-Gaussian transmission function of the collimator. Fits of the Gaussian function to the simulation results showed that residuals of the fits within the Gaussian core region were below 1\%. Outside the Gaussian core region, whose span in the spin angle varied with assumed bulk velocity and temperature, the elevated non-Gaussian wings were visible in the residuals as power-law increase in the residuals magnitudes. They were present for the collimator-integrated flux values $F_{\mathrm{He}}\left(\psi\right) \lesssim 0.01 F_{\mathrm{He}}\left(\psi_{max}\right)$, where $\psi_{\mathrm{max}}$ is the spin phase angle of the peak flux, as illustrated in Fig.~\ref{figSimuGauss}.

		\begin{figure*}[t]
		\centering
		\begin{tabular}{ccc}
		\includegraphics[scale=0.85]{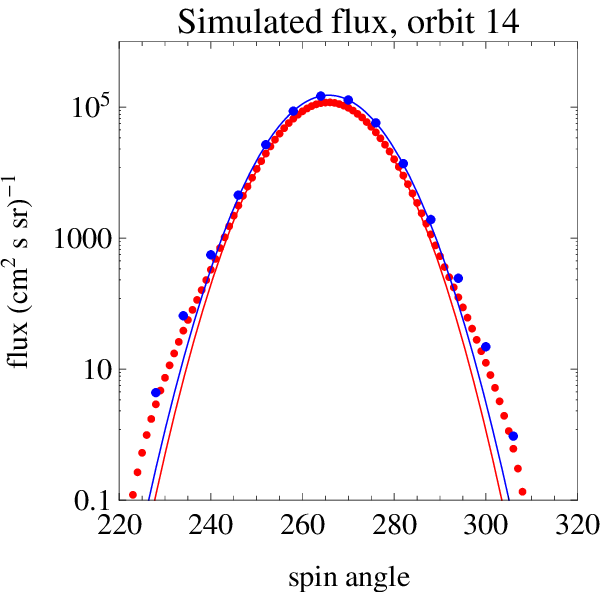}&\includegraphics[scale=0.85]{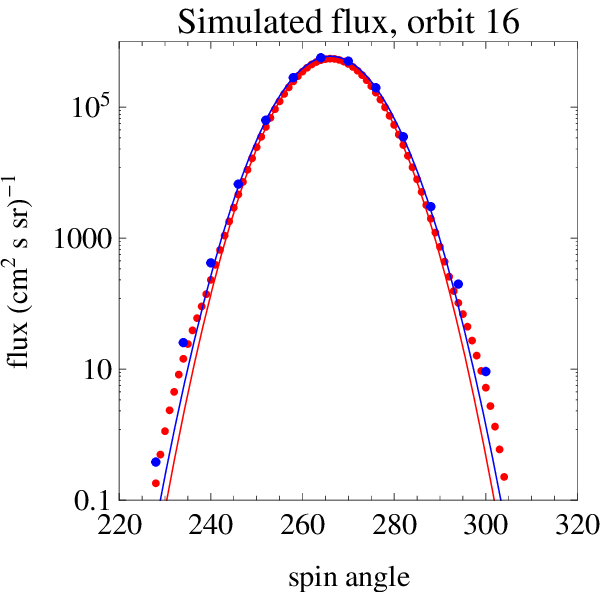}&\includegraphics[scale=0.85]{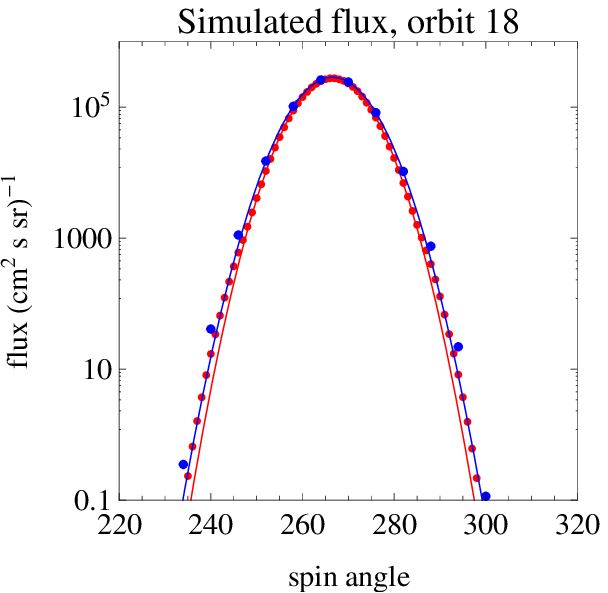}\\
		\end{tabular}
		\caption{Effect of the width of the binning in spin phase of the simulated NISHe flux at orbits 14 (left panel), 16 (middle panel), and 18 (right panel). Red dots are simulation results averaged over the select ISM flow observation times with the flux binned at $1\degr$ resolution and thick blue dots are for simulations binned $6\degr$. The lines are the Gaussian formula given in Eq (\ref{eqGaussBeam}) fitted to the simulations.}
		\label{fig1DegVs6Deg}
		\end{figure*}
		\clearpage
		
		\begin{figure*}[t]
		\centering
		\epsscale{2}
		\plottwo{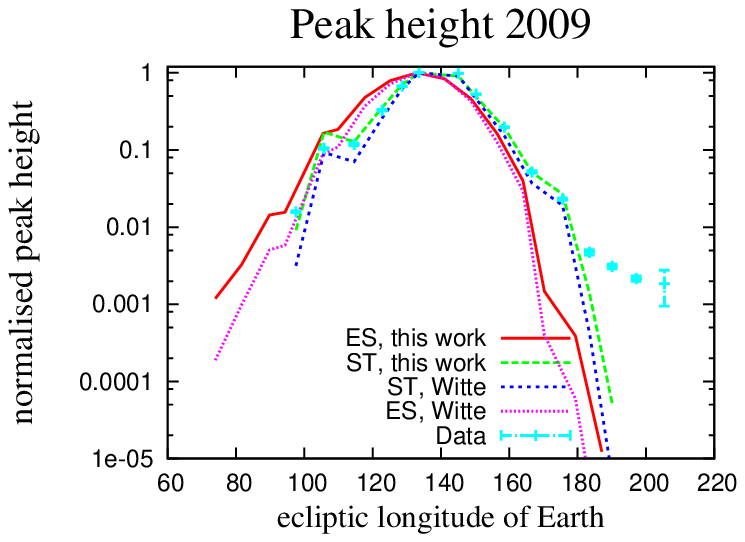}{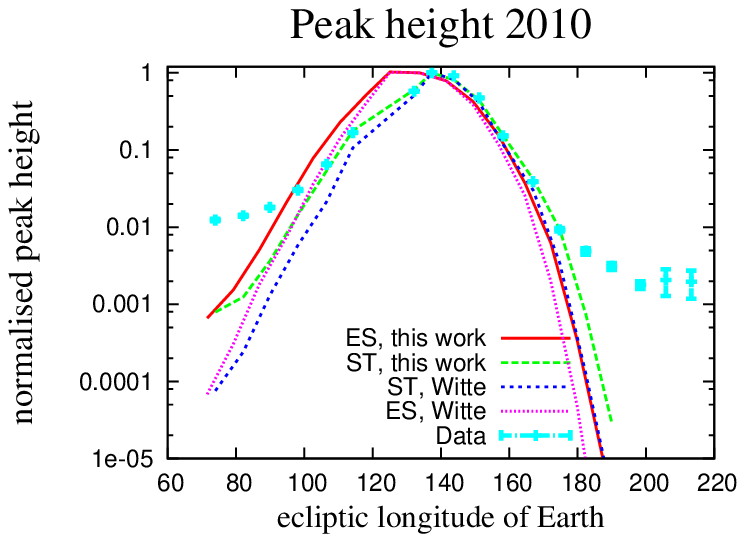}\\
	  \epsscale{2}
	 	\plottwo{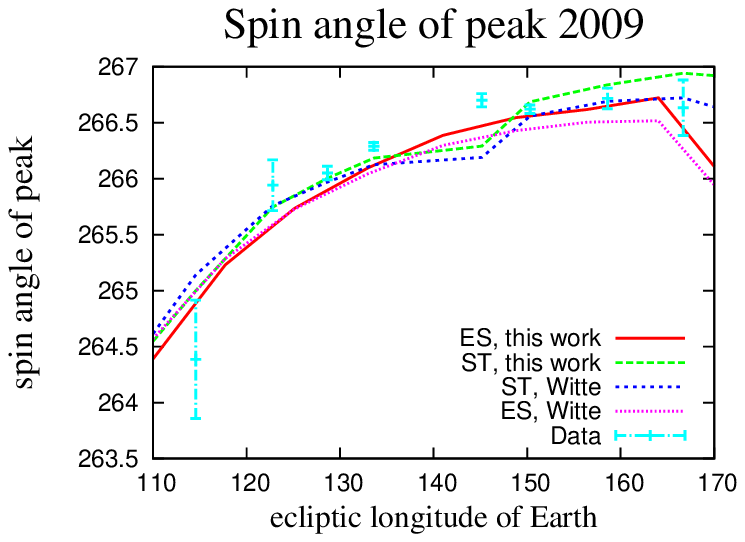}{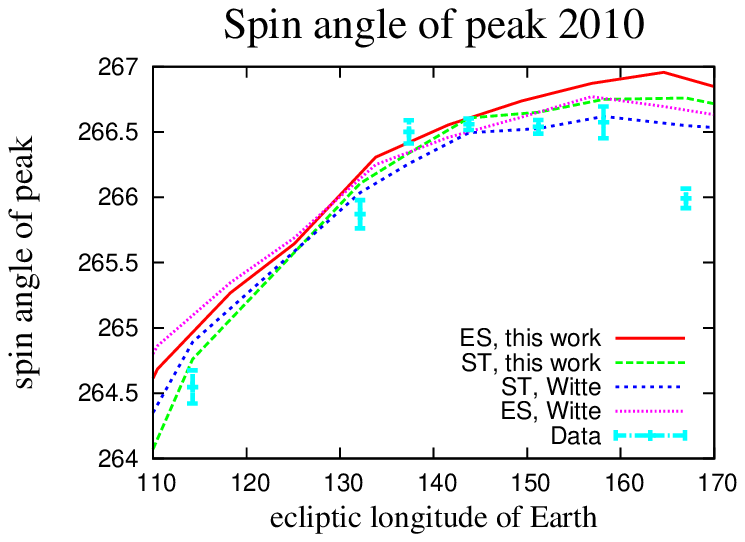}\\
		\epsscale{2}
		\plottwo{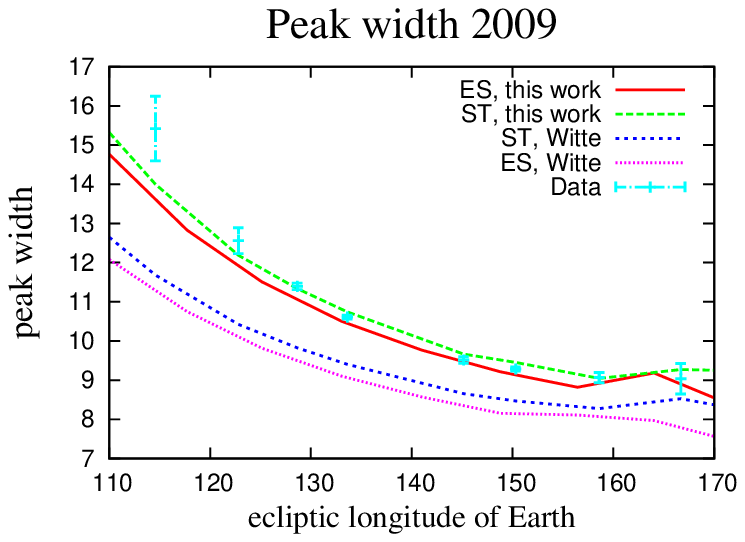}{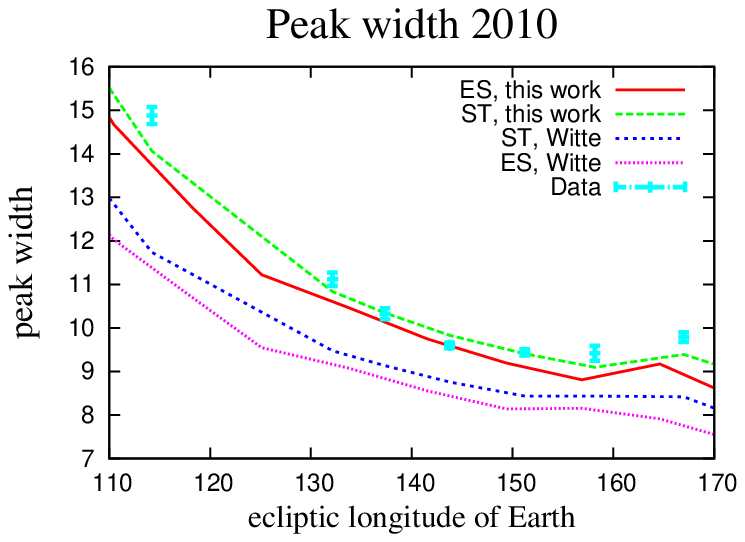}\\
		\caption{Parameters of the NISHe beam: peak height (upper row), peak position (middle row), and peak width (lower row) during the 2009 season (left column) and 2010 season (right column). Beam parameters for the Exact Sun-Pointing longitude of the spin axis (the ES best fit case, red) differ from beam parameters averaged over select ISM flow observation times (best fit case, green). The simulations in the ES and select ISM flow times cases are shown for comparison as dotted purple and blue lines, respectively. Cyan dots with error bars show beam parameters of the data averaged over the select ISM flow observation times. Peak heights are normalized to values for the 16-th and 64-th orbits for the 2009 and 2010 seasons, respectively. Step-like features in the peak heights visible during the 2009 seasons both in the observations and simulations are due to characteristics of the spin axis pointing.}
		\label{figESvsGT}
		\end{figure*}
		\clearpage

We also found that for orbits earlier than 12 (and equivalent in 2010) the simulated collimator-averaged signal increasingly deviates from the Gaussian shape with the decrease of Earth's ecliptic longitude. The flux profiles as function of spin angle become increasingly asymmetric relative to the peak even though the Maxwellian distribution function in the LIC is assumed, as shown by the blue line in the upper left panel of Fig.~\ref{figHvsHeFlux} (orbit 11). Despite the non-Gaussianity of the profiles, their peaks are well defined and can be easily compared with observations. Such comparisons were in fact done and used as basis to formulate the hypothesis that the excess signal observed by IBEX at these Earth's longitude interval is due to an additional population of neutral He in or near the heliosphere. 

We further verified that binning data into $6\degr$ bins does not remove the Gaussian character of the signal, as shown in Fig.~\ref{fig1DegVs6Deg}. Similarly, averaging of the signal over the entire duration of the select ISM flow observation times maintains the Gaussian shape, but the parameters of the Gaussians (peak height, peak width and peak location) are changed, as illustrated in Fig. \ref{figESvsGT}, where simulations performed for the ES conditions are compared with simulations performed for the actual select ISM flow observation intervals.

\begin{figure}[t]
		\centering
		\epsscale{1}
		\plotone{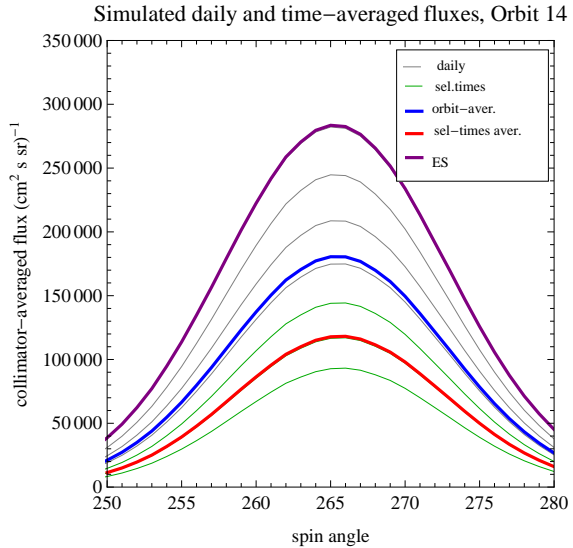}
		\caption{Simulated collimator-averaged flux of NISHe gas at IBEX-Lo, Orbit 14. Thin lines correspond to the flux at midnight for each day during orbit Science Operations. The gray color marks the days outside the select ISM flow observation times, green marks the days within these times. The flux systematically decreases with time over the orbit. Thick blue line marks the average flux over the entire duration of Science Operations and the thick red line marks the average flux over the select ISM flow observation times only. The thick purple line marks the flux for the instant when the ecliptic longitude of the spin axis is exactly equal to the longitude of the Sun (the ES conditions). The parameters of the NISHe gas from \citet{witte:04} were used in the simulations.}
		\label{figDailyOrb14}
		\end{figure}

\begin{figure}[t]
		\centering
		\epsscale{1}
		\plotone{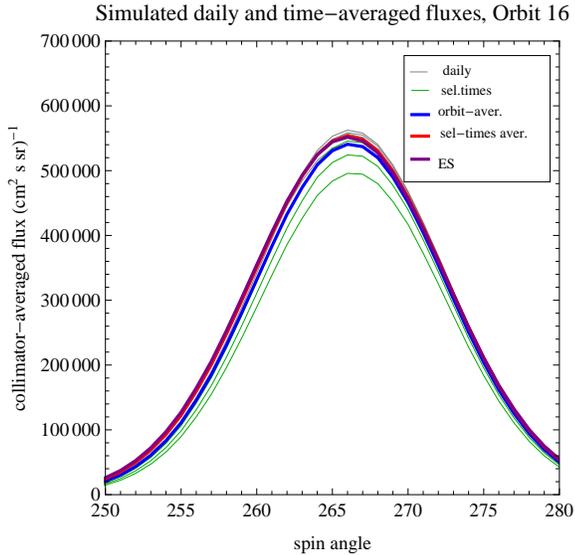}
		\caption{Simulated collimator-averaged flux of NISHe gas at IBEX-Lo, Orbit 16. The color/line style code and the parameter set used in the simulations are the same as in Fig. \ref{figDailyOrb14}.}
		\label{figDailyOrb16}
		\end{figure}

\begin{figure}[t]
		\centering
		\epsscale{1}
		\plotone{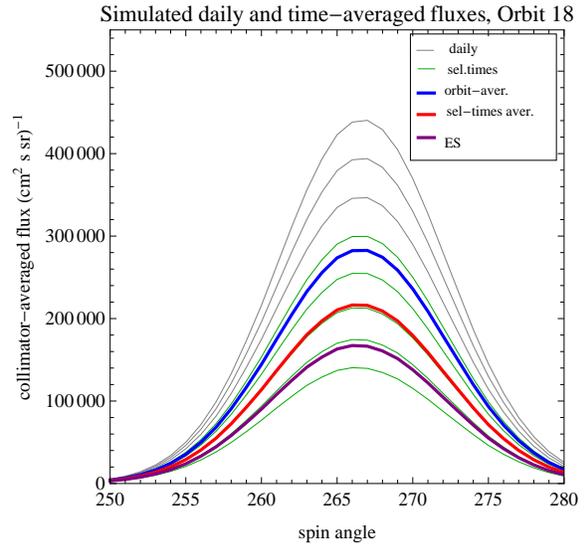}		
		\caption{Simulated collimator-averaged flux of NISHe gas at IBEX-Lo, Orbit 18. The color/line style code and the parameter set used in the simulations are identical as in Fig. \ref{figDailyOrb14}. The flux systematically increases with time.}		
		\label{figDailyOrb18}
		\end{figure}

The reason for the differences between the select ISM flow observation times and ES beam parameters is that because the spin axis of the spacecraft, which is never aligned with the Sun,  does not change during an orbit \citep{scherrer_etal:09a, hlond_etal:12a}, the beam of the NISHe gas, which in the solar inertial frame is invariant relative to the distant stars, wanders through the FOV of the sensor, changing gradually its angular size, peak location, and height. This effect is especially visible in the orbits before or after orbits 16 and 64 and is illustrated in Figs~\ref{figDailyOrb14}, \ref{figDailyOrb16}, and \ref{figDailyOrb18}. These figures demonstrate the importance of an exact determination of select ISM flow observation times in order to have a faithful representation of the data in the simulations. These figures also demonstrate why \citet{mobius_etal:12a} had to extrapolate their observations to the ES conditions for the comparison with their analytic model. 

If the select ISM flow observation times extended over the entire orbit, then IBEX-Lo would have observed daily fluxes marked by the thin lines, which, when averaged, would equal the thick blue lines. However, these times do not extend over the entire orbit. In Fig. \ref{figDailyOrb14}, the flux of the incoming interstellar He atoms is most intense during the first days of the orbit and with time the beam moves away from the field of view of the collimator. Since the select ISM flow observation times cover the last portion of the orbit, a lower average flux is observed, as illustrated with the thick red line. However, the spin axis pointed toward the Sun at the beginning of the orbit, so the flux relevant for the ES conditions, marked with the thick purple line, is much higher than the flux actually measured. 

During orbit 16 (Fig. \ref{figDailyOrb16}) IBEX observed the peak NISHe flux. The select ISM flow observation times occur during the $\sim 3$ days at the beginning of the orbit. However, since this is the peak flux and the NISHe beam is directed into the sensor, the flux varies little with time and the observed mean flux is very similar to the flux averaged over the entire Science Operations for this orbit. Thus in Fig. \ref{figDailyOrb16} the ES flux (purple), the observed average flux (red) and the orbit average flux (blue) are very similar. 

In Orbit 18 (Fig. \ref{figDailyOrb18}) IBEX is beyond the peak NISHe flux and viewing the beam edge. For this orbit, select ISM flow observation times occurred during the first $\sim 5$ days of Science Operations. The beam moves into the field of view near the middle of the orbit, but IBEX views the beam when it is off the peak and the average flux is lower for the selected times than for the full orbit. The spin axis pointed exactly to the Sun at the beginning of the orbit, so the flux at the ES time is lower than the flux averaged over the Select Times and lower than the flux averaged over the entire orbit. 

From this analysis we conclude that the portions of the observed count rates that are Gaussian in shape correspond to the NISHe population from the LIC and the portions that cannot be fitted by a Gaussian must correspond to something different, probably another source of neutral helium in or near the heliosphere. In either case, these non-Gaussian components are eliminated from the analysis of the NISHe population for now. We also conclude that care must be taken to accurately reproduce the flux observed during select ISM flow observation times, especially for orbits that are not near the peak flux.

	\subsection{Role of spin axis pointing, IBEX orbital motion, and ellipticity of Earth's orbit }
Finally, before starting the parameter fitting procedure, we discuss miscellaneous effects that should be included in the simulations. These effects are listed at the beginning of Section 2. 

The ellipticity of Earth's orbit results in a small deflection of the direction of the Earth velocity vector from the right angle to the Earth radius vector, which slightly modifies the aberration of the NISHe beam. Further, an additional change in the aberration and relative velocity of the beam and the detector is caused by the small radial component of the Earth's velocity (on the order of $1\; \mathrm{km} \, \mathrm{s}^{-1}$). Also of the order of a few km~s$^{-1}$ is the proper motion of IBEX relative to the Earth. In the simulations we used actual Earth ephemeris, which accounts for the Earth location. The total velocity vector of the spacecraft plus the Earth is accounted for by using the proper motion of the satellite along its orbit and the total velocity vector of the Earth.

\begin{figure*}[t]
		\centering
		\epsscale{1.9}
		\plottwo{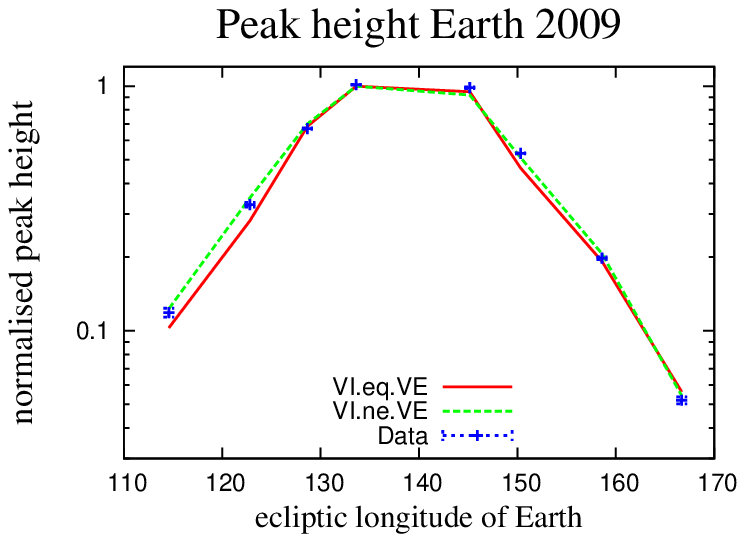}{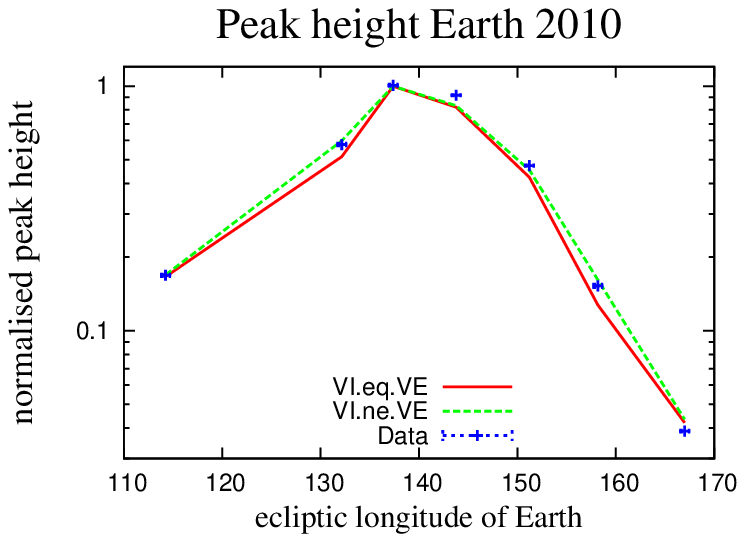}\\
		\epsscale{2}
		\plottwo{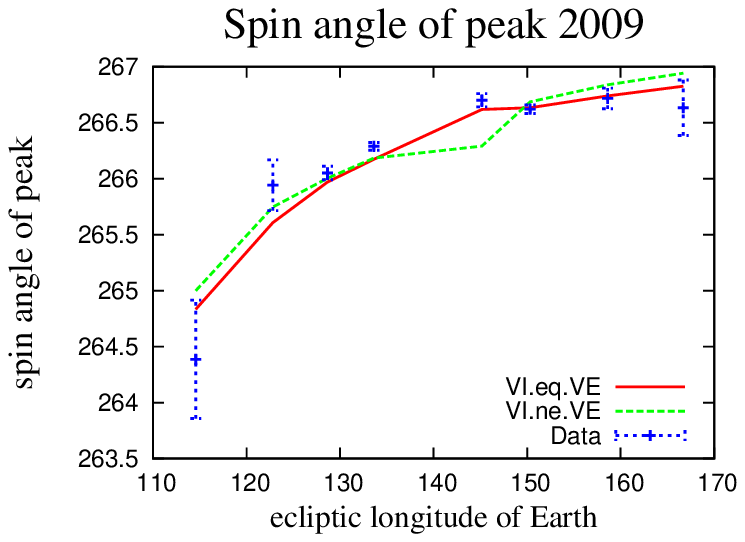}{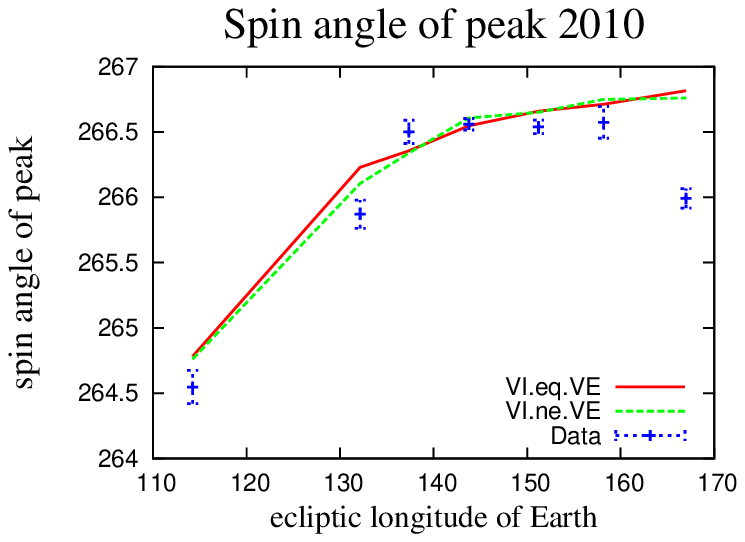}\\
		\epsscale{2}
		\plottwo{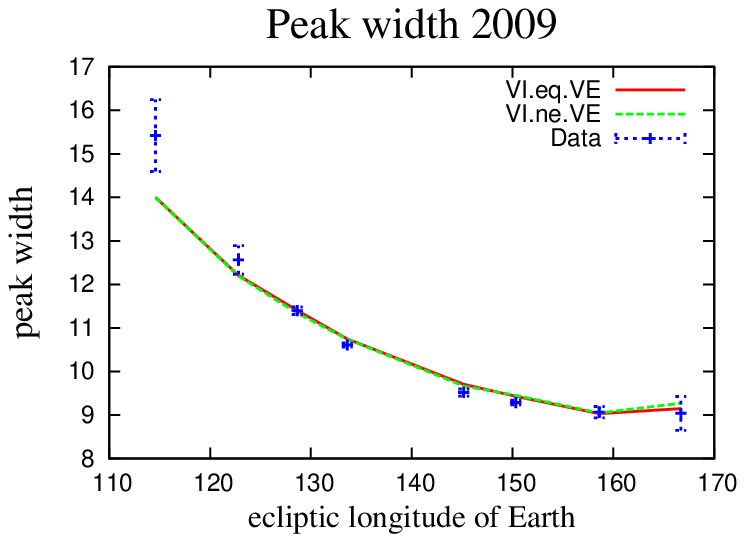}{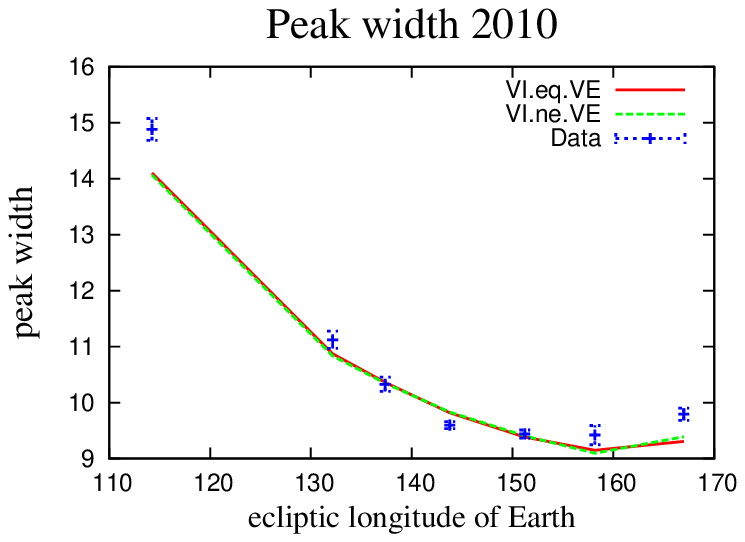}\\
		\caption{Illustration of the effect of the proper motion of IBEX on the simulated observations of NISHe gas for the select ISM flow observation times for 2009 (left column) and 2010 (right column). Shown are peak heights (upper panel), peak spin angle (middle panel), and peak width (lower panel). The results of simulations performed assuming the actual IBEX velocity relative to the Sun are in green, while the simulations for the IBEX velocity assumed to be equal to the Earth velocity are in red. Observed values are the blue dots with error bars. The exact magnitude of this proper motion effect depends on the duration of the select ISM flow observation times and their position on the orbit. Shown are simulations for the parameters of the NISHe flow as established in this paper.}
		\label{figIBEXMotion}
		\end{figure*}

As shown in Fig. \ref{figIBEXMotion}, the IBEX motion relative to the Earth has its strongest effect on the peak height of the observed NISHe beam. Only the peak height effect exceeds the measurement uncertainty. The effect on peak width is, understandably, negligible, and the effect on peak spin angle is comparable to the measurement uncertainty. Since the effect on the magnitude of the flux cannot be neglected, the satellite proper motion was included in the simulations. The velocity vector of the spacecraft in the inertial frame of the Sun was taken as a vector sum of the Earth velocity relative to the Sun and IBEX's velocity about the Earth, and was calculated using the software developed by the ISOC \citep{schwadron_etal:09a} based on the SPICE toolkit \citep{acton:96a}. It should be noted that the aberration effect is stronger during the ES time for each orbit, because that occurs during the ascent of IBEX to apogee when the spacecraft speed is still substantial and therefore the effect is also to be taken into account in the analysis by \citet{mobius_etal:12a}.

The simulations shown in Fig.~\ref{figIBEXMotion} were done for the NISHe flow parameters established in this paper based on fitting of the model with all the effects included. It is not surprising then that the simulations without the IBEX orbital velocity fit the data less well. Since we know that IBEX is moving in its orbit and we know from Fig.~\ref{figIBEXMotion} that the influence of this effect on the observed fluxes is small, but not negligible, we include these effects in the simulations.  

\begin{figure*}[t]
		\centering

		\plottwo{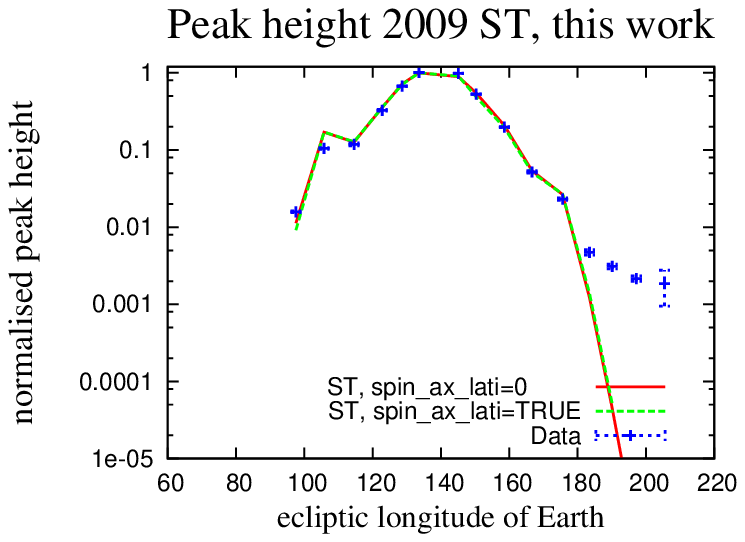}{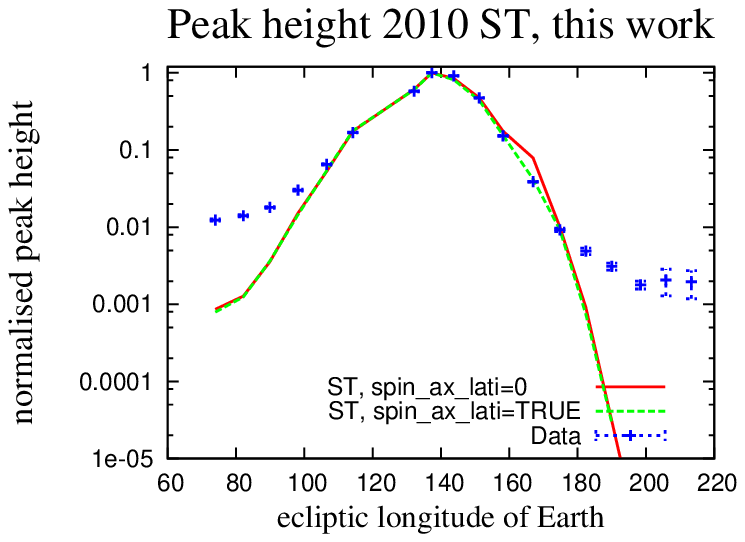}\\

		\plottwo{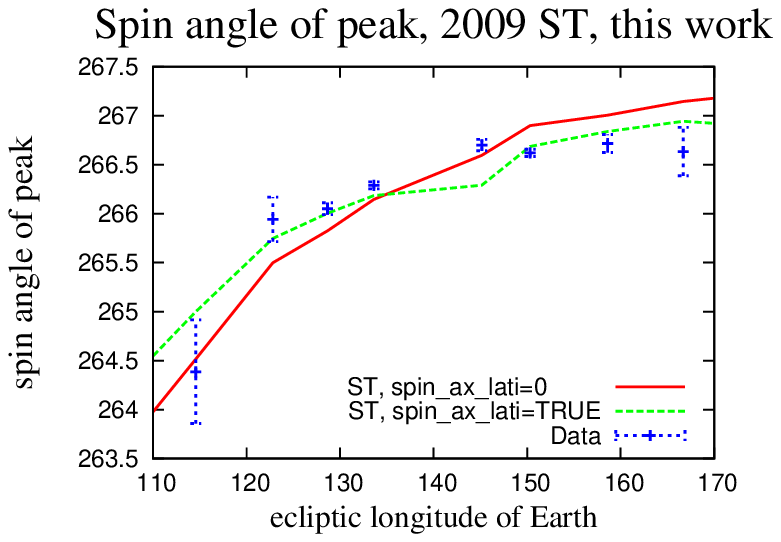}{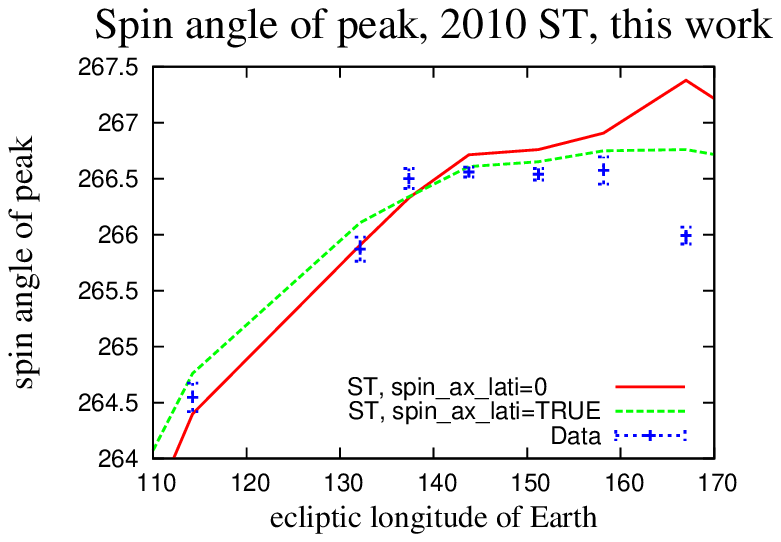}\\

		\plottwo{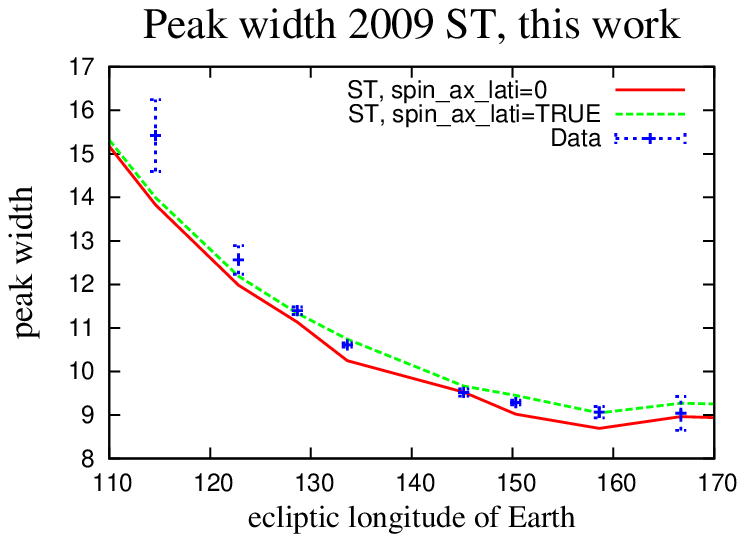}{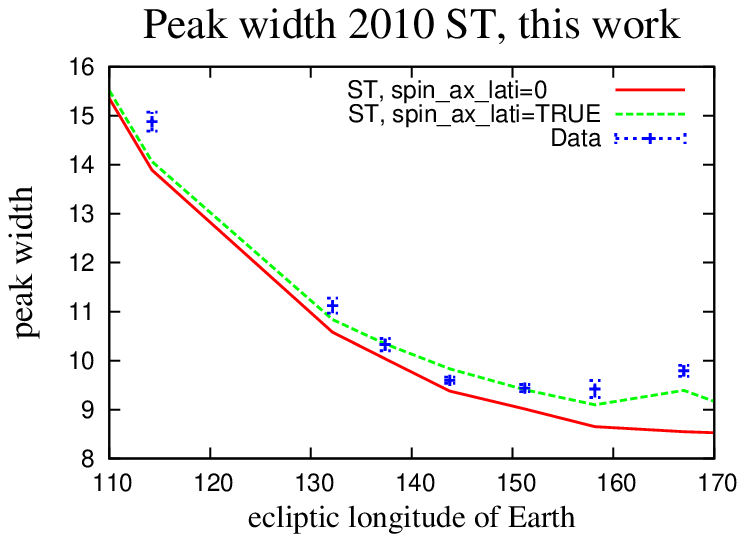}\\
		\caption{Influence of the IBEX spin axis latitude on the observed flux. Shown are the peak heights (upper panel), peak spin angle (middle panel), and peak width (lower panel) for the 2009 (left-hand column) and 2010 (right-hand column) observing seasons. The simulations were done for Select Observations Times and the best fitting parameter set as established in this paper and for the IBEX spin axis latitude either as provided by ISOC (see Fig. \ref{figAxisPointing}) or assumed to be 0. Blue dots with error bars represent the beam parameters obtained from the data.}
		\label{figSpinAxEffect}
		\end{figure*}

The small tilt of the spin axis out of the ecliptic also affects the observed flux because it excludes a small portion of the beam while accepting a different part compared to the situation when the spin axis is exactly in the ecliptic plane. This effect is especially pronounced in early orbits before the crossing of the ISM flow peak (Fig.~\ref{figSpinAxEffect}). Both the width and spin phase of the peak maximum are affected with offsets clearly larger than the error bars. In contrast, the peak height is only weakly affected.

The exact magnitude of the effects discussed in this section depend on the details and are challenging to plan in advance, i.e., on the actual select ISM flow observation times, which are determined by a combination of operational aspects, stochastic backgrounds, particle events, actual realizations of spin-axis repointing maneuvers, etc. We decided that instead of attempting to correct the observations for all of these issues, it was better to simply include them in the simulation. We emphasize that the magnitude of various effects can vary depending on the adopted parameter set and also from orbit to orbit. This supports the decision to complicate the simulation pipeline for the sake of fidelity of the model rather than try to correct the observations.

		\clearpage

		\clearpage

		\section{Data}
Observations used in this analysis are discussed by \citet{mobius_etal:12a} and the ground calibration of the IBEX-Lo instrument by \citet{mobius_etal:09a, bochsler_etal:12a, saul_etal:12a}. We used data collected in Energy Step 2 (center energy 27 eV) of IBEX Lo (hydrogen). The hydrogen atoms that are observed were sputtered off the conversion surface of the IBEX-Lo instrument by the incoming NISHe atoms. The fact that the observed signal is actually due to helium was verified by comparing the H to O ratio observed in flight with the ratio observed in laboratory calibration using a neutral helium beam of the same energy as the NISHe beam. 

To compare with simulations, counts $c_k$ accumulated at an orbit $k$ during select ISM flow observation times $\Delta T_{ki}$ in the $6\degr$ bins were converted into averaged count rates $d_k$ using the following relation:
		\begin{equation}							
		d_k=8\times60\frac{c_k}{\sum \limits_{i=1}^{N_k}{\Delta T_{ki}}}
		\label{eqCounts2CntRates}
		\end{equation}
where the sum in the denominator is the total length of the $N_k$ intervals of select ISM flow observation times at the $k$-th orbit and the $8 \times 60$ factor reflects the fact that IBEX-Lo observes at 8 energy channels (thus 1 channel is active for $1/8$-th of the time) and each of the 60 $6\degr$ bins is observed during $1/60$-th of the time.

The data counts are subject to the Poisson statistics with uncertainties of square root of the total counts registered in a given data bin. Statistical errors in counts are converted into the errors in count rates using Eq.~(\ref{eqCounts2CntRates}). 

Before starting the search for flow parameters of NISHe we performed data selection based on insight obtained from the modeling. Analysis of the expected NISHe beam peak heights as function of the ecliptic longitude of IBEX showed that for no reasonable set of parameters we are able to reproduce the peak heights in the orbits before Orbit 60 during the 2010 season. Orbits 11 and 12 from the 2009 season showed a similar behavior as illustrated in the upper-right panel of Fig.\ref{figSpinAxEffect}.  Thus we concluded that the flux observed at these orbits must have a strong component different from the NISHe gas and removed these orbits for later, separate analysis. Similarly, profiles of the count rates from orbits 21 and 69 could not be fitted and a similar conclusion was adopted, supported by the predicted presence of a component from neutral interstellar hydrogen, confirmed by \citet{saul_etal:12a}. Consequently, we were left with orbits 13--20 from the 2009 season and 60, 61, and 63--68 from the 2010 season. Regrettably, there are no data from orbit 62 because of a spacecraft reset. 

\begin{figure*}[t]
		\centering
		\begin{tabular}{ccc}
		\includegraphics[scale=.5]{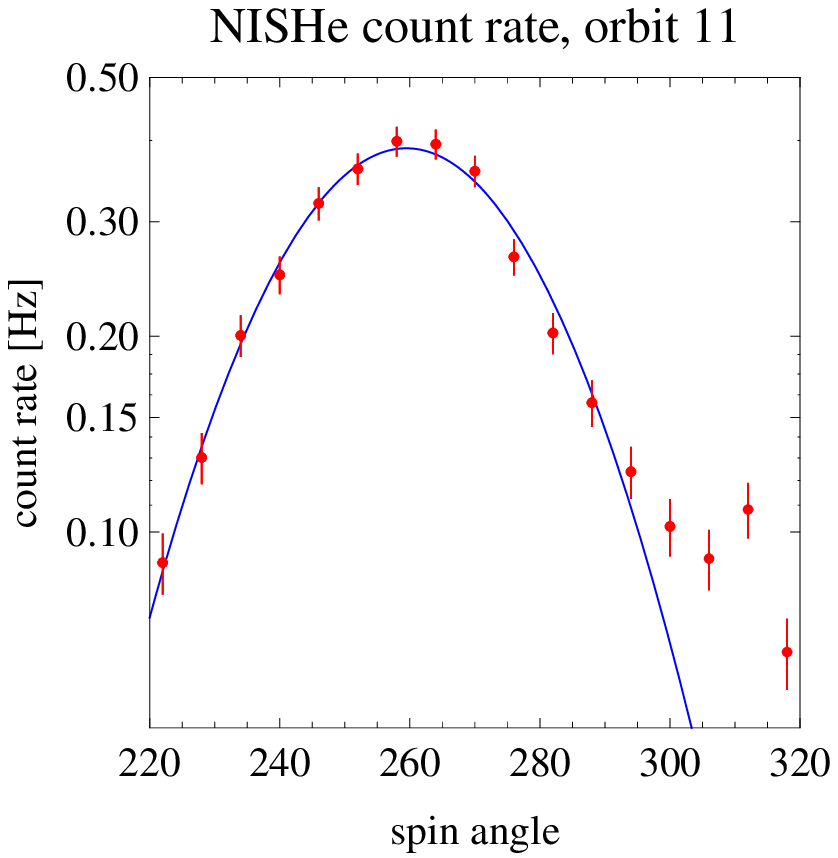}&\includegraphics[scale=.5]{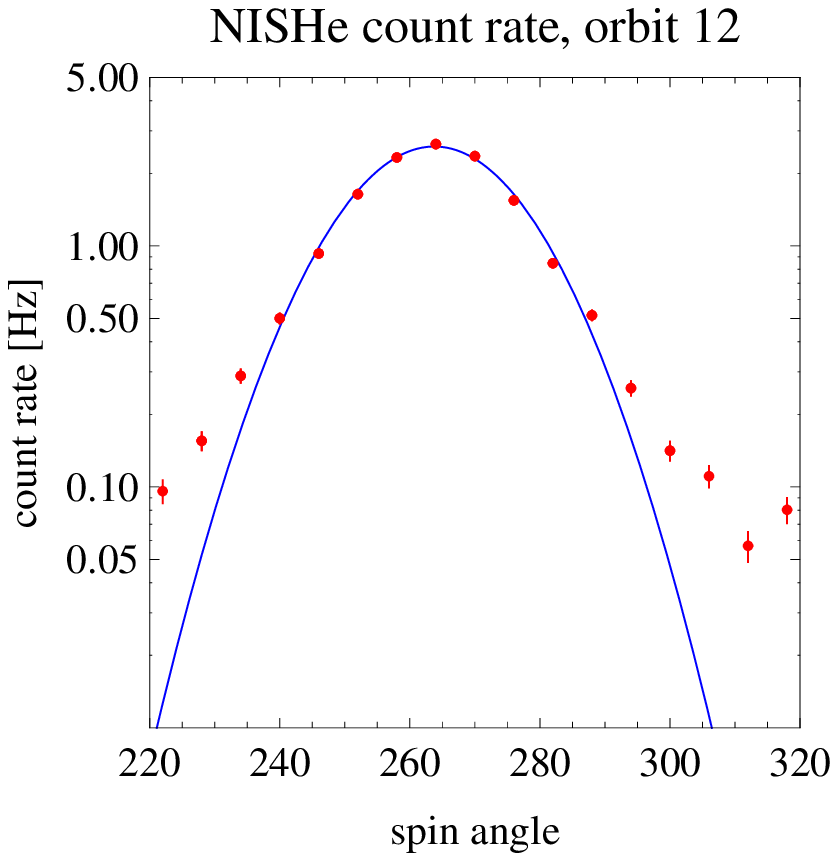}&\includegraphics[scale=.5]{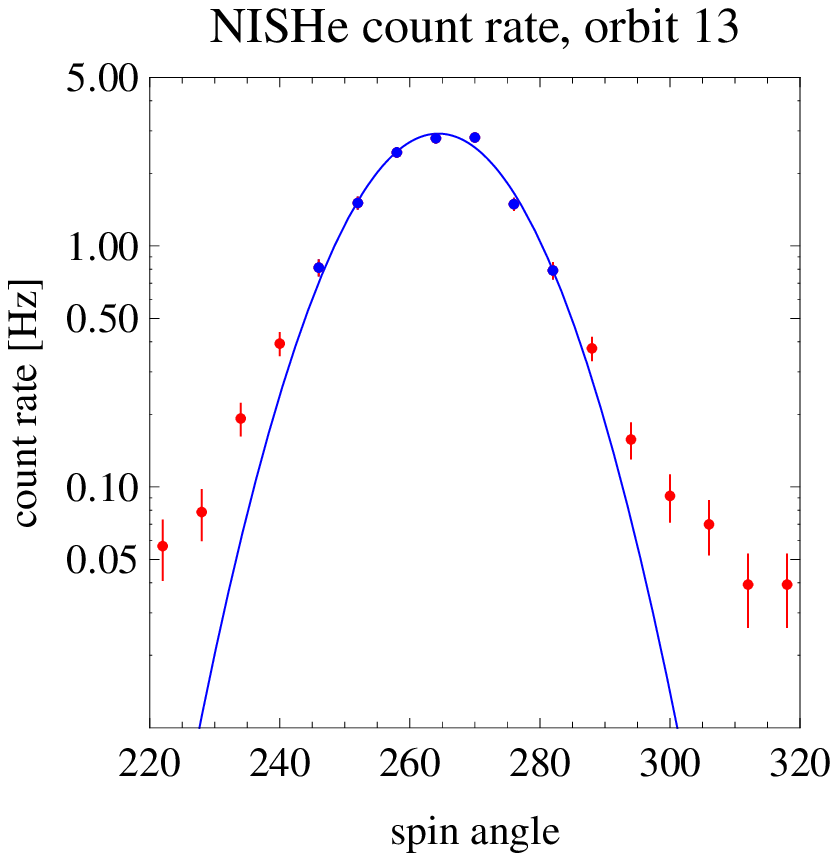}\\
		\includegraphics[scale=.5]{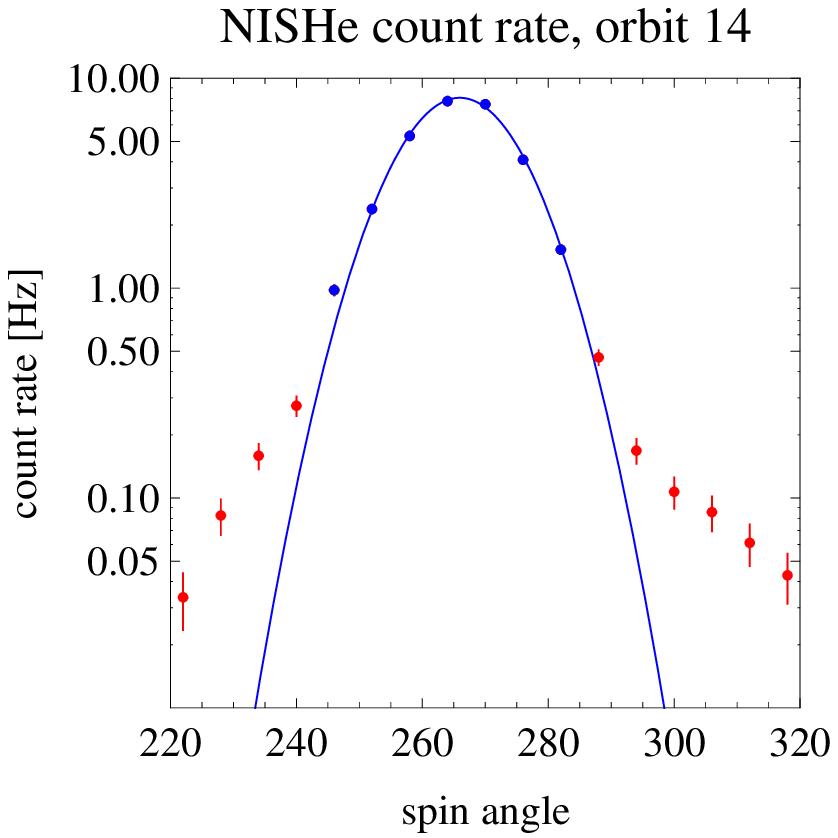}&\includegraphics[scale=.5]{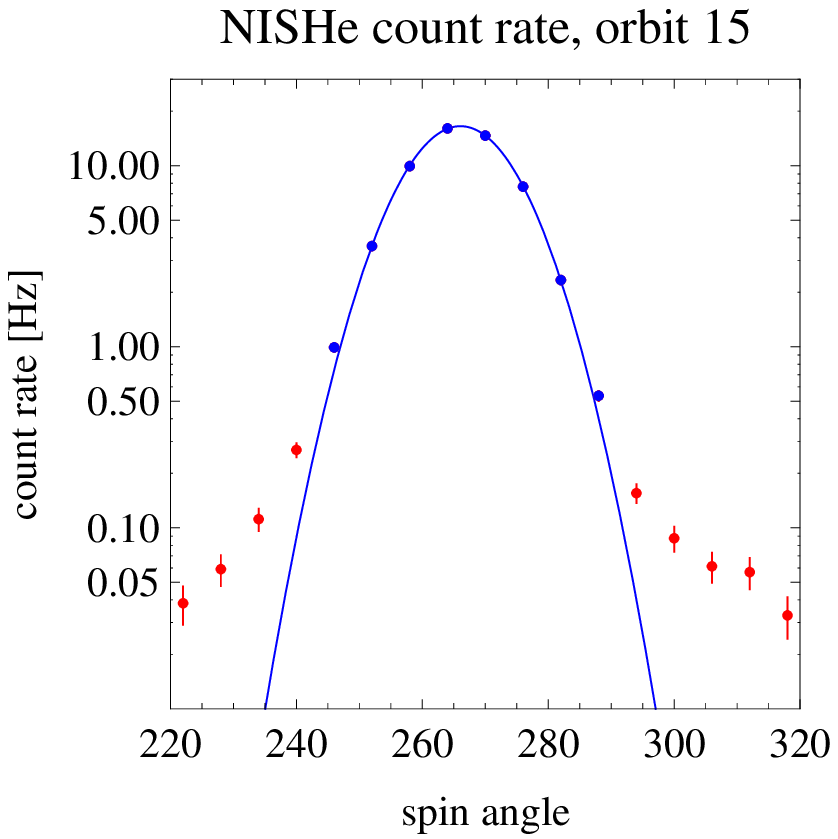}&\includegraphics[scale=.5]{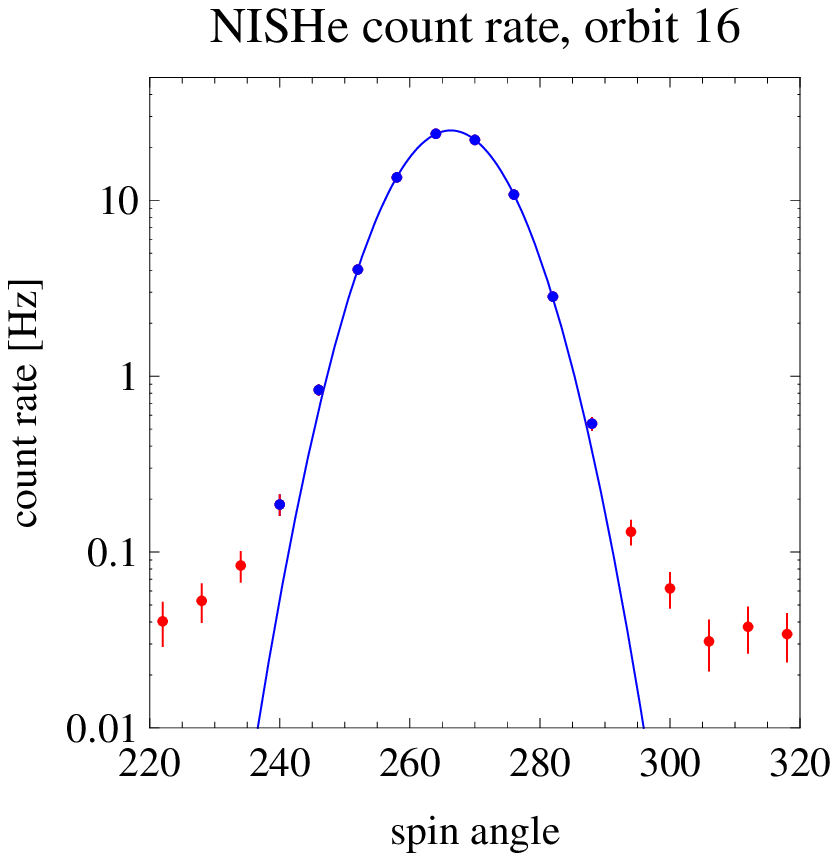}\\
		\includegraphics[scale=.5]{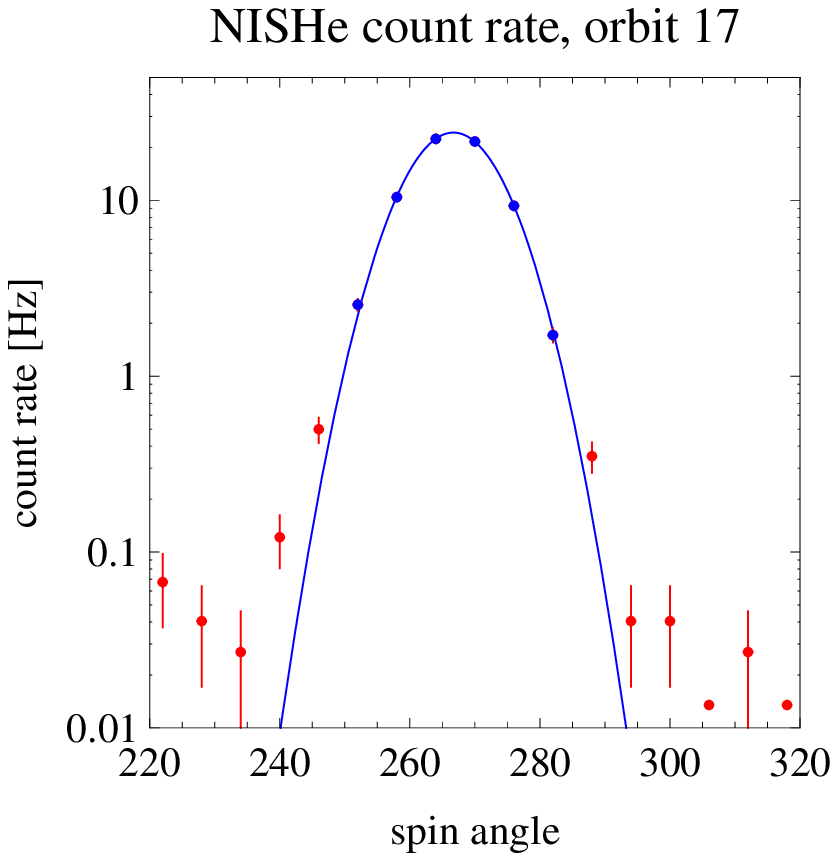}&\includegraphics[scale=.5]{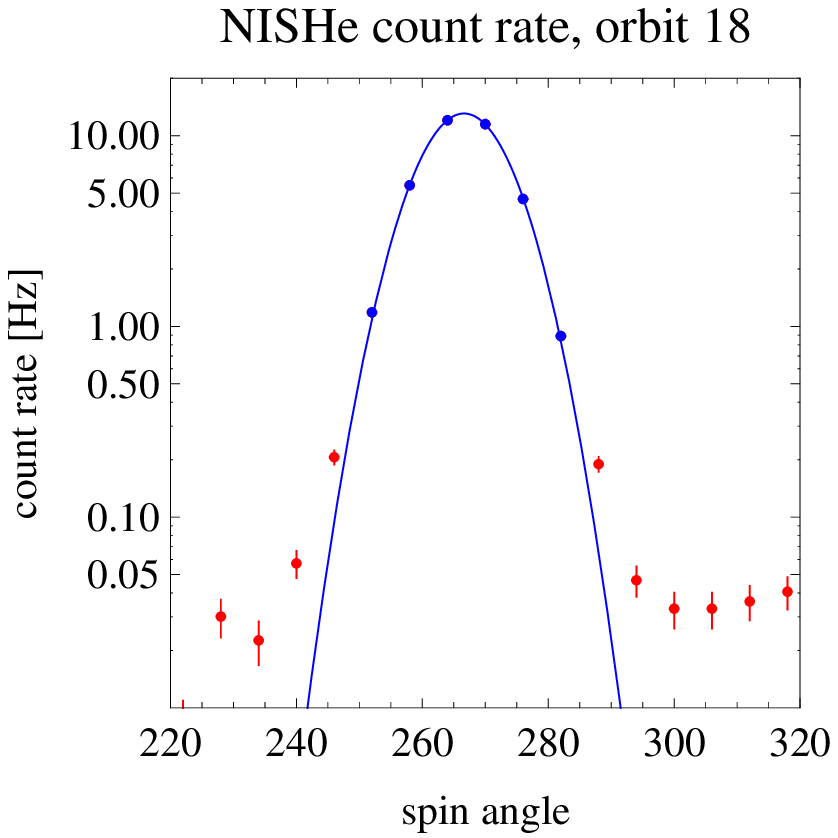}&\includegraphics[scale=.5]{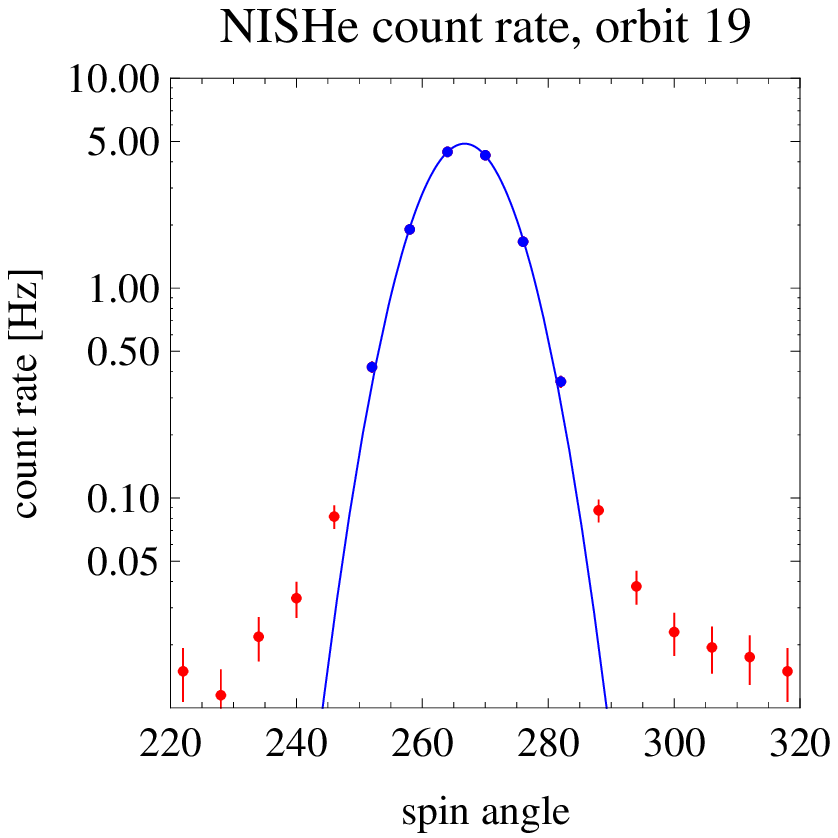}\\
		\includegraphics[scale=.5]{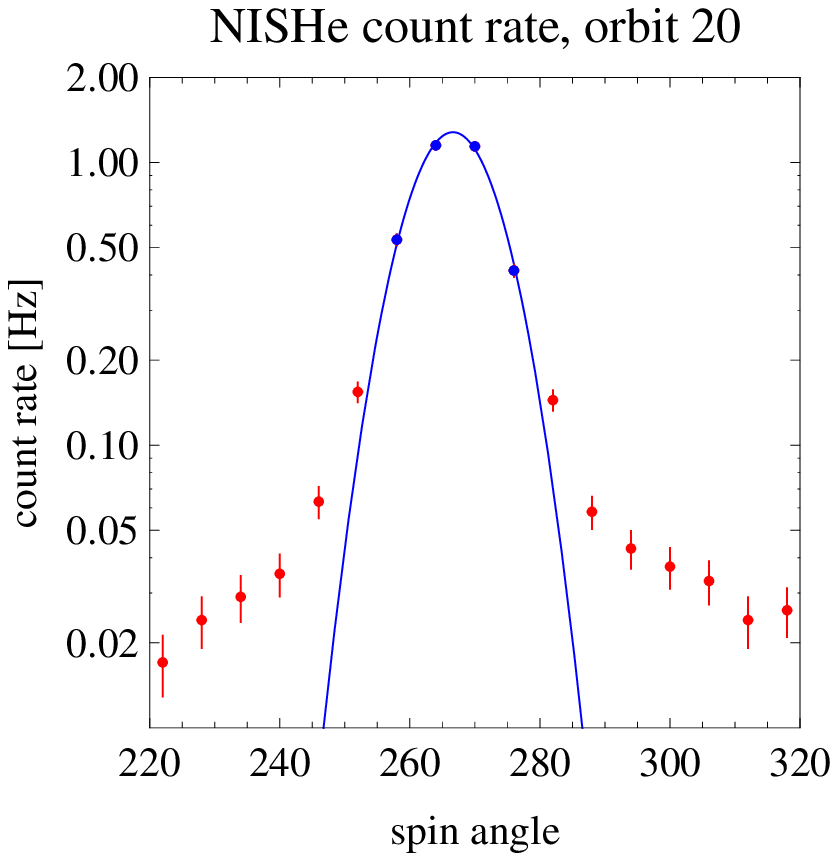}&\includegraphics[scale=.5]{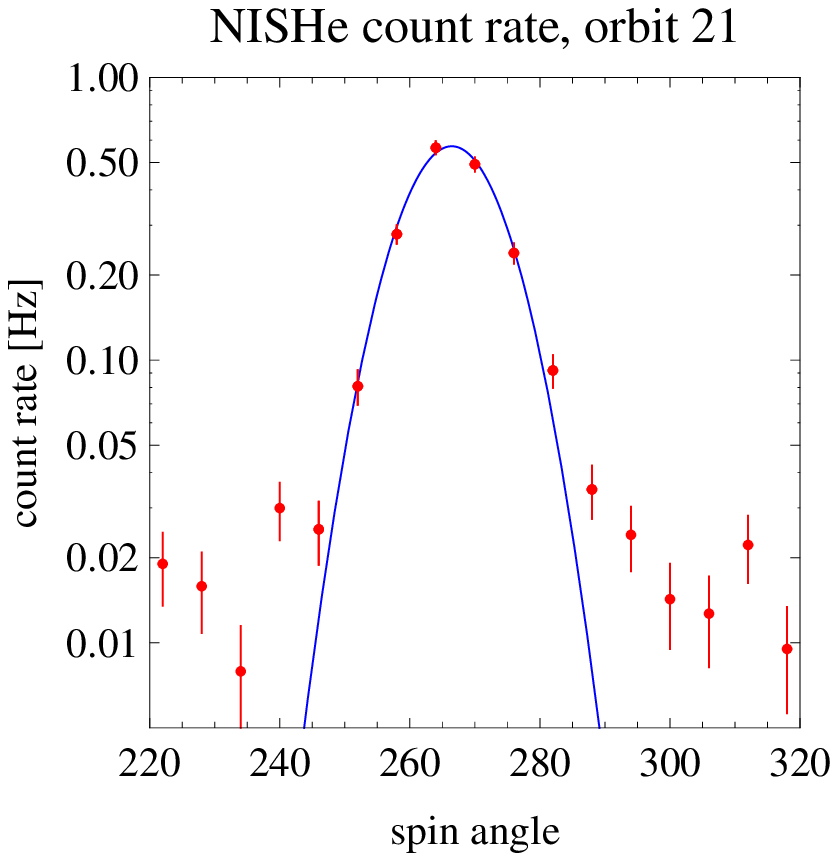}&\includegraphics[scale=.5]{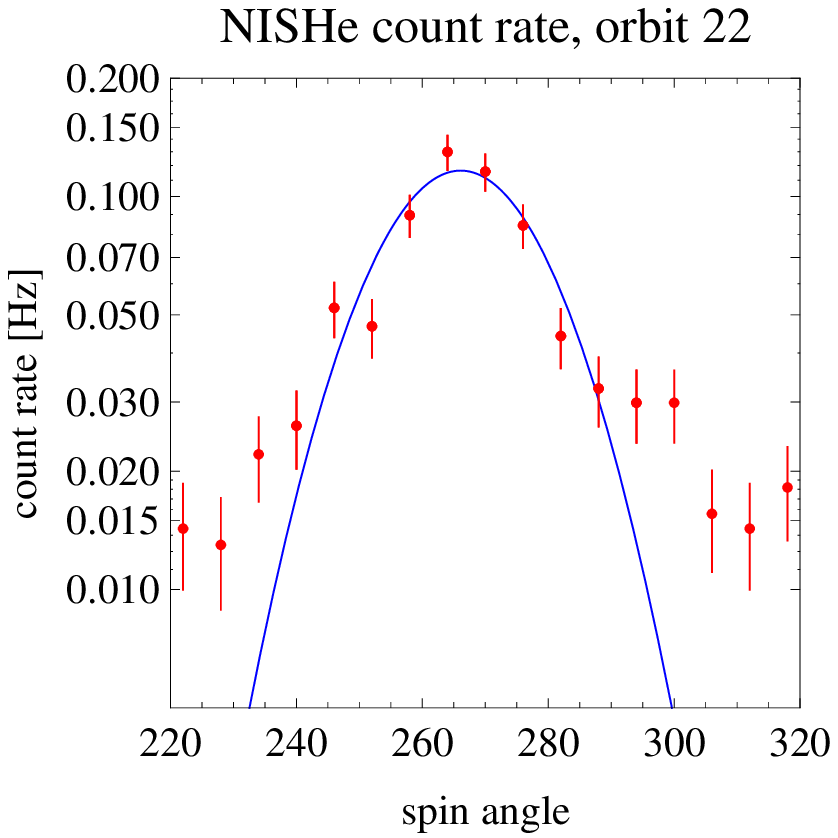}\\
		\end{tabular}
		\caption{Count rates averaged over the select ISM flow observation times, observed by IBEX-Lo in Energy Step 2 for orbits 11 through 22 in 2009. Blue dots with error bars mark the portion of the data that fits a Gaussian well. The fitted Gaussians are drawn in blue lines. Red dots show data that do not fit to the Gaussian and have been excluded from the analysis. The data from Orbits 11, 21, and 22 are all excluded from analysis as explained in the text and consequently are drawn in red. The orbits used in the NISHe parameter search are 13 through 20.}
		\label{figData2009}
		\end{figure*}

\begin{figure*}[t]
		\centering
		\begin{tabular}{ccc}
		\includegraphics[scale=.5]{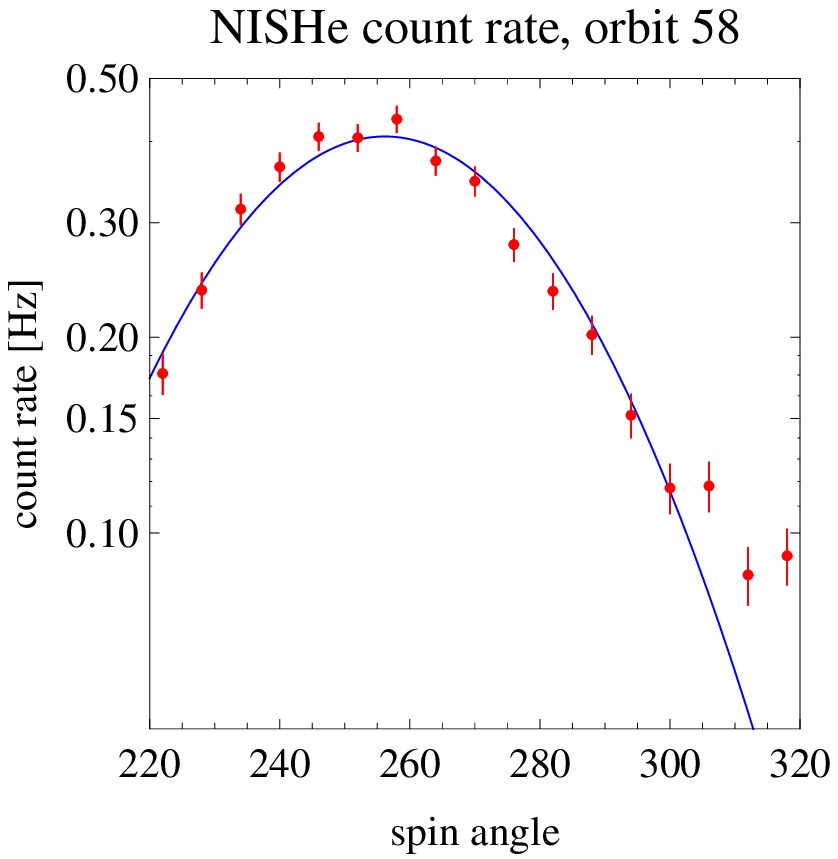}&\includegraphics[scale=.5]{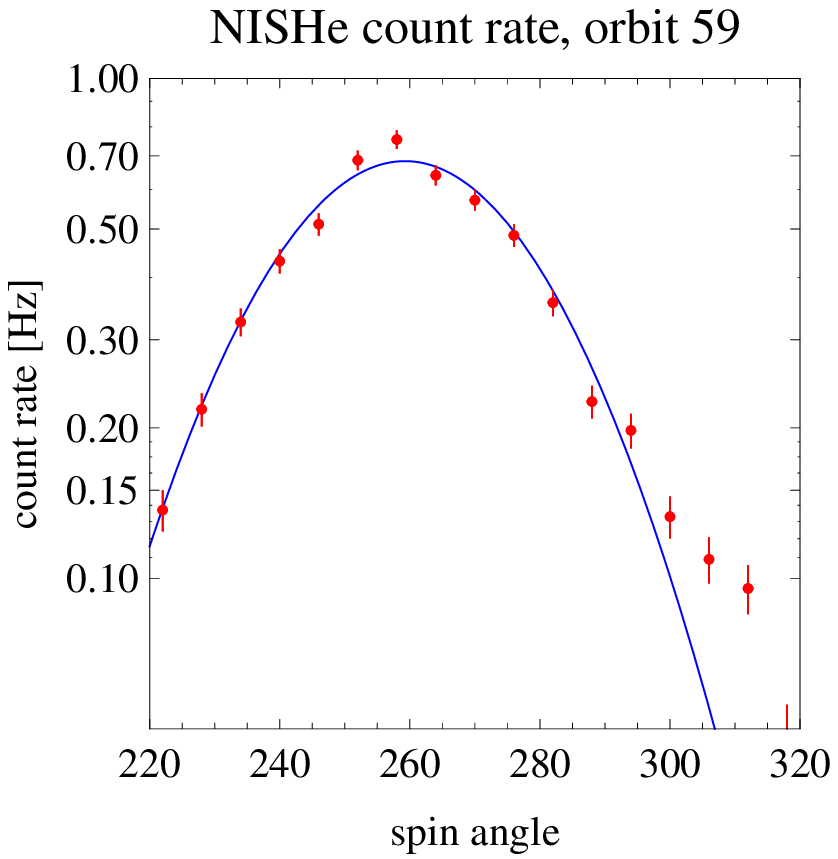}&\includegraphics[scale=.5]{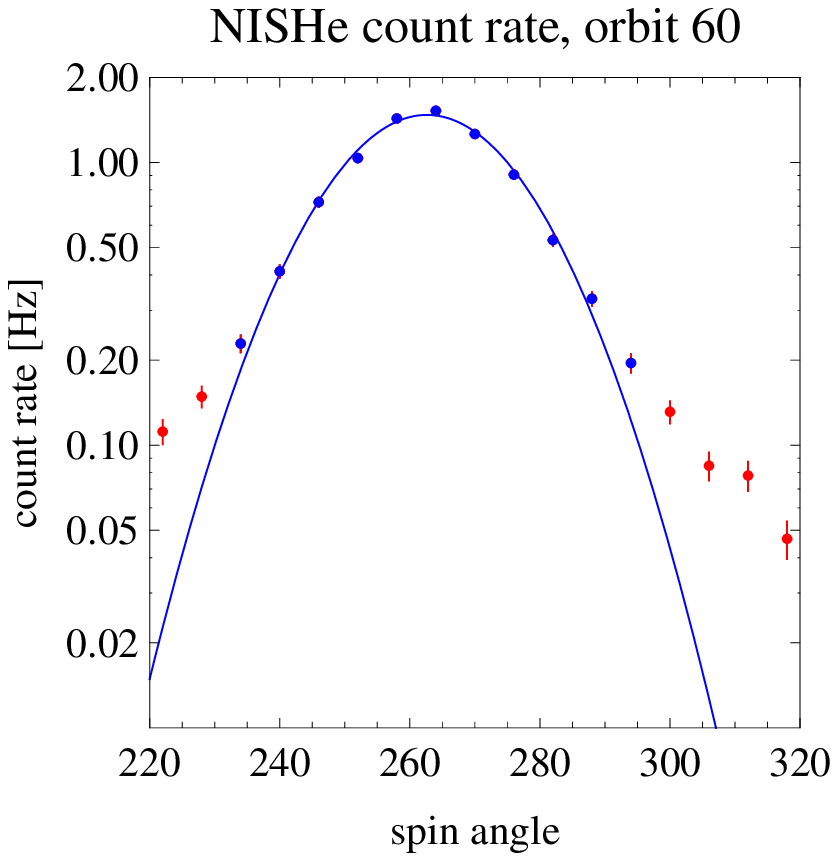}\\
		\includegraphics[scale=.5]{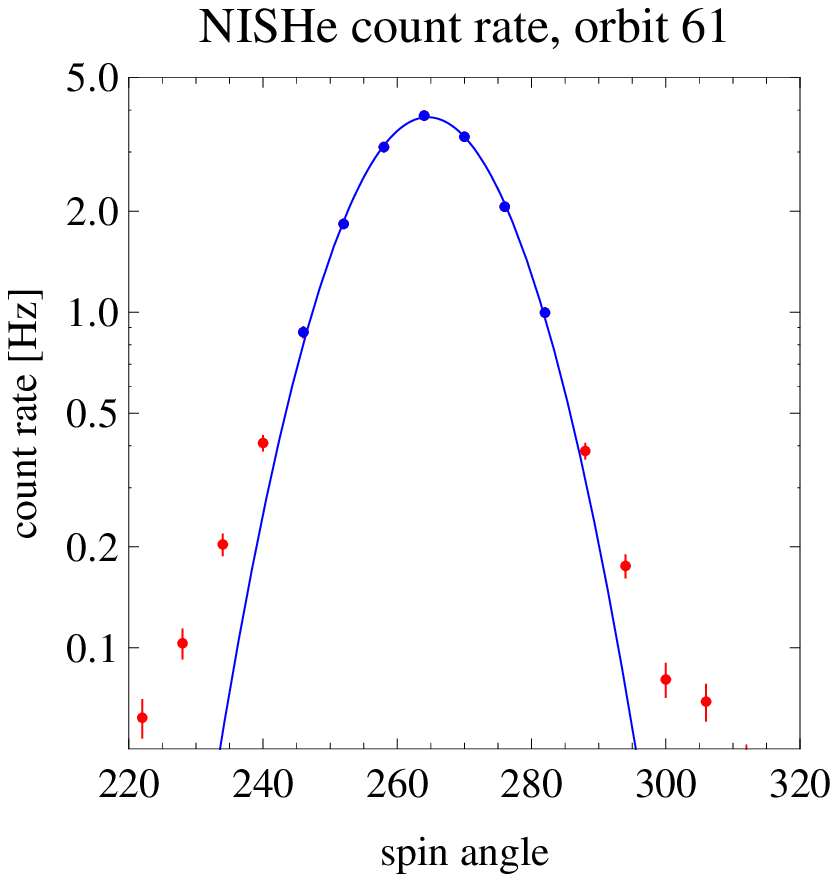}&\includegraphics[scale=.5]{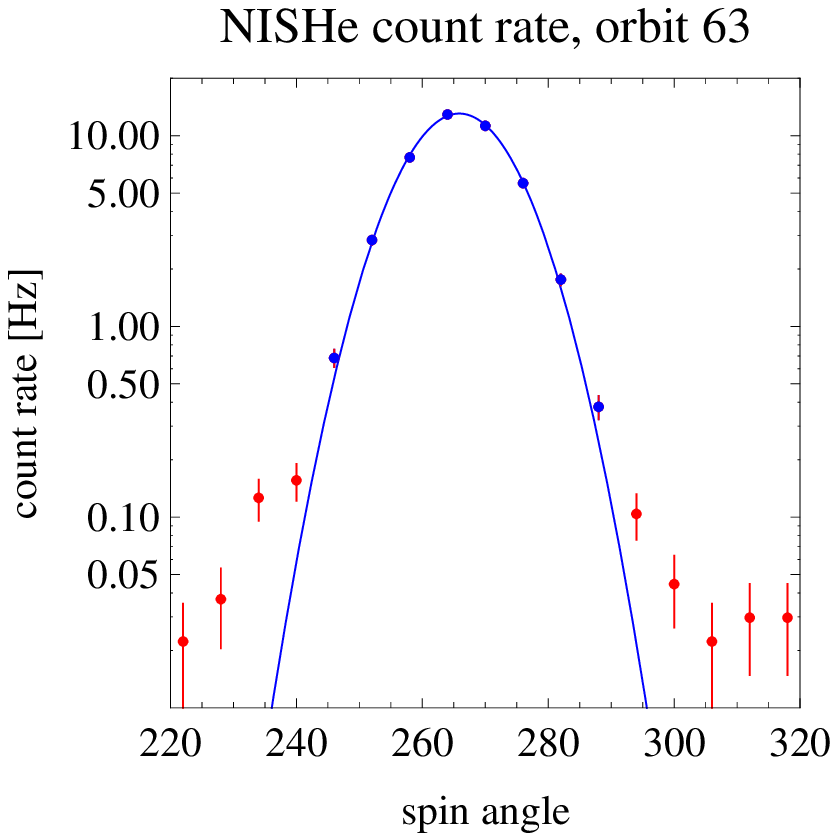}&\includegraphics[scale=.5]{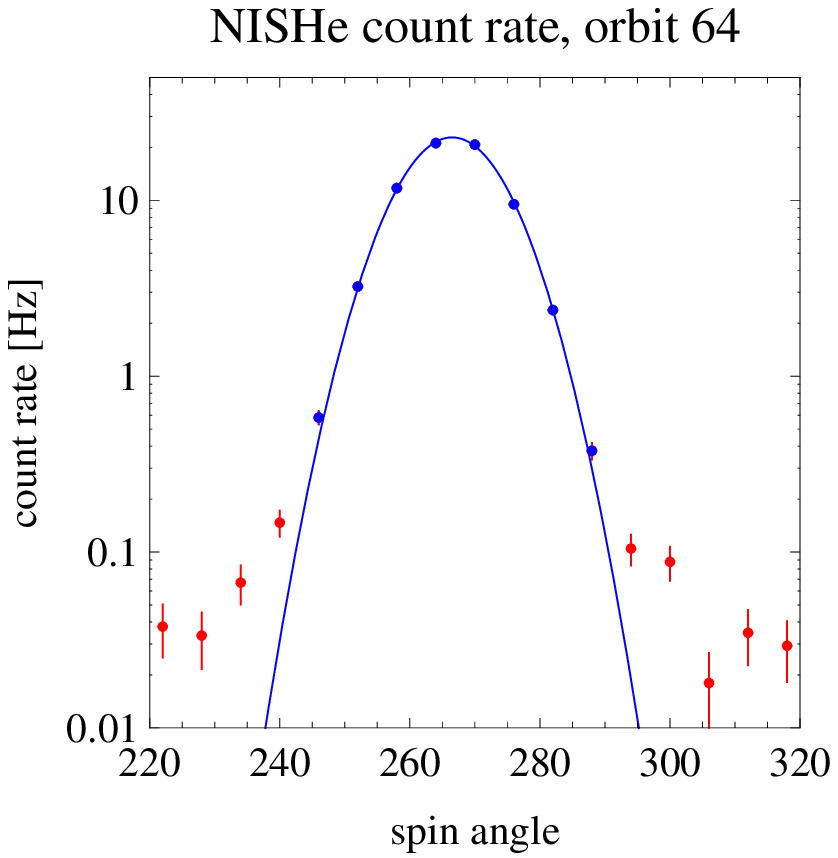}\\
		\includegraphics[scale=.5]{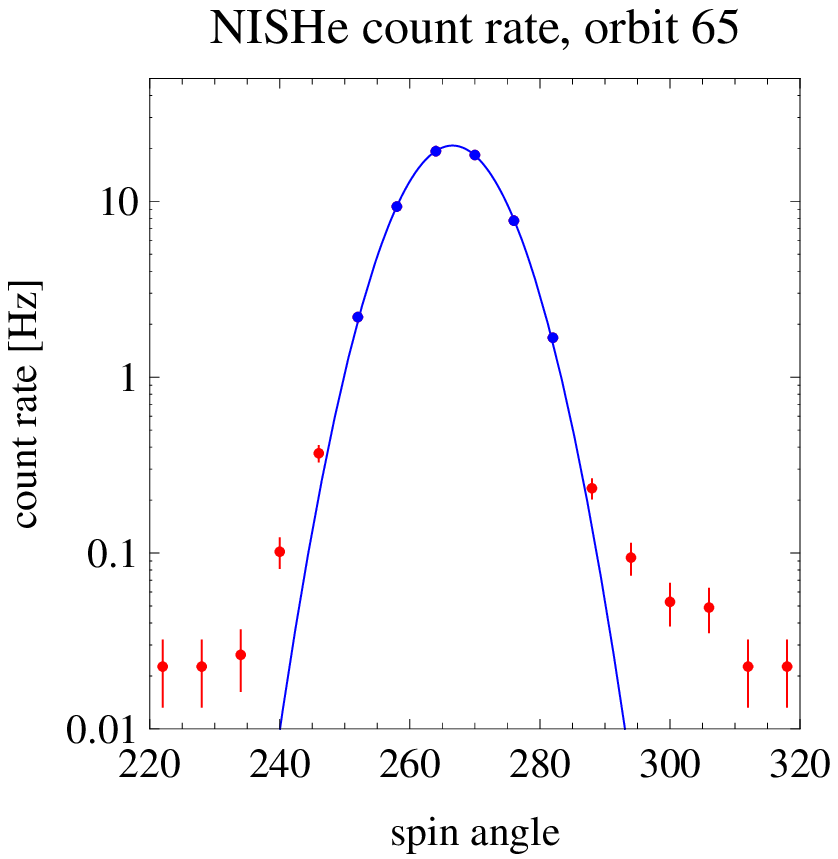}&\includegraphics[scale=.5]{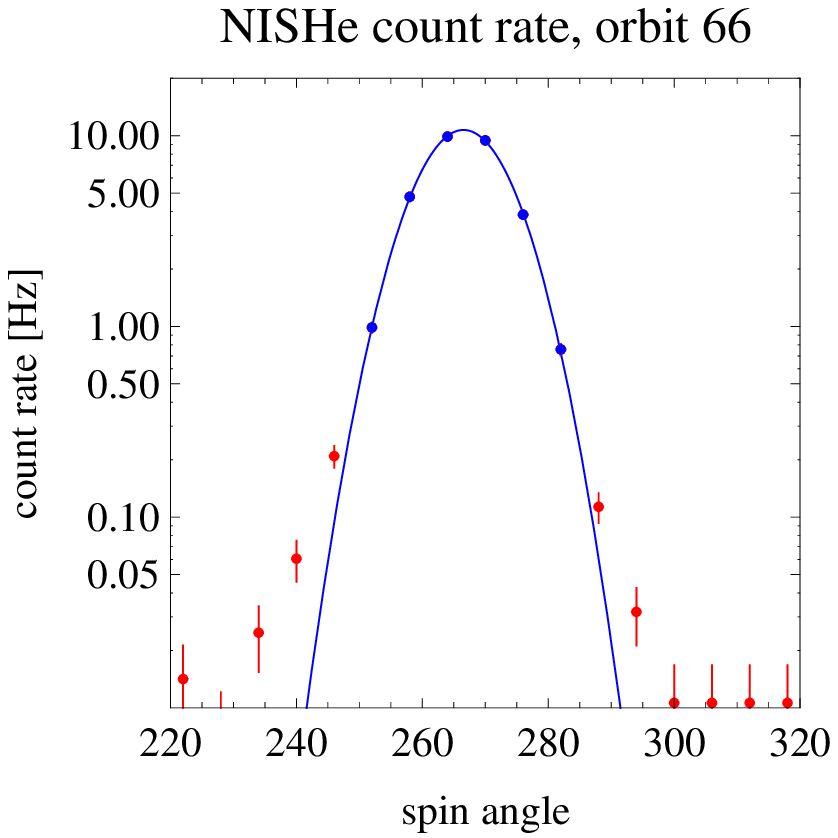}&\includegraphics[scale=.5]{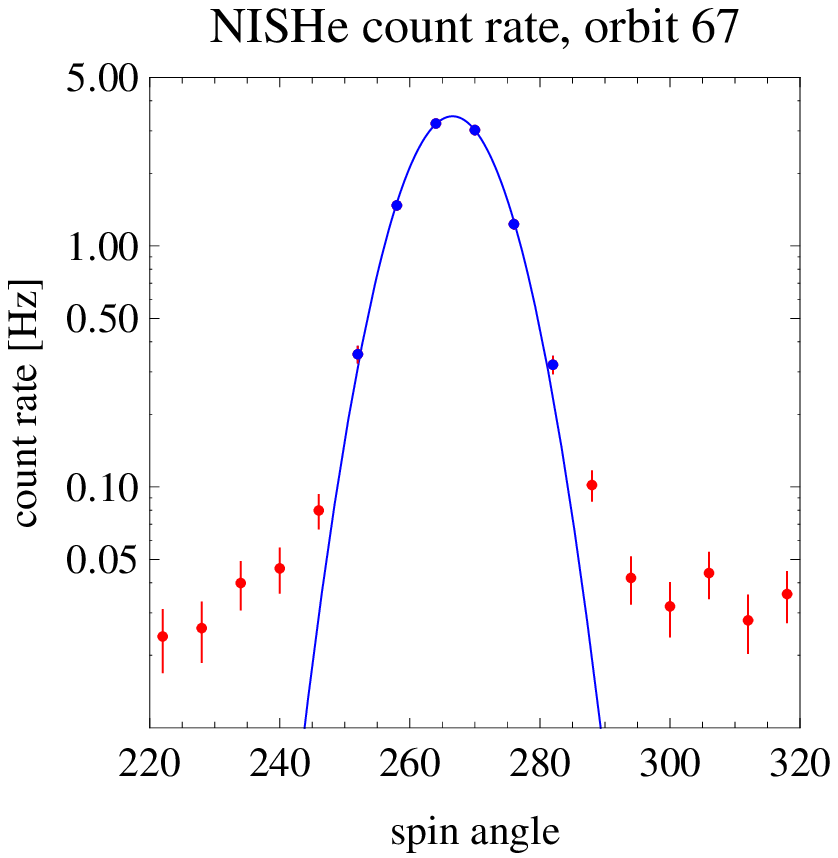}\\
		\includegraphics[scale=.5]{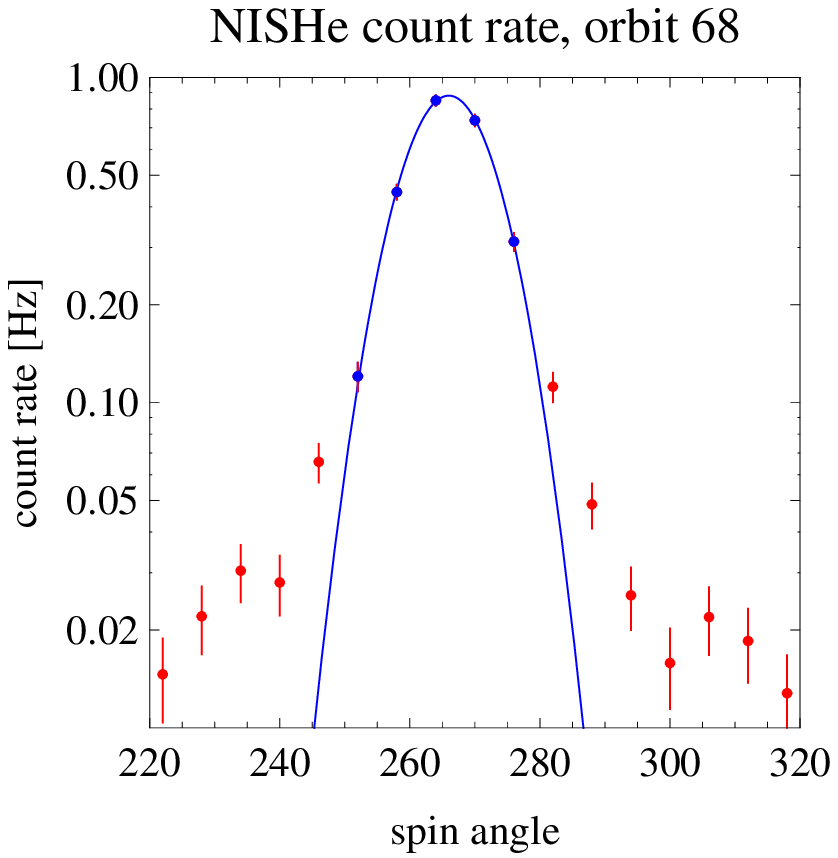}&\includegraphics[scale=.5]{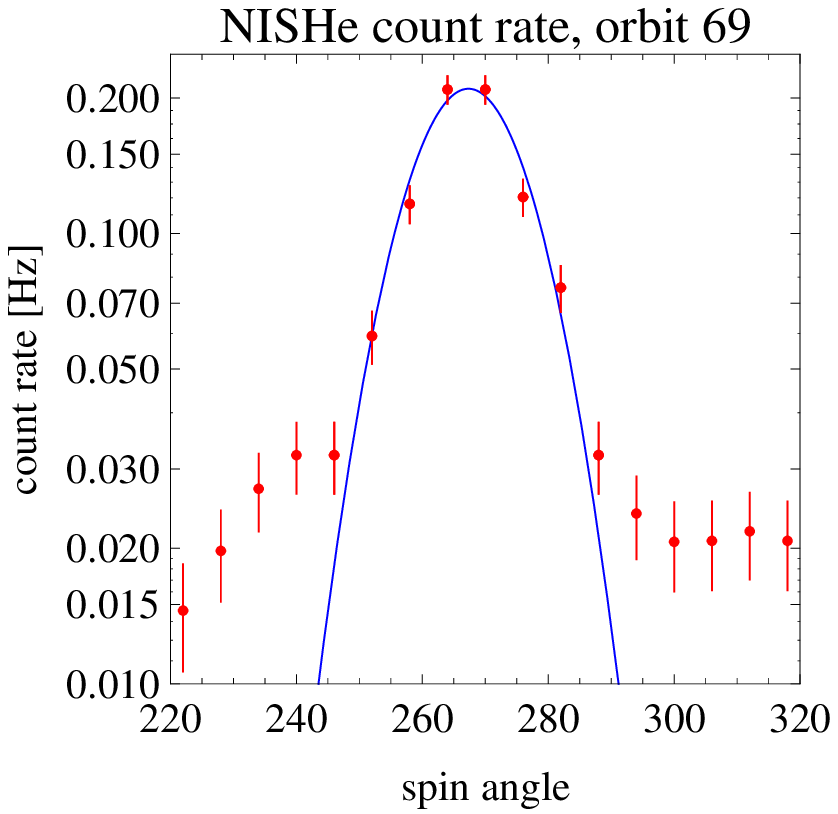}\\
		\end{tabular}
		\caption{Count rates averaged by select ISM flow observation times, observed by IBEX-Lo in Energy Step 2 for IBEX orbits 58 through 69 during 2010. As for Fig.~\ref{figData2009}, blue dots with error bars mark the portion of the data that fits a Gaussian shape well. The fitted Gaussians are drawn in blue lines. Red dots show data that do not fit to the Gaussian and have been excluded from the analysis. The data from Orbits 58, 59, and 69 are all excluded from analysis as explained in the text and consequently are drawn in red. The orbits used in the NISHe parameter search are 60 through 68.}
		\label{figData2010}
		\end{figure*}

In the data from these orbits, based on the prediction that the observed NISHe beams should be Gaussian in shape, we fitted Gaussian functions defined in Eq.~(\ref{eqGaussBeam}) and removed the non-Gaussian wings. The original data and the portion left for the analysis are shown in Fig.~\ref{figData2009} for the 2009 season and \ref{figData2010} for the 2010 season. The fitted Gaussian functions are also shown in the figure. 

		\clearpage
	
		\clearpage

	\section{Parameter fit for the NISHe flow}
	\subsection{Method}

The goal of our analysis is to determine the flow direction, velocity, and temperature of the neutral interstellar helium gas in the Local Interstellar Cloud ahead of the heliosphere. We accomplished this by fitting simulations of the NISHe flux to the data, with the ecliptic longitude and latitude of inflow direction, inflow speed, and gas temperature in the LIC as free parameters. Optimizing a multi-parameter model fit to data usually involves selecting a merit function whose free parameters are the fitted model parameters, and finding its minimum in the multi-dimensional parameter space. Colloquially speaking, the merit function describes the ``distance'' of the model predictions from the data in the observation $N$-space and searching for the best parameters requires finding the parameter set for which this distance is minimum.

A well tested and widely used method of fitting parameters of a model to a data set is the maximum likelihood method. In this method, the merit function is the likelihood function. To use it, one needs to know probability distributions $f_{p,i}\left(x_{p,i}, d_i, \vec{p}\right)$ of all the data points $d_i$, parameterized by the model parameters $\vec{p}$. In principle, probability distributions for different data points can be described by different probability distribution functions, but in our case we assume that for all data points they are identical, i.e., for all $i$, $f_{p,i}\left(x_{p,i}, d_i, \vec{p}\right) = f_{p}\left(x_{p,i}, d_i, \vec{p}\right)$. 

With these definitions we calculate the conditional probability $P_i$ that if the model with a given parameter set $\vec{p}$ is correct, then our experiment in case $i$ provides measurement $d_i$, given by the formula:
		\begin{equation}							
		P_{i}\left(x_{p,i}\right)=f_{p}\left(x_{p,i},d_i,\vec{p}\right)
		\label{eqProb1}
		\end{equation}
The series $x_{p,i}$ is the series of model predictions of the measurements for parameters $\vec{p}$. The probability $P$ that our series of $N$ measurements returns a series of results $d_i$, $i = 1, \ldots ,N$ is, of course, a product of all $N$ probabilities $P_i$:
		\begin{eqnarray}	
		P\left(x_{p,1},\ldots,x_{p,N},d_1,\ldots,d_N\right)=\prod_{i=1}^{N}{P_i}=\nonumber \\
		=\prod_{i=1}^{N}{f_{p}\left( x_{p,i},d_i,\vec{p} \right)}
		\label{eqProb2}
		\end{eqnarray}	
Fitting the parameters $\vec{p}$ is equivalent to finding the parameters $\vec{p}_{\mathrm{best}}$ for which absolute maximum of $P$ is achieved. Finding this absolute maximum is the basis of the maximum likelihood method. Remaining details determine how to best accomplish the goal and the mathematical methods to apply depend on the nature of the problem on hand. 

The IBEX-Lo detector actually counts incoming NISHe atoms in $6\degr$ spin angle bins, so the number of atoms in each bin is subject to Poisson statistics. Hence we immediately have estimates of the measurement errors according to:
		\begin{equation}							
		\sigma_{i}=\sqrt{d_i}
		\label{eqErr}
		\end{equation}
But the counts are relatively high and in this case the Poisson statistics asymptotically transforms into the Gaussian. Thus the likelihood function in Eq.~(\ref{eqProb2}) becomes:
		\begin{eqnarray}	
		P\left(x_{p,1},\ldots,x_{p,N},d_1,\ldots,d_N\right)=\nonumber \\
		=\prod_{i=1}^{N}{\left(\frac{1}{\sqrt{\pi}\sigma_i}\exp\left[-{\left(\frac{d_i-x_{p,i}}{\sigma{i}}\right)}^2\right]\right)}
		\label{eqProbGauss}
		\end{eqnarray}		
which we must maximize. Since all the probabilities are positive numbers, we can take natural logarithm of both sides of this equation and obtain:
		\begin{eqnarray}	
		\ln\left[P\left(x_{p,1},\ldots,x_{p,N},d_1,\ldots,d_N\right)\right] = \nonumber \\
	= \sum_{i=1}^{N}\left[\ln \left(\frac{1}{\sqrt{\pi}\sigma_i}\right)-{\left(\frac{d_i-x_{p,i}}{\sigma{i}}^2 \right)}\right].
		\label{eqLogProbGauss}
		\end{eqnarray}		
Since for a given measurement series the first term under the logarithm in the sum in Eq. (\ref{eqLogProbGauss}) is a constant, we can remove it because our goal is to find the parameter set for which the likelihood function will be maximum, and not the maximum value itself. Thus we define the following merit function $-L\left(\vec{p}\right)$:
		\begin{equation}							
		-L\left(\vec{p}\right)=-\sum_{i-1}^{N}{{\left(\frac{d_i-x_{p,i}}{\sigma_i}\right)}^2}
		\label{eqMeritFun1}
		\end{equation}
which takes negative values. We can omit the minus signs and then instead of maximizing the term at the right-hand side we have to minimize it. The parameter set $\vec{p}$ for which function $L(\vec{p})$ is minimal will not change when we divide it by the number of degrees of freedom in the problem, equal to $N - n_p$, where $n_p$ is the number of parameters in the parameter set $\vec{p}$. In our case $n_p = 4$. Dividing by the number of degrees of freedom converts this function into the chi-squared function and enables direct comparison of the quality of approximation between data series with different numbers of degrees of freedom. Effectively, the merit function in the form 
		\begin{equation}							
		L\left(\vec{p}\right)=\frac{1}{N-n_p}\sum_{i-1}^{N}{{\left(\frac{d_i-x_{p,i}}{\sigma_i}\right)}^2}
		\label{eqMeritFun2}
		\end{equation}
is the mean distance between data and simulations in the measurements $N$-space, normalized by the number of degrees of freedom and by the uncertainties of the measurements.

The simulations return count rates of NISHe atoms, while the observations are total counts accumulated during the select ISM flow observation times. We had to make the two quantities compatible and the choice was either to convert the model count rate into total counts or to convert the counts into the average count rate by dividing the counts and their errors by the duration of the select ISM flow observation times. We decided to adopt the second solution for practical reasons: recalculation could be done only once, and another selection would require converting all of the simulation cases, adding an unnecessary computational burden.
	
Although extensive pre-launch calibrations were conducted on the sensor \citep{fuselier_etal:09b, mobius_etal:12a}, we chose to avoid possible systematic changes in the observation conditions due to changes in the instrument functions from year to year by comparing observations and simulations separately for the 2009 and 2010 seasons.

From the simulations, we knew that count rate profiles as function of spin angle should be Gaussian. For each season we selected a reference orbit and fitted its data with a Gaussian function specified in Eq.~(\ref{eqGaussBeam}). The fitted peak height $f_0$ from the reference orbit was used as scaling factor for all the data points from that season. Effectively, this returned observed count rates relative to the fitted peak value at the reference orbit. To facilitate comparison, simulated count rates were scaled using a similar procedure. As the reference orbits we selected those with the highest count rates: Orbit 16 in 2009 and Orbit 64 in 2010.

Having brought simulations and observations to the common scale, we could look for the minimum of the 4-parameter function given by Eq.~(\ref{eqMeritFun2}). This function is purely numerical, because simulations results $x_i\left(\vec{p}\right)$ used to evaluate the merit function $L$ are purely numerical. In such a situation, calculating derivatives in the parameter space is problematic. Hence the numerical method used to minimize this function used finite differences instead of derivatives of the merit function in the parameter space. In addition, since individual simulations are very time consuming, the number of evaluations of the merit function had to be kept as low as possible. We decided to adapt the quasi-Newton method suggested by \citet{press_etal:07a}.

In this method, a starting point $\vec{p}_{\mathrm{start}}$ in the parameter space is adopted and then, using finite-differences to approximate partial derivatives, the gradient of the merit function is calculated. The gradient provides the direction of the slope of the merit function in the parameter space. This direction is sampled by calculation of a number of points $\vec{p}_1,\ldots,\vec{p}_n$ and a minimum at a point $\vec{p}_{\mathrm{loc}}$  along this direction is estimated from a parabola fit to the sampled points. The $\vec{p}_{\mathrm{loc}}$ that is found by this method becomes a new starting point. This iteration continues until the change in the magnitude of the function being minimized is considered sufficiently small. The final parameter set is then retrieved as $\vec{p}_{\mathrm{best}}$. It is essential in this method to specify initial steps in all coordinates of the parameter space such that the magnitudes of partial derivatives are similar. 

	\subsection{Calculations}
Following this method, we computed $\chi^2$ (merit function) values defined in Eq.~(\ref{eqMeritFun2}) for various sets $\vec{p}$ of flow direction longitude $\lambda$, latitude $\beta$, velocity $v$ and temperature $T$, starting from the values reported by \citet{witte:04}. Resulting from the simulations were five-dimensional ``landscapes'' of $\chi^2\left(\lambda,\beta,v,T\right)$. To find the best-fitting set of parameters $\vec{p}_{\mathrm{best}}$, we performed numerical minimizations of $\chi^2\left(\lambda,\beta,v,T\right)$ function separately on observations from each of the two seasons and collectively on combined observations from both seasons. 

The minimizations were performed on a mesh of flow longitudes $\lambda$, with the longitude fixed and latitude, speed, and temperature being free parameters. The values of the 3 free parameters in the simulations were selected by the algorithm based on the local gradient of the minimized function and hence they are not on a regular mesh. The robustness of the minima of $\chi^2$ values for all the $\lambda$ mesh points was checked by restarting the minimization algorithm from the parameter set the algorithm had reported as optimum, until no further improvements could be obtained. Usually, such a restart of the procedure did not result in a significant reduction of $\chi^2$.

After the absolute solution $\vec{p}_{\mathrm{best}}$ was found, we extended the simulations on a regular grid of parameters centered at the best solution $\vec{p}_{\mathrm{best}}$ in order to check on the covariance of parameters and to better illustrate the acceptable parameter region. This mesh regularization resulted in a very small correction of the best fitting parameter set. The improvement in $\chi^2$ value was only $\sim0.001$ with the changes in parameters of $\sim50$~K in temperature and $\sim 0.1 \, \mathrm{km} \, \mathrm{s}^{-1}$ in the flow velocity. 

During the minimization process we performed simulations using a total of about 4000 sets of the NISHe gas parameters. The minimization algorithm kept track of its own steps, so that results of simulations for individual parameter sets could be used in as many minimization processes as needed. 

\begin{figure}[t]
		\centering
		\epsscale{1}
		\plotone{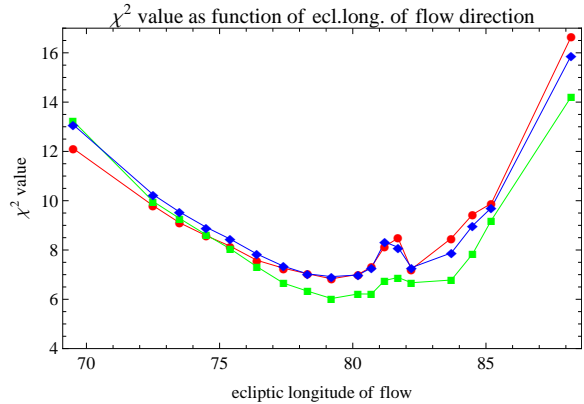}		
		\caption{Values of $\chi^2$ statistic defined in Eq.~(\ref{eqMeritFun2}) shown as function of ecliptic longitude of the flow direction of the NISHe gas obtained as a result of our fitting procedure. The statistics of the fits performed separately for the observation campaigns 2009 and 2010 are shown in red (dots) and green (squares), respectively; the statistic for the fitting performed collectively on the data from both campaigns is shown in blue (diamonds). The value of $\chi^2$ for the parameter set obtained by \citet{witte:04} is 143.9.}		
		\label{figChi2tD00}
		\end{figure}

The values of the merit function $\chi^2$ minimized for the mesh values of the flow direction $\lambda$ are shown in Fig. \ref{figChi2tD00}. The results of the minimization performed separately on the data from 2009 and 2010 (red and green, respectively) are consistent with each other and with the results of the minimization performed collectively on the data from 2009 and 2010 (blue). They are presented and discussed in greater detail in the following section.\

	\section{Results}
The fitting procedure results in a new set of the parameters of neutral interstellar helium in the LIC, which differs from the previously obtained by \citet{mobius_etal:04a, witte:04, gloeckler_etal:04b, vallerga_etal:04a}. The downwind direction of the NISHe gas in the LIC best fitting to the data is longitude $\lambda = 79.2\degr$ , latitude $\beta = -5.12\degr$. The bulk speed is $v = 22.756 \,\mathrm{km}\,\mathrm{s}^{-1}$ and the temperature $T = 6165$ K.

The flow parameters fitted separately to the observations from the 2009 season are $\lambda = 79.2\degr$, $\beta = -5.06\degr$, $v = 22.831 \,\mathrm{km}\,\mathrm{s}^{-1}$, $T = 6094$ K and to the observations from the 2010 season $\lambda = 79.2\degr$, $\beta = -5.12\degr$, $v = 22.710 \,\mathrm{km}\,\mathrm{s}^{-1}$, $T = 6254$~K. It is clear that the solutions obtained separately from the two seasons are consistent with each other and with the solution obtained for the two seasons collectively.

\begin{figure*}[t]
		\centering
		\begin{tabular}{ccc}		\includegraphics[scale=.6]{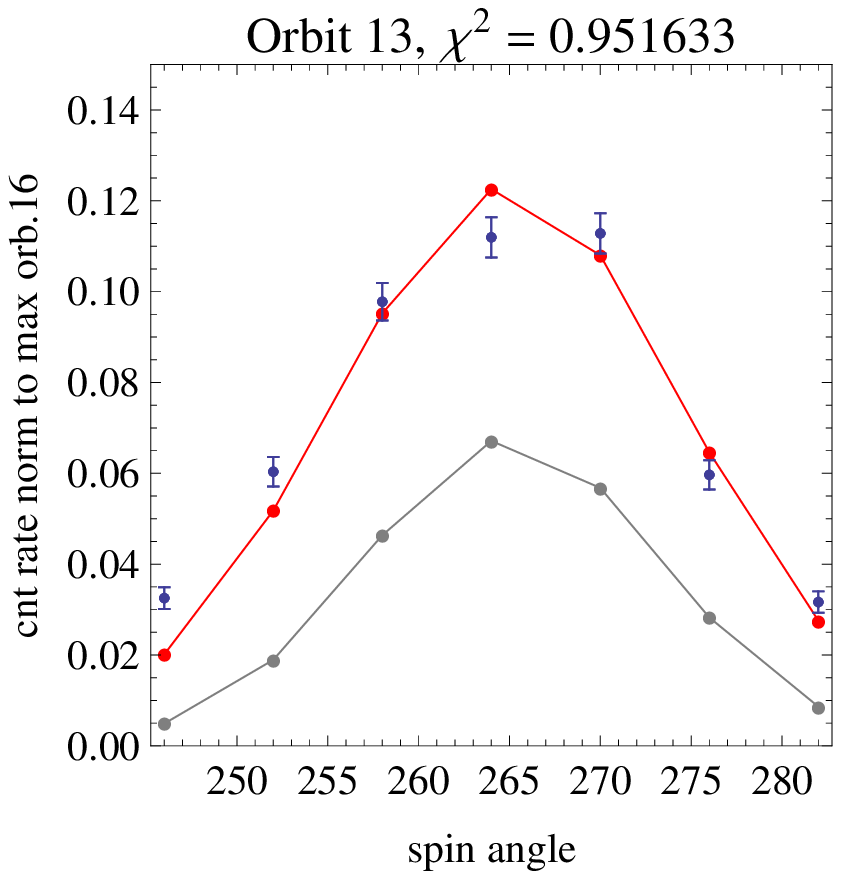}&\includegraphics[scale=.6]{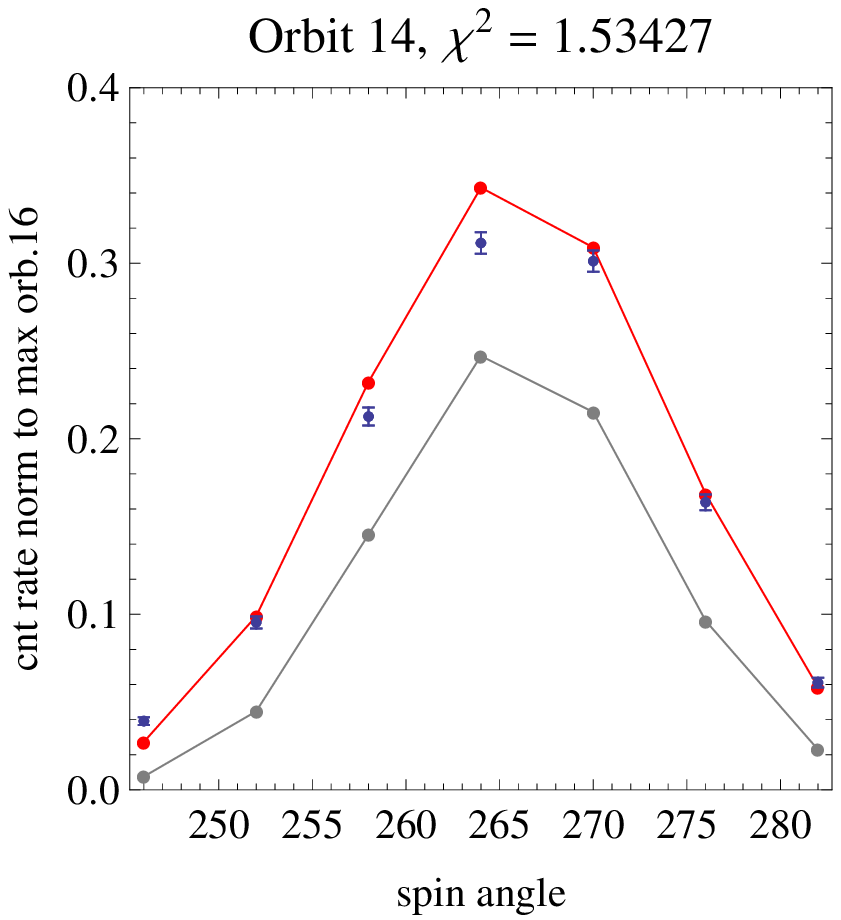}&\includegraphics[scale=.6]{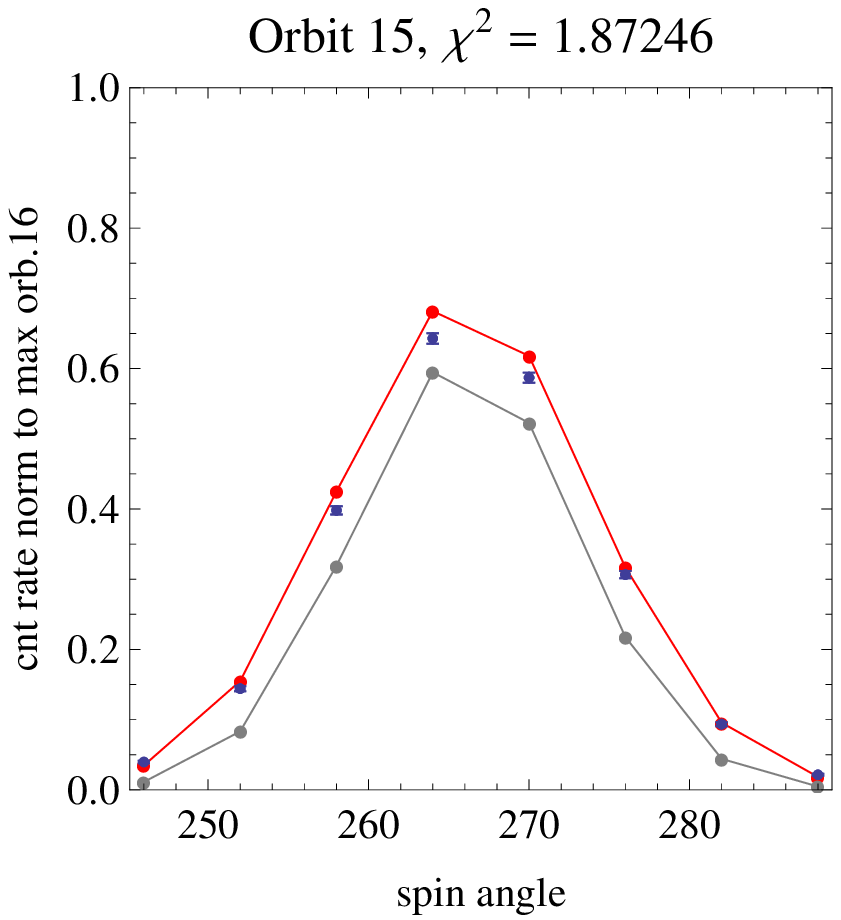}\\		\includegraphics[scale=.6]{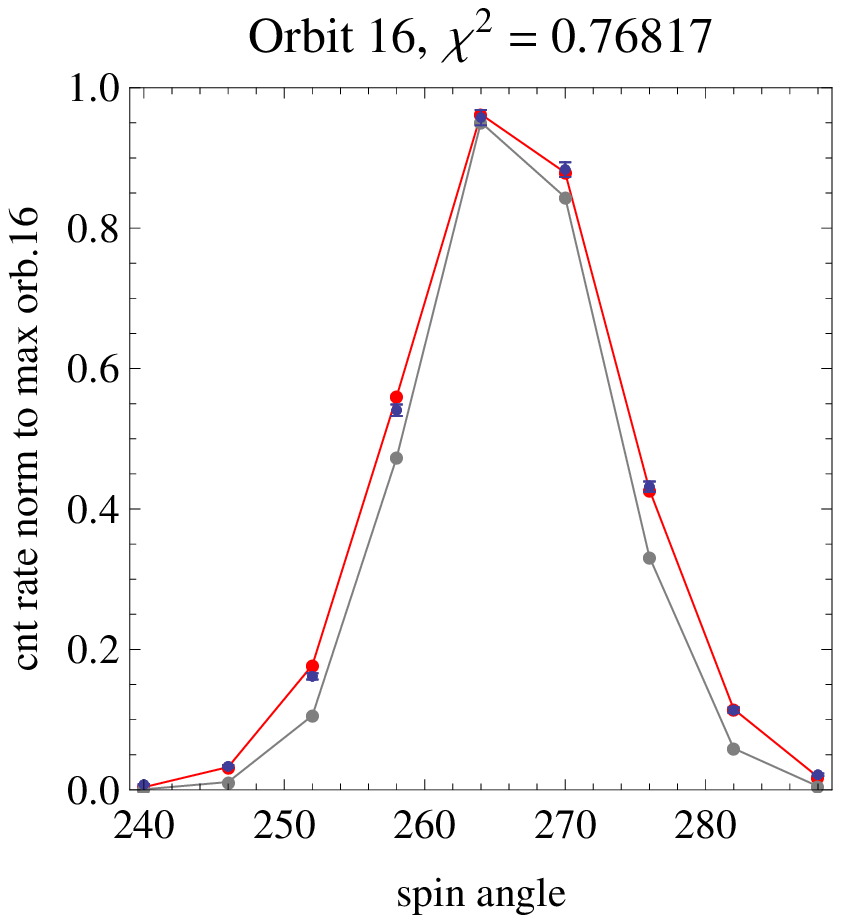}&\includegraphics[scale=.6]{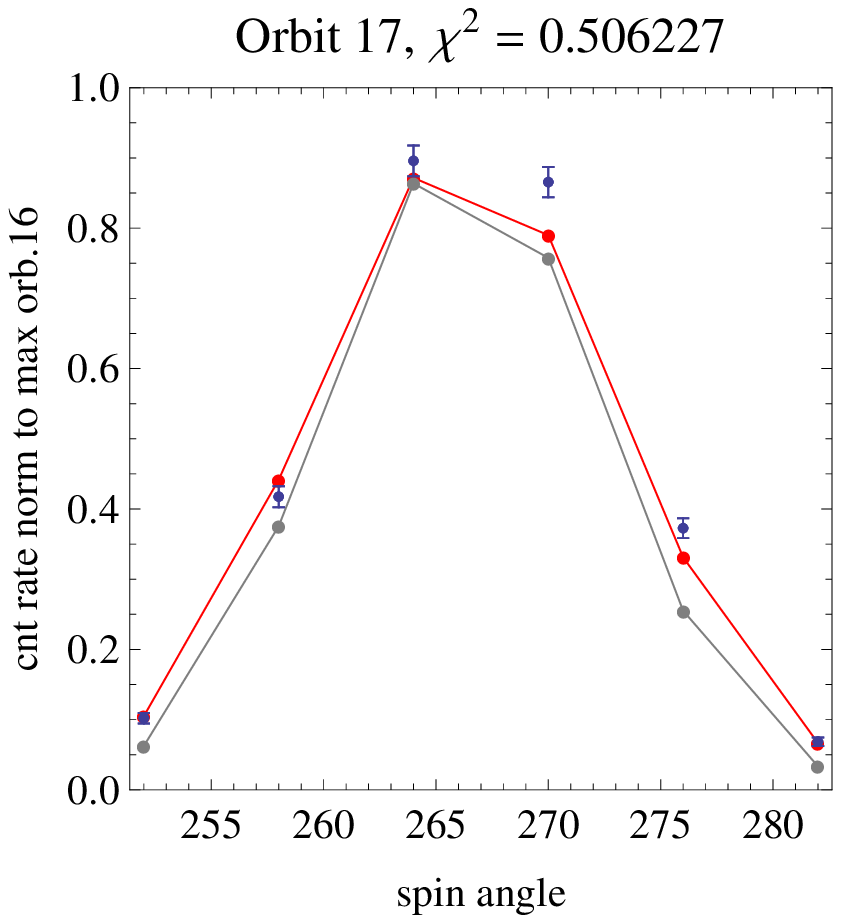}&\includegraphics[scale=.6]{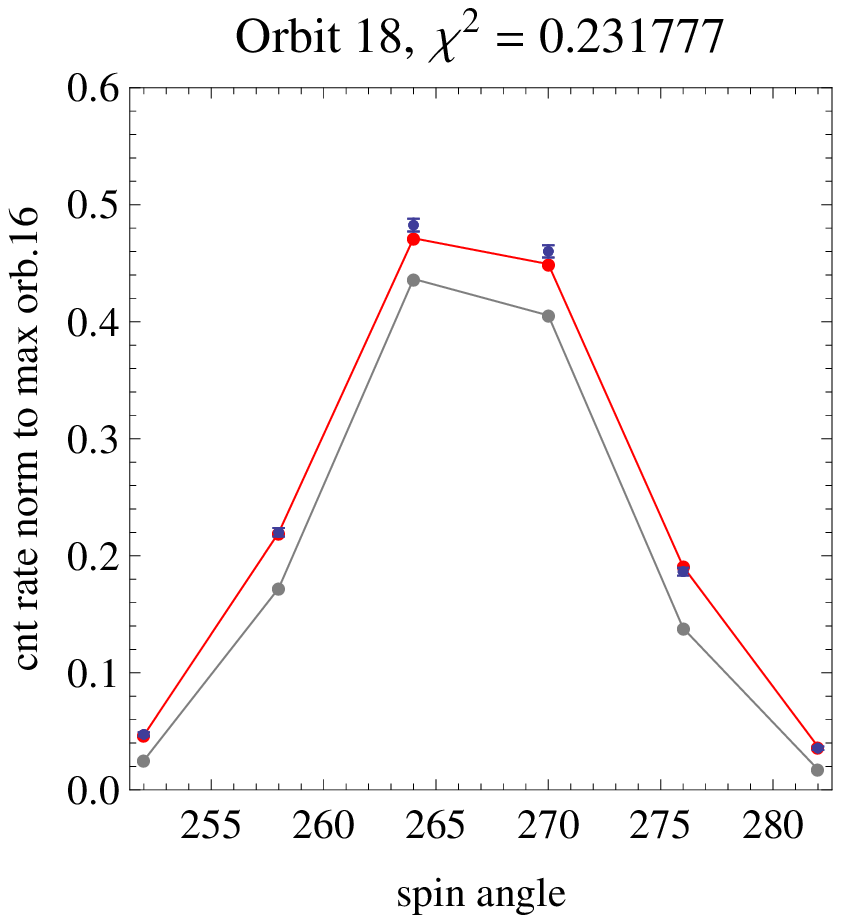}\\
		\includegraphics[scale=.6]{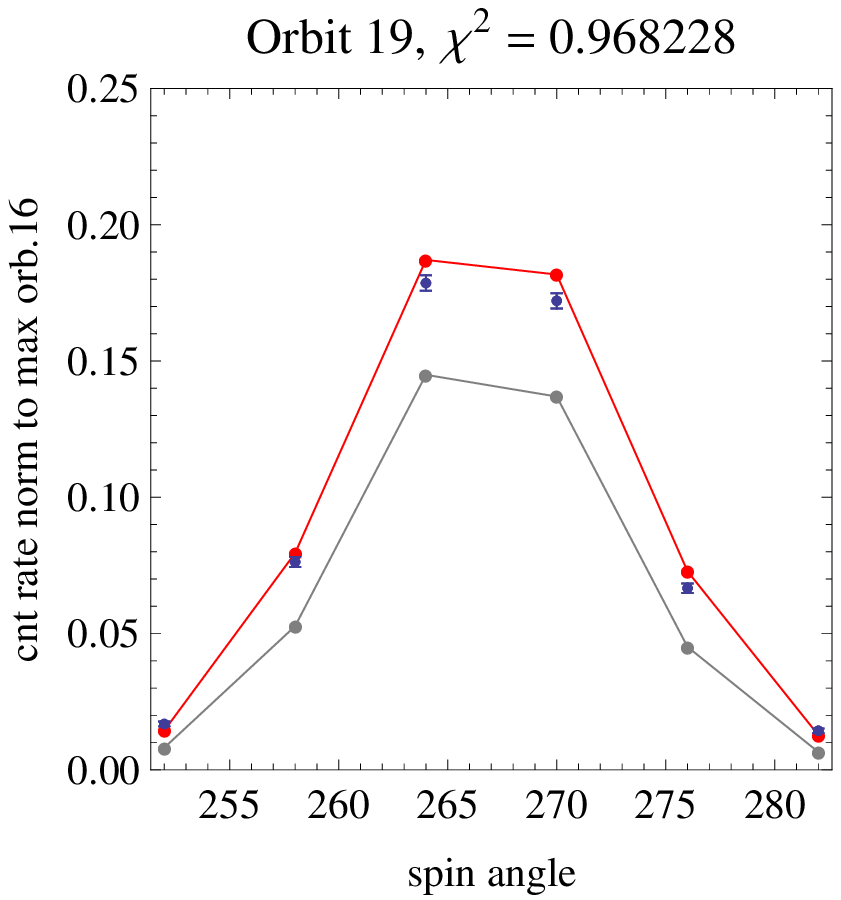}&\includegraphics[scale=.6]{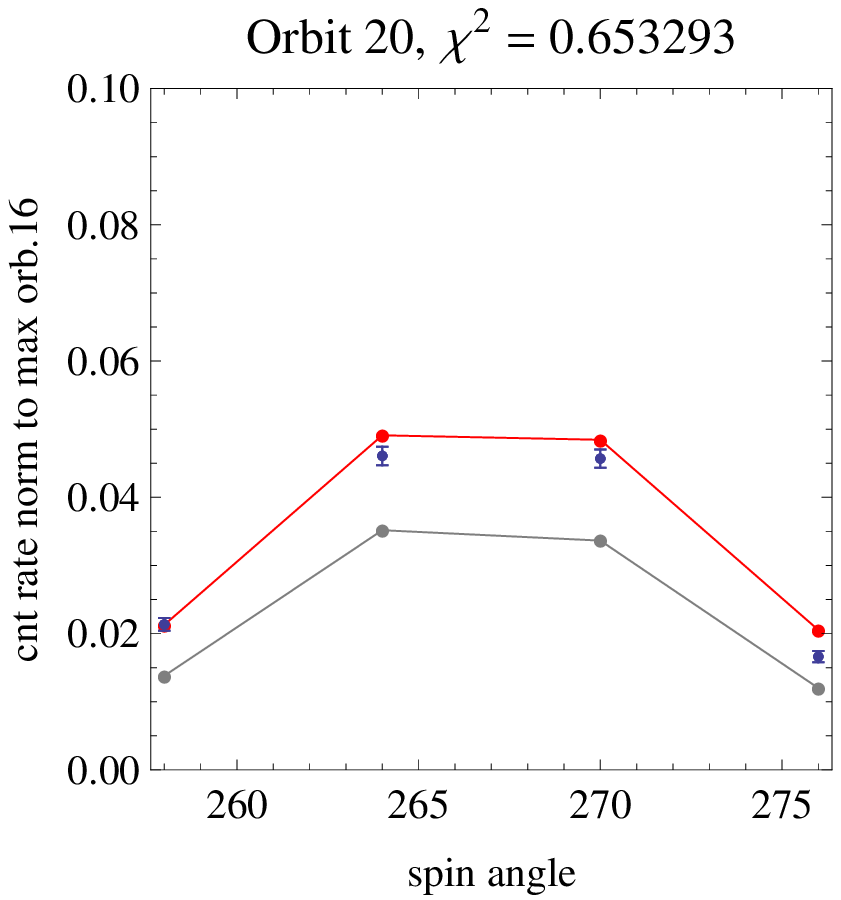}\\
		\end{tabular}
		\caption{Comparison of count rates of NISHe atoms observed by IBEX-Lo for orbits 13 through 20 during the 2009 NISHe observation campaign (blue dots with error bars) with the simulated count rates calculated for the set of parameters best fitting to the data from both seasons (red lines) and for the parameter set suggested by \citet{witte:04} (gray lines). Both observations and simulations are normalized to their respective peak values at Orbit 16, as discussed in the text. The value of $\chi^2$ at the given orbit for the best case is listed in the headers.}
		\label{figBestFitVsObsS01}
		\end{figure*}

\begin{figure*}[t]
		\centering
		\begin{tabular}{ccc}	\includegraphics[scale=.6]{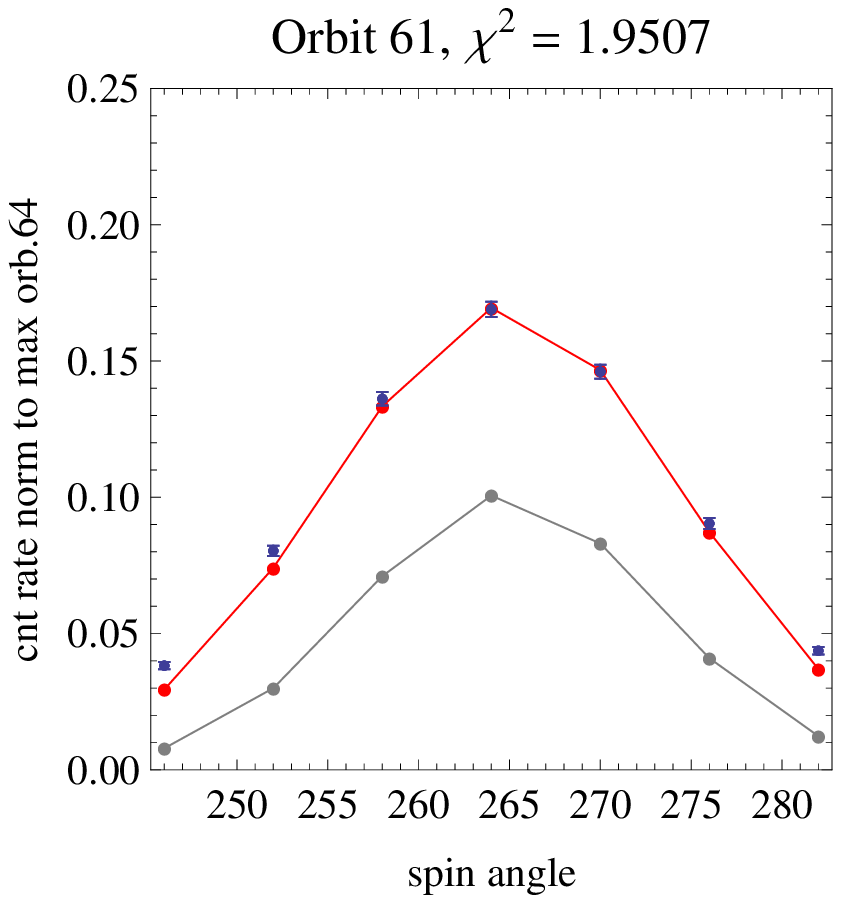}&\includegraphics[scale=.6]{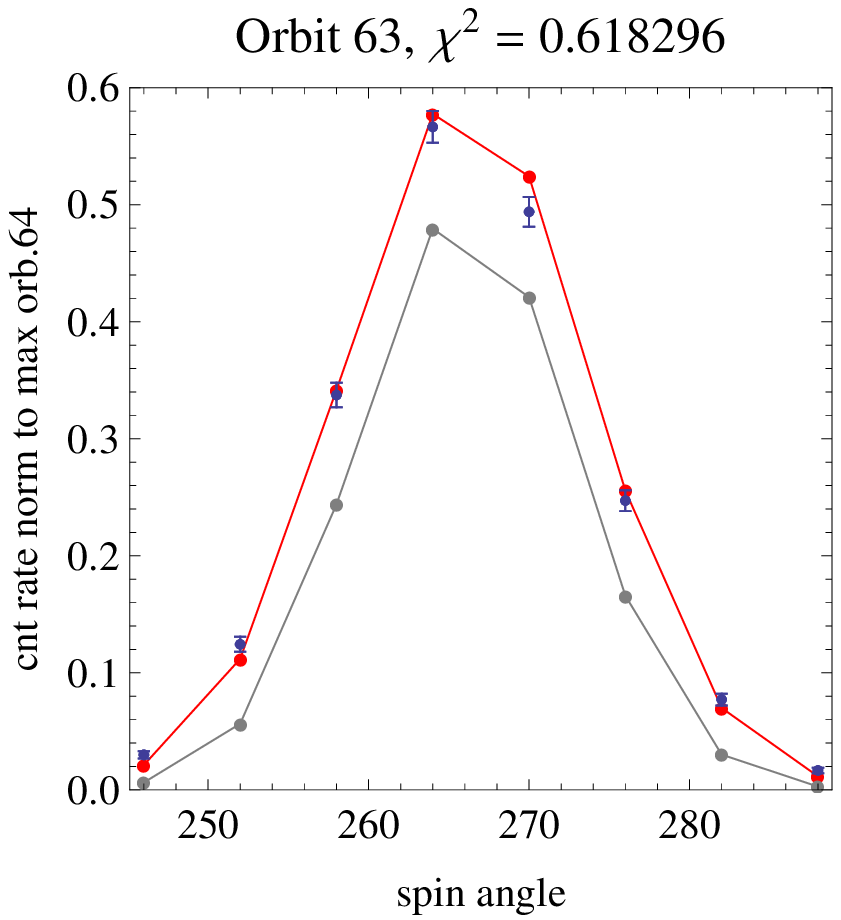}&\includegraphics[scale=.6]{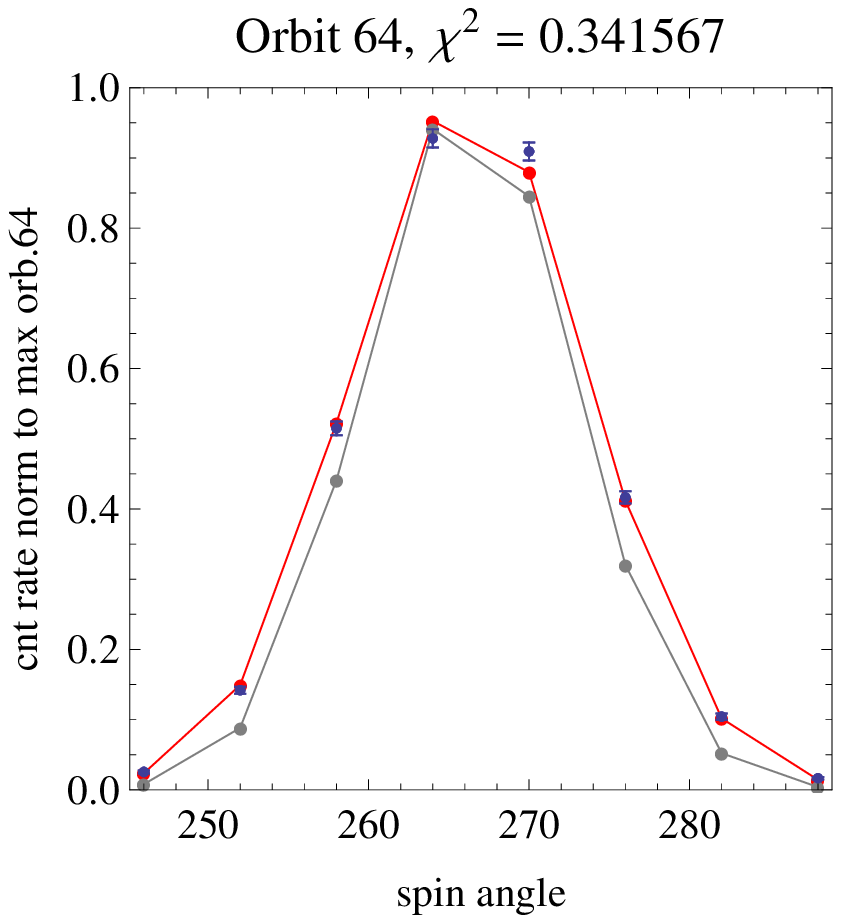}\\
\includegraphics[scale=.6]{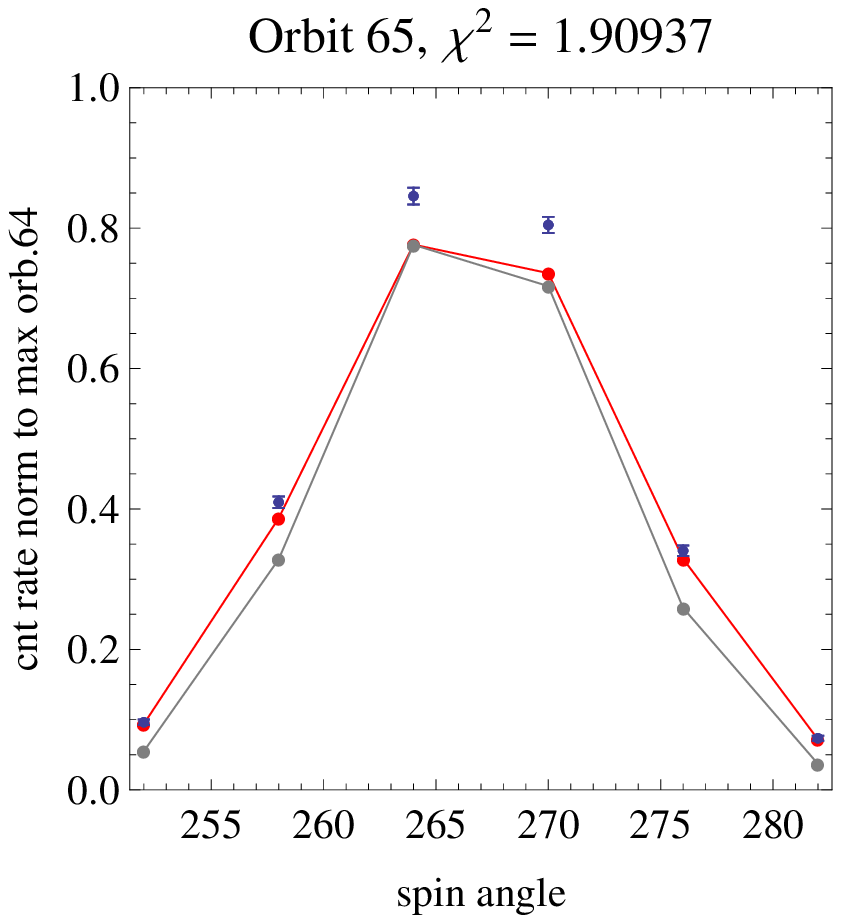}&\includegraphics[scale=.6]{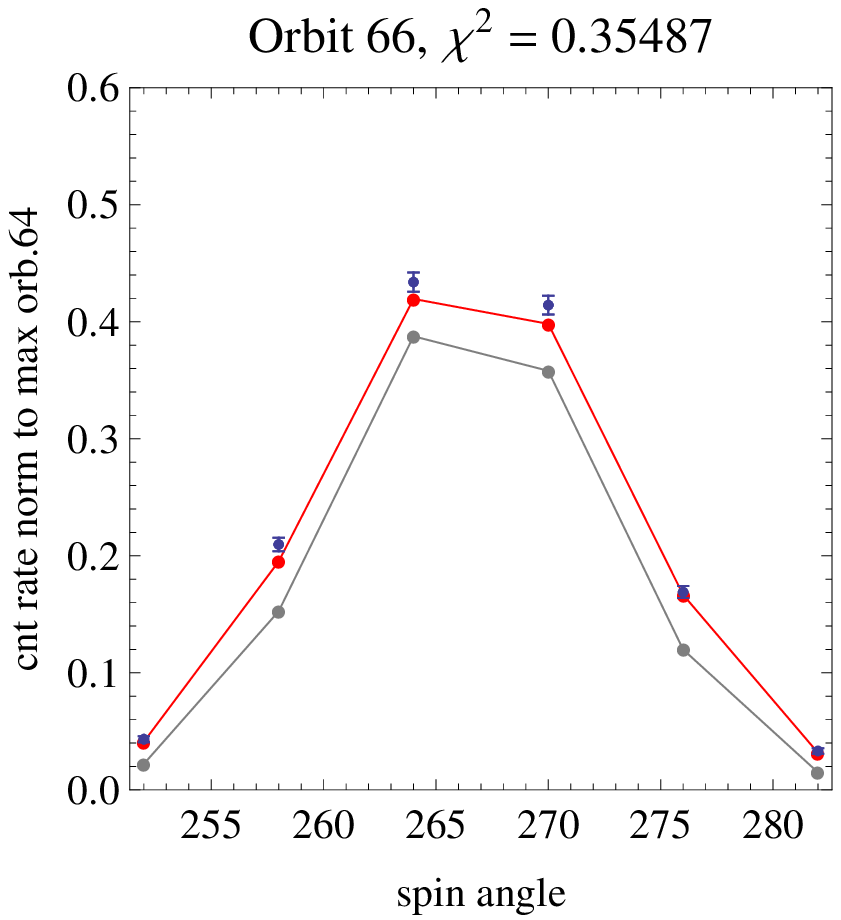}&\includegraphics[scale=.6]{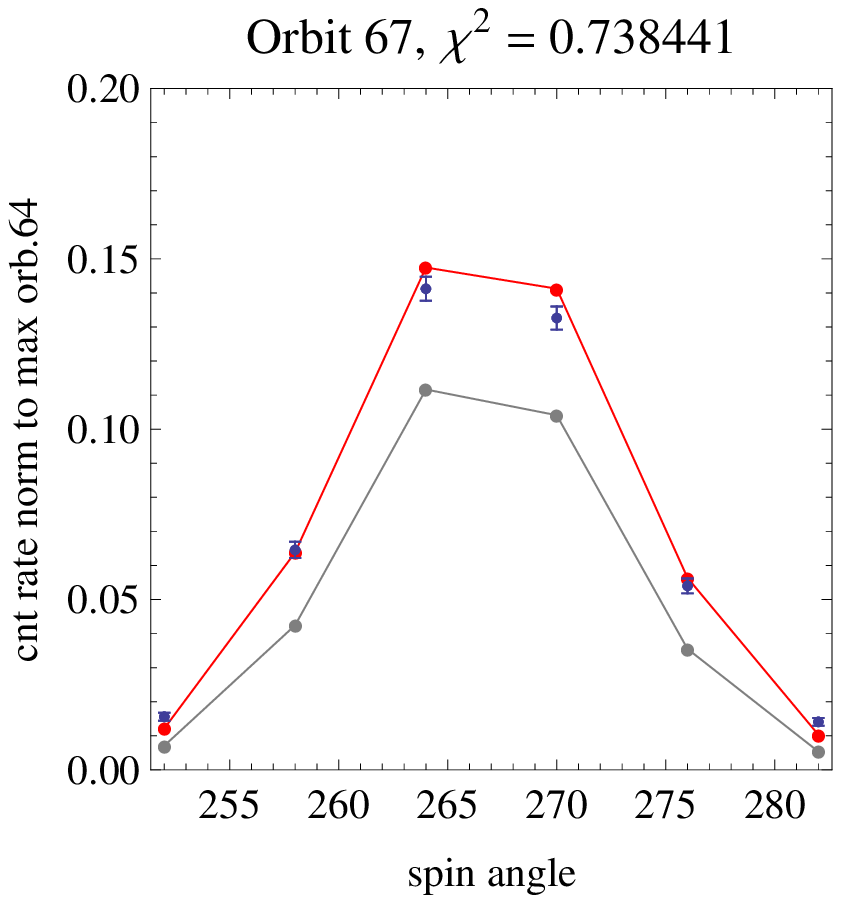}\\
\includegraphics[scale=.6]{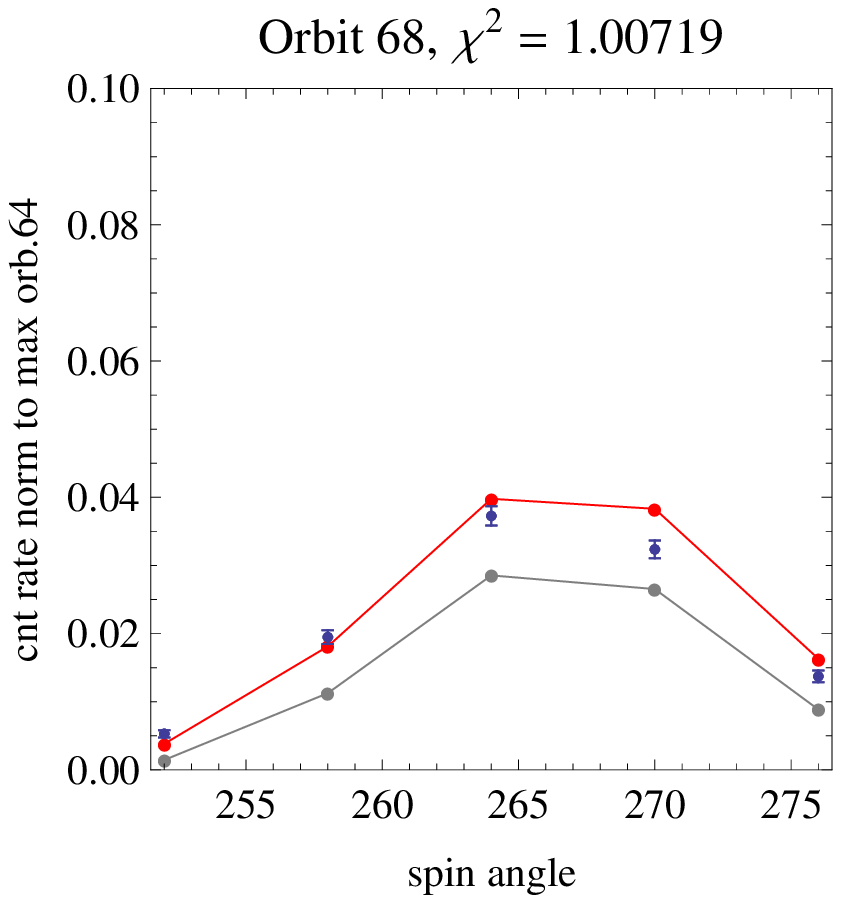}\\
\end{tabular}
		\caption{Comparison of the count rates of NISHe atoms observed by IBEX-Lo for orbits 61 through 68 during the 2010 NISHe observation campaign (blue dots with error bars) with the simulated count rates calculated for the set of parameters best fitting to the observations from both seasons (red lines) and for the parameter set suggested by \citet{witte:04} (gray lines). Both observations and simulations are normalized to their respective peak values at Orbit 64, as discussed in the text. The value of $\chi^2$ at the given orbit for the best case is listed in the headers.}
		\label{figBestFitVsObsS02}
		\end{figure*}

\begin{figure*}
		\centering
		\epsscale{2.3}
		\plottwo{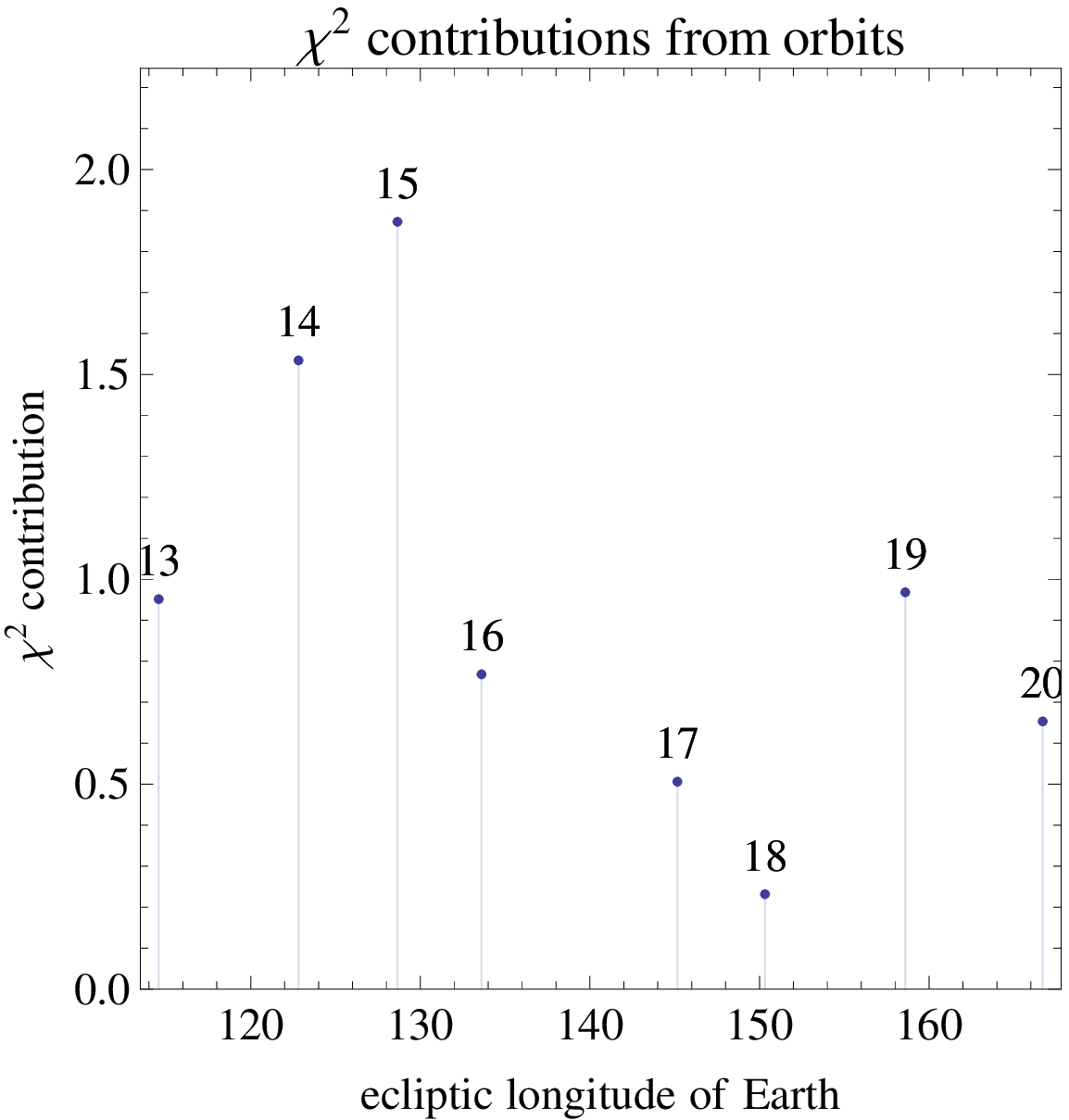}{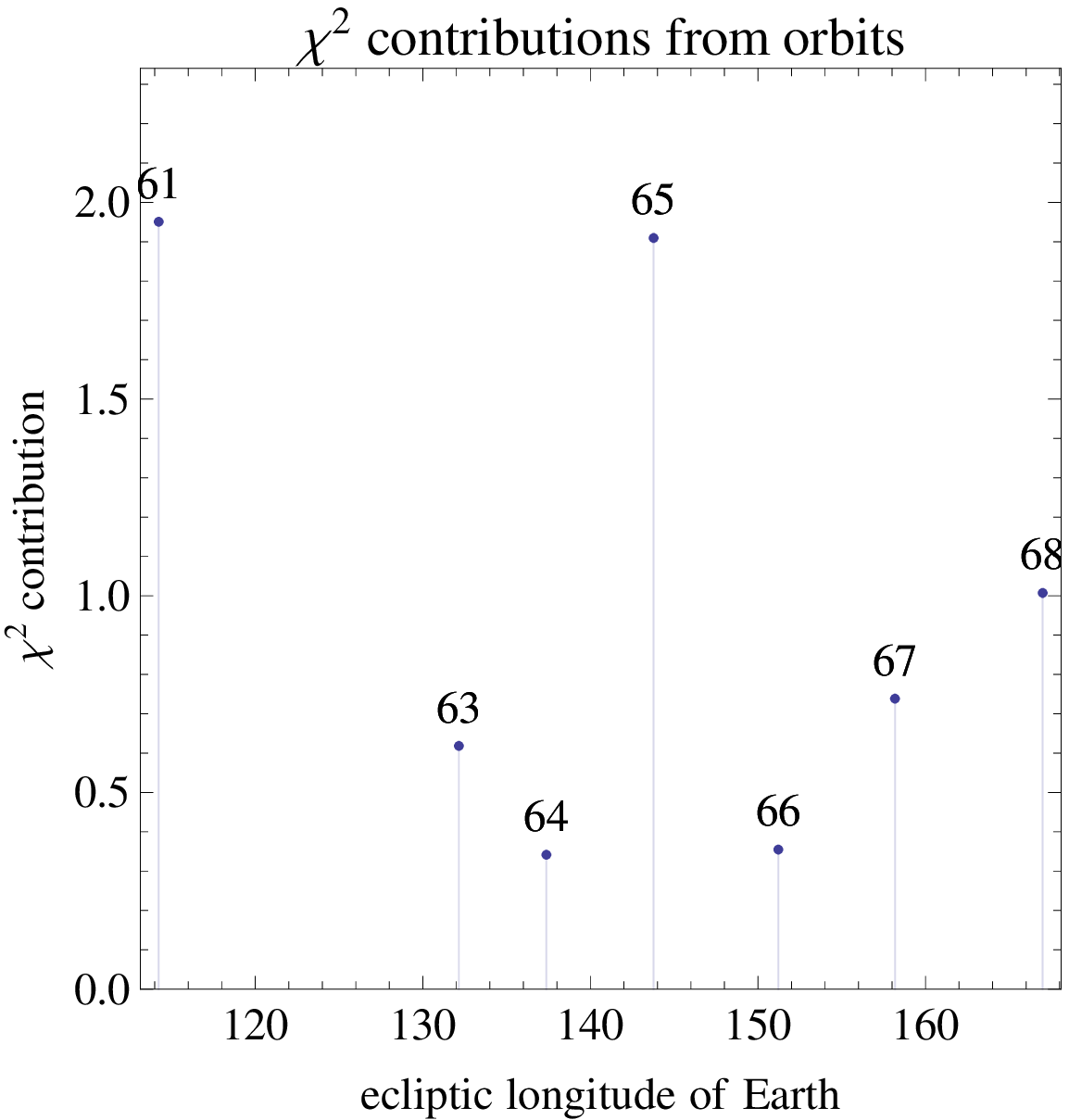}\\
		\caption{Contributions to $\chi^2$ values from the individual orbits during the 2009 (left) and 2010 seasons (right). The horizontal axes show the mean ecliptic longitudes of Earth for individual orbits, calculated as averages over the select ISM flow observation time intervals. The numbers on the points indicate orbit numbers.}
		\label{figChi2Contrib}
		\end{figure*}

The values of $\chi^2$ calculated for the two seasons together and separately are shown in Fig.~\ref{figChi2tD00} as function of the downwind longitude. The quality of the fits for each orbit can be assessed in Fig.~\ref{figBestFitVsObsS01} for the 2009 season and in Fig.~\ref{figBestFitVsObsS02} for the 2010 season; it is apparent that the quality of our fit is much better than the solution from \citet{witte:04}. Contributions from individual orbits to the total $\chi^2$ value for the 2009 and 2010 seasons are presented in Fig.~\ref{figChi2Contrib}.
				
		\clearpage

		\clearpage
	
Inspection of Figs \ref{figChi2tD00}, \ref{figBestFitVsObsS01}, \ref{figBestFitVsObsS02}, and \ref{figChi2Contrib} shows that contributions of various orbits to $\chi^2$ vary appreciably among the orbits and that $\chi^2\left(\lambda\right)$ exhibits humps and traces of secondary minima. We tried to identify the causes of non-smooth features in $\chi^2$ curves. The obvious candidates were the strongest contributors to $\chi^2$ total. Therefore we repeated minimizations without Orbits 14 and 15 for the 2009 season and without orbit 65 for the 2010 season. 

\begin{figure*}[t]
		\centering
		\epsscale{2.3}
		\plottwo{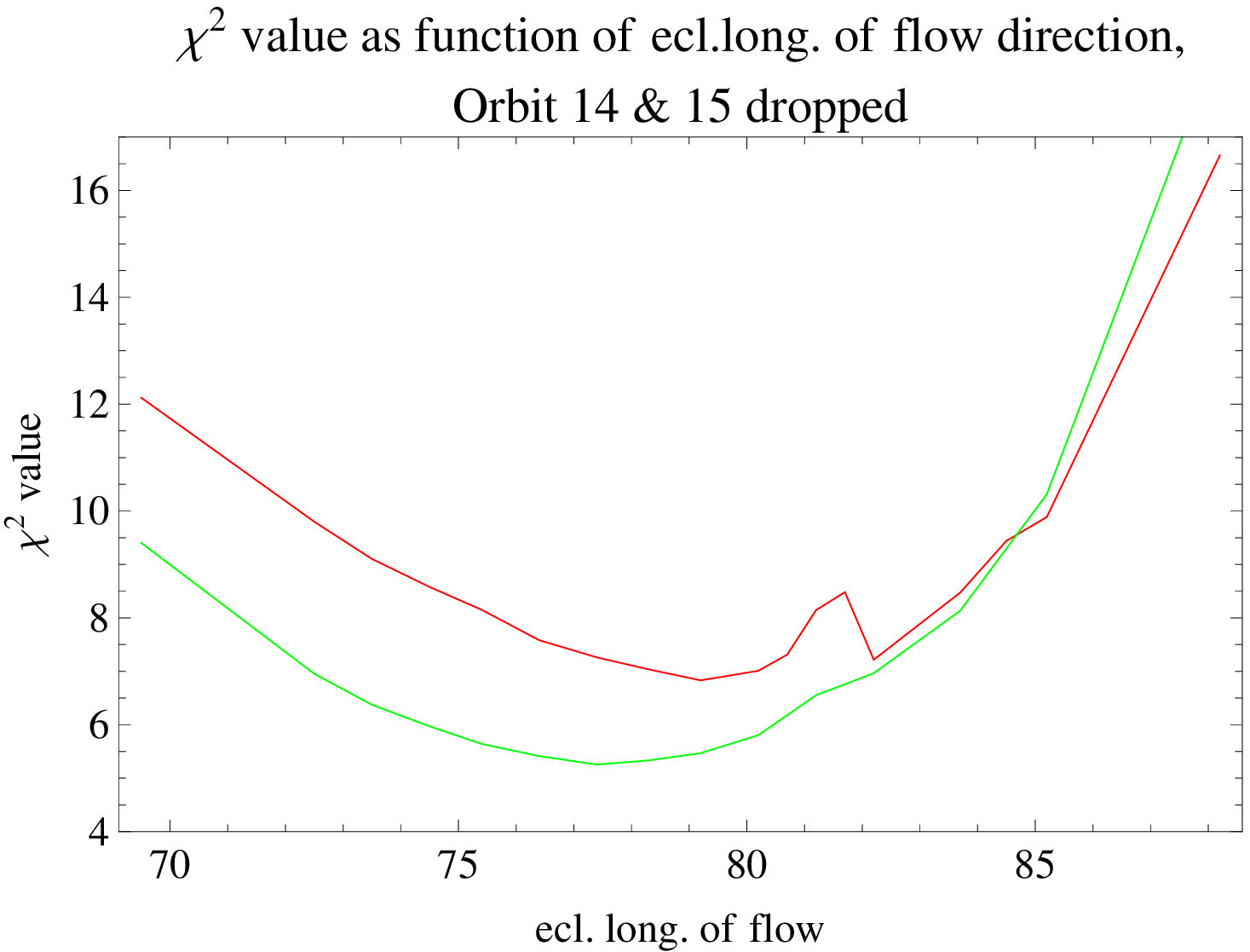}{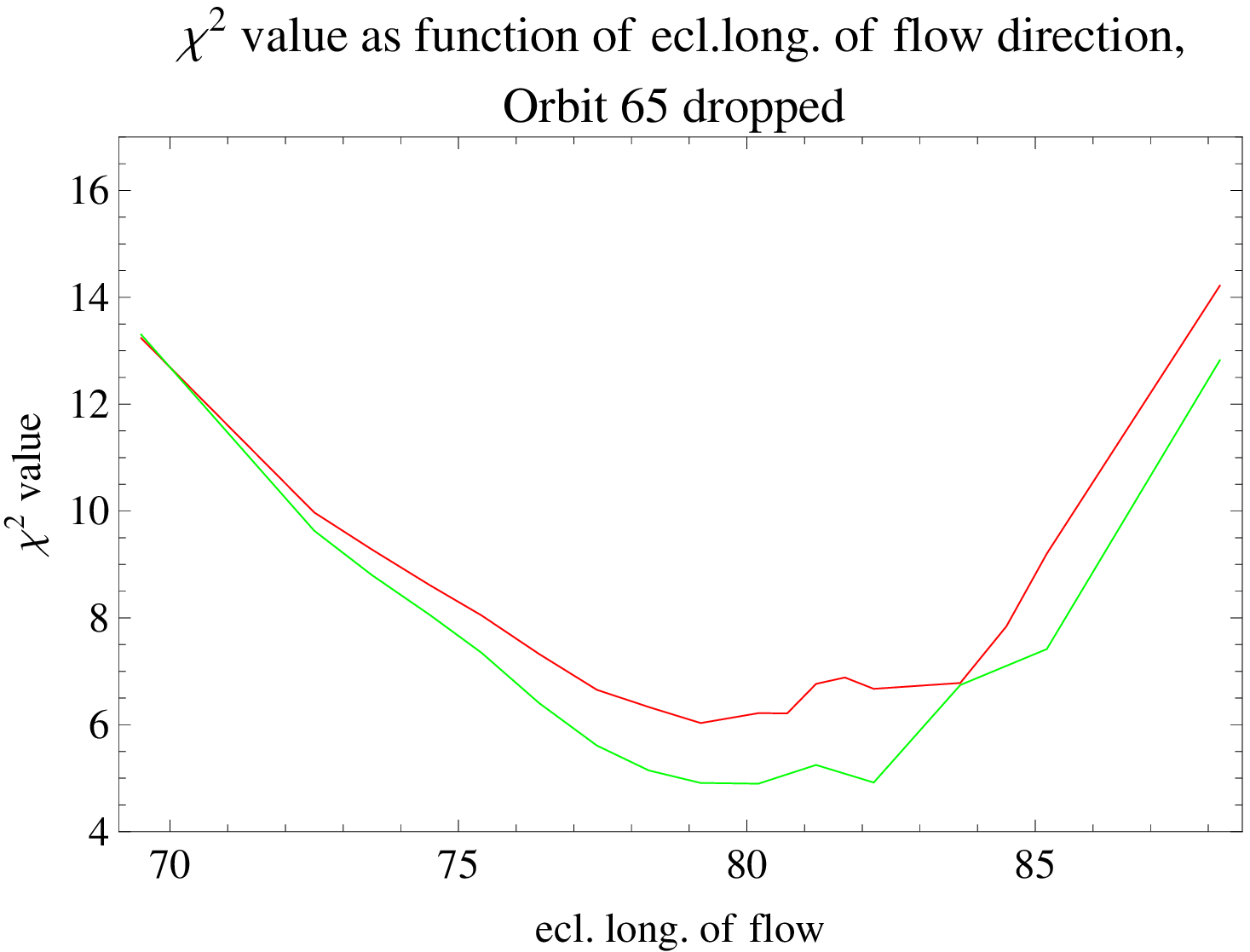}\\
		\caption{Values of $\chi^2$ for the 2009 (left) and 2010 seasons (right) with the contributions from orbits 14 and 15 and 65 removed, shown as green lines. Red lines are repeated from Fig.~\ref{figChi2tD00} for comparison.}
		\label{figChi2OrbDropped}
		\end{figure*}

The results of this procedure are shown in Fig.~\ref{figChi2OrbDropped}. Removing Orbits 14 and 15 eliminates the hump seen in $\chi^2$ vs $\lambda$ plots about $\lambda = 82\degr$, and removing Orbit 65 results in a lowering of minimum values and shifting position toward greater $\lambda$, even though it does not remove the correlation of the $\chi^2$ values from individual orbits with the angular separation of the Earth from the position of the peak flux, shown in Fig.~\ref{figChi2Contrib}. Thus, the source of the unexpected features in the $\chi^2$ vs $\lambda$ plots was identified. However, inspection of the observations, select ISM flow observation times, and the entire measurement process for these orbits did not reveal any reasons why the quality of the data from these orbits should be suspect. Hence there is no obvious reason to reject the three orbits. Instead, we adopted the size of the irregularities as an indicator of the uncertainty in $\chi^2$ values and used it to constrain the regions of acceptable values of the gas flow parameters. 

\begin{figure*}[t]
		\centering
		\begin{tabular}{cc}
		\includegraphics[scale=0.6]{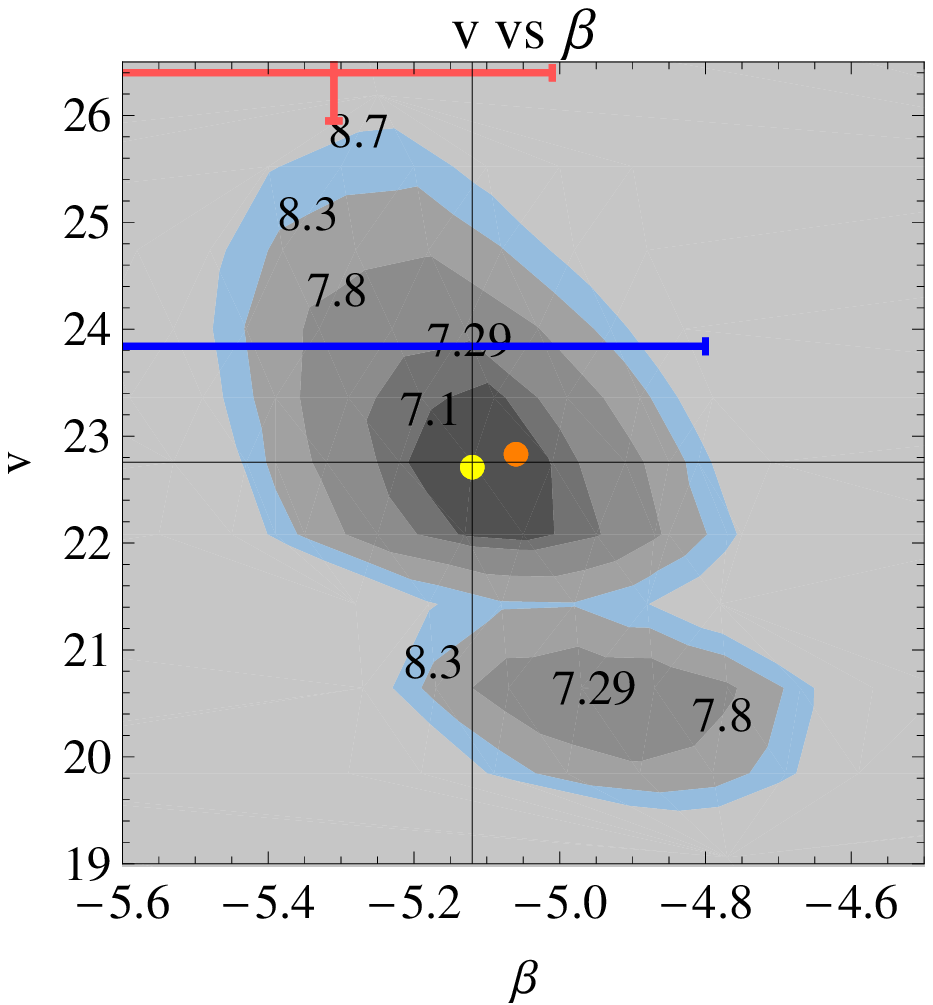} &	\includegraphics[scale=0.6]{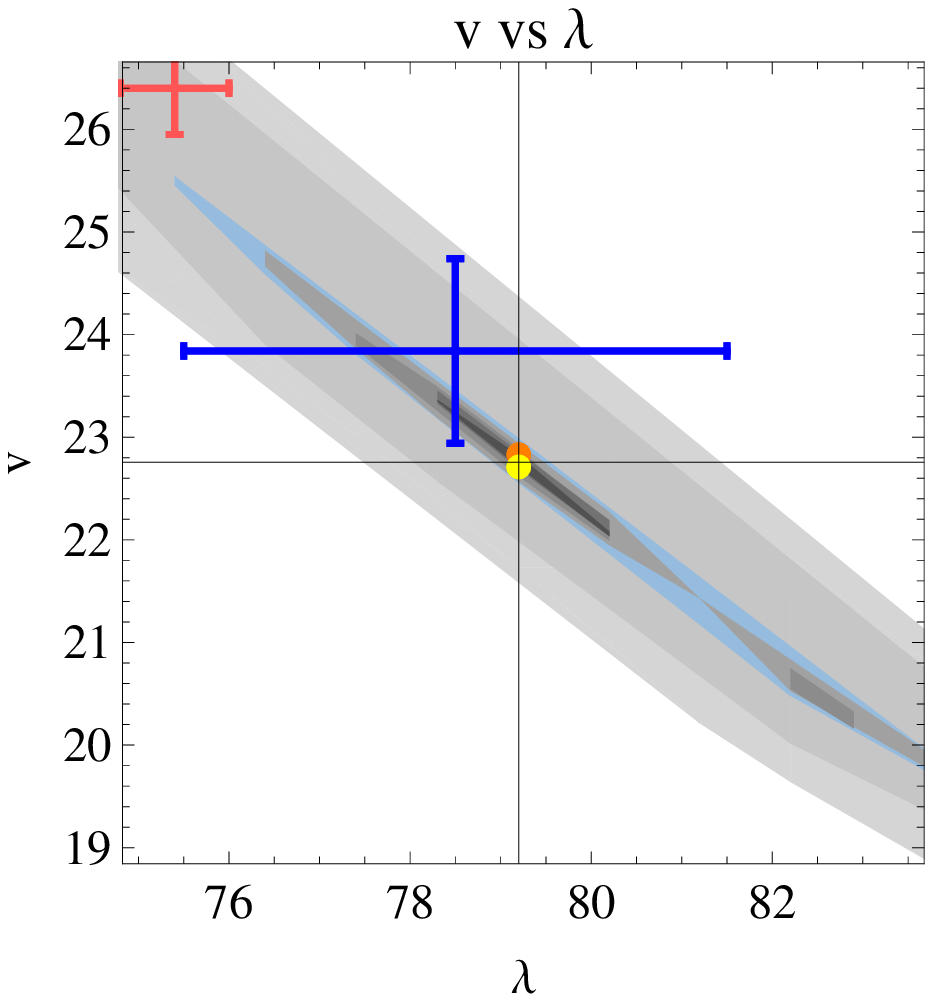}\\
  	\includegraphics[scale=0.6]{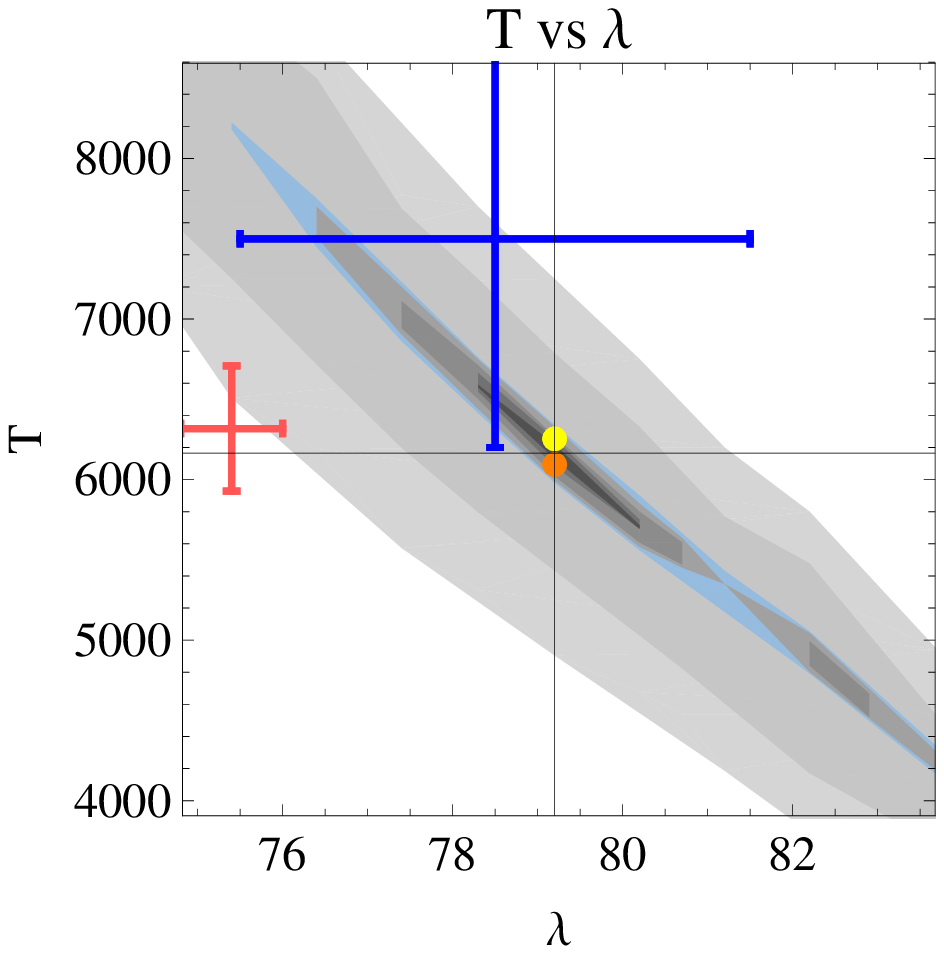} &	\includegraphics[scale=0.6]{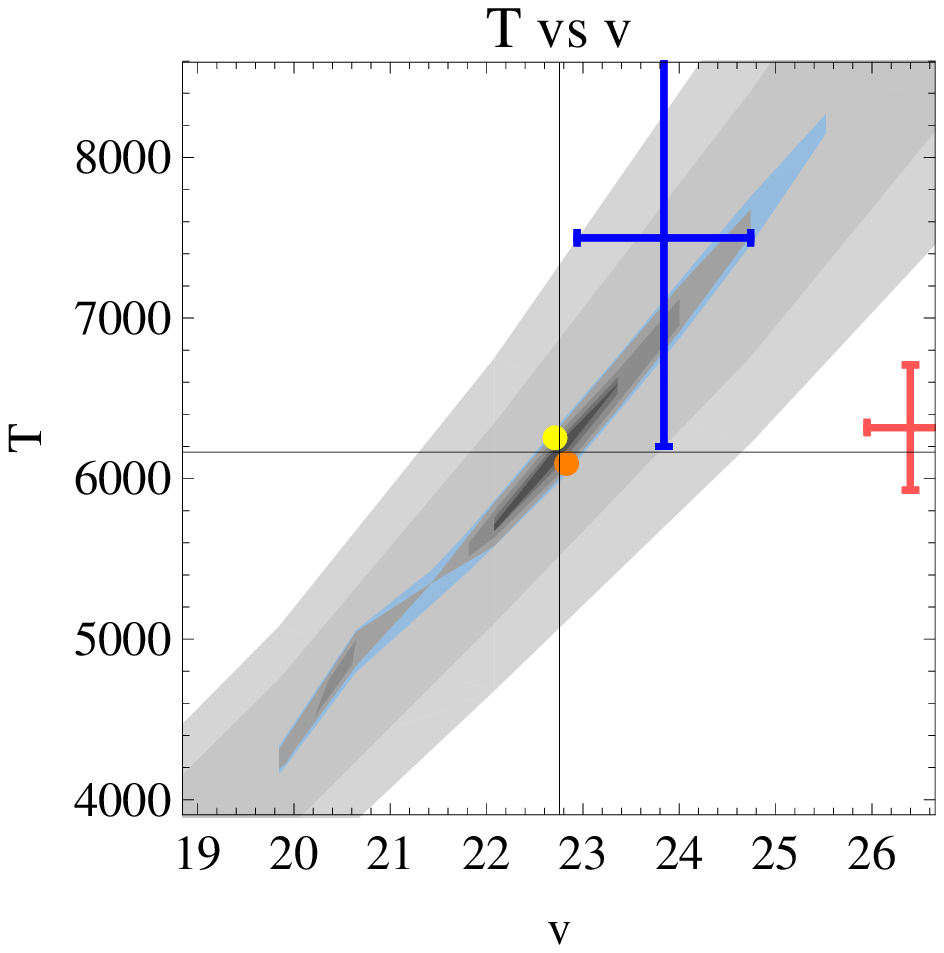}\\
		\includegraphics[scale=0.6]{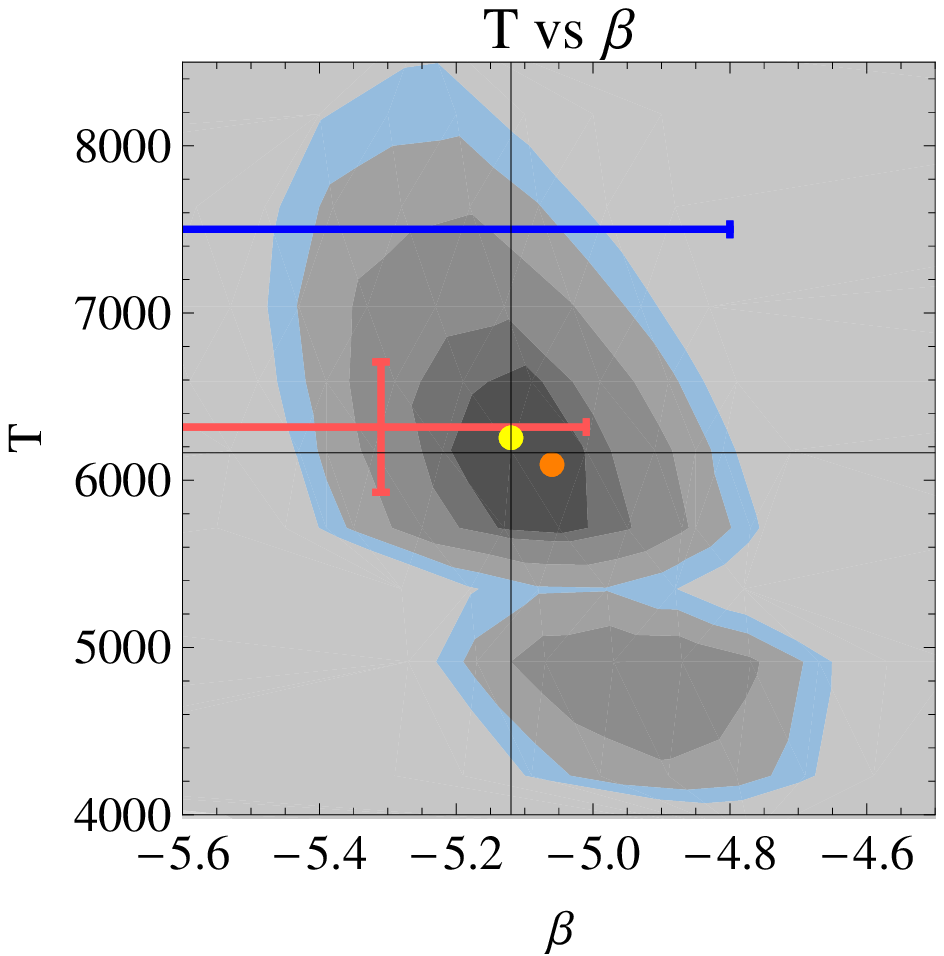} &	\includegraphics[scale=0.6]{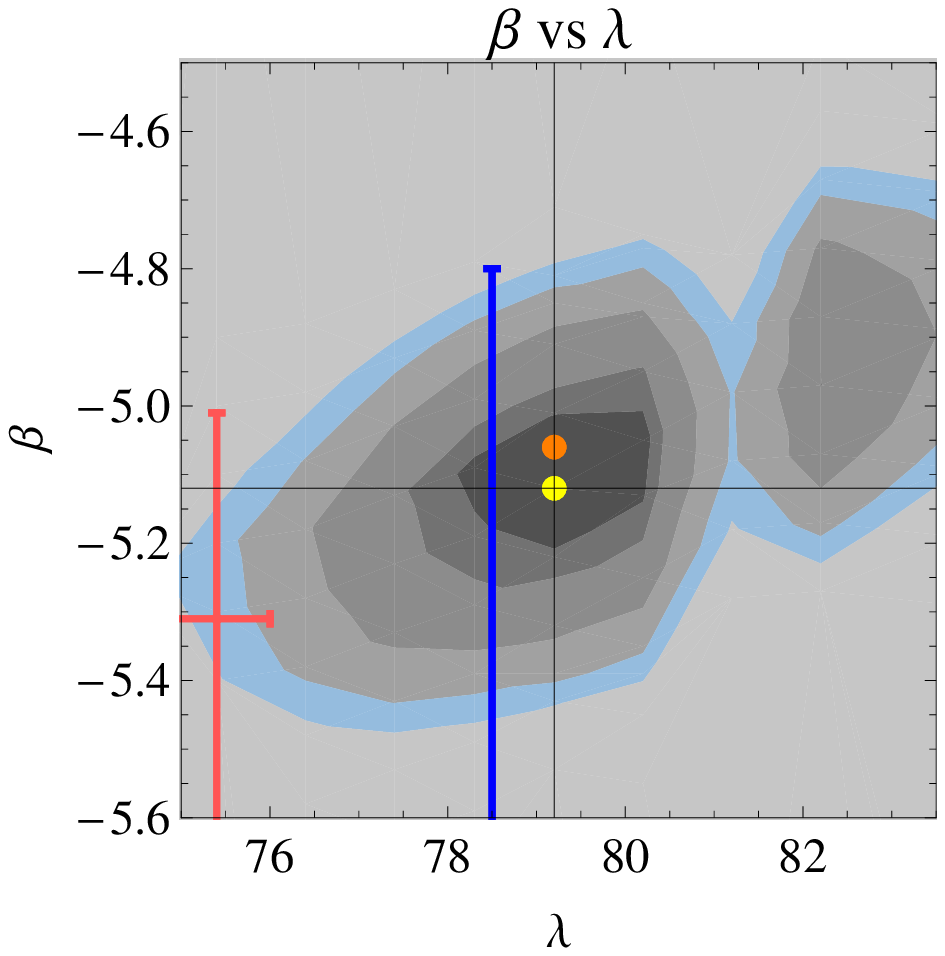}\\
		\end{tabular}
		\caption{{\tiny Parameters of the NISHe flow that yield $\chi^2$ values below the limit of 8.7 (inside the blue contour and including the darker contours), shown in cuts through the two-parameter sub-spaces of the 4D parameter space. The best-fit solution is indicated by the black cross-hairs, while the solution obtained by \citet{witte:04} is marked with the smaller error bar cross. The yellow and orange points are the solutions obtained separately for the 2009 and 2010 seasons. The larger error bar cross (only fragments visible in some of the panels) marks the LIC flow vector from \citet{redfield_linsky:08a}. $\chi^2$ values for the contours are indicated in the upper left panel; the values for the two lightest contours, from which the darker one makes the background in $v$ vs $\beta$, $T$ vs $\beta$, and $\beta$ vs $\lambda$ panels, are 30 and 87. Note that the boundaries of the two lightest panels are very approximate because of the sparse coverage of theses regions of the 4D $\chi^2$ space.}  }
		\label{figParamCuts}
		\end{figure*}

To this end, we selected $\chi^2 = 8.7$ value as the upper limit, we used all the points that return $\chi^2\leq8.7$ and we plotted blue contours surrounding the geometric location of those simulation points in two-parameter cuts of the $\chi^2$ space in Fig.~\ref{figParamCuts}. They form well-defined regions in the two-parameter sub-spaces. We consider these regions as regions of acceptable values of NISHe gas parameters. 

To be acceptable, the components $(\lambda, \beta, v, T)$ of a parameter set $\vec{p}$ must be within the acceptable regions in all panels of Fig. \ref{figParamCuts} with no exceptions. Even one exception invalidates a given solution. The acceptable values of the flow longitude vary from $75.2\degr$ to $83.6\degr$, but there are only limited ranges of the other parameters possible for a given value of the flow longitude.  Similarly, the acceptable velocities range from $\sim20$ to $25.5\; \mathrm{km}\, \mathrm{s}^{-1}$, but for a given velocity value only a narrow range of the remaining parameters is acceptable. It is clear, for example, that the solution for the NISHe flow parameters obtained by \citet{witte:04}, shown as the smaller of the two error bar crosses in Fig. \ref{figParamCuts}, is outside the region permitted by our analysis. If we take the flow longitude identical as obtained by \citet{witte:04}, then the gas temperature must be in a narrow range about 8000~K, and velocity must be $\sim 25.5 \, \mathrm{km} \, \mathrm{s}^{-1}$, both outside the error space. Thus we conclude that our solution differs from the solution obtained by \citet{witte:04} and from the consensus solution from \citet{mobius_etal:04a} on a statistically significant level. 

By contrast, the LIC flow vector that \citet{redfield_linsky:08a} obtained from a careful analysis of all available lines of sight toward nearby stars agrees with our solution. These error bars, which appear large on the scale of the figures, intersect the acceptable region in all panels. 

We also show approximate contours for a few other select levels of $\chi^2$, as indicated in the upper left panel of Fig.~\ref{figParamCuts}. A level of 7.29 corresponds to the depth of the secondary minimum in $\chi^2$ space (cf. Fig.~\ref{figChi2tD00}); the secondary minimum is then plotted as a tiny dot. We plot two contours between the values of the primary and secondary minimum in the $\chi^2$ space, to better illuminate the 3D shape between the two minima. We also show two contours outside the acceptable region: they are for the $\chi^2$ levels above 35 and above 80. While the sampling of these regions in the $\chi^2$ space is much sparser, we are confident we have not missed any significant secondary minimum. The figure illustrates how deep the valley is in the $\chi^2$ space.

		\clearpage
	
		\clearpage
		
		\begin{figure*}[t]
		\centering
		\epsscale{2}
		\plotone{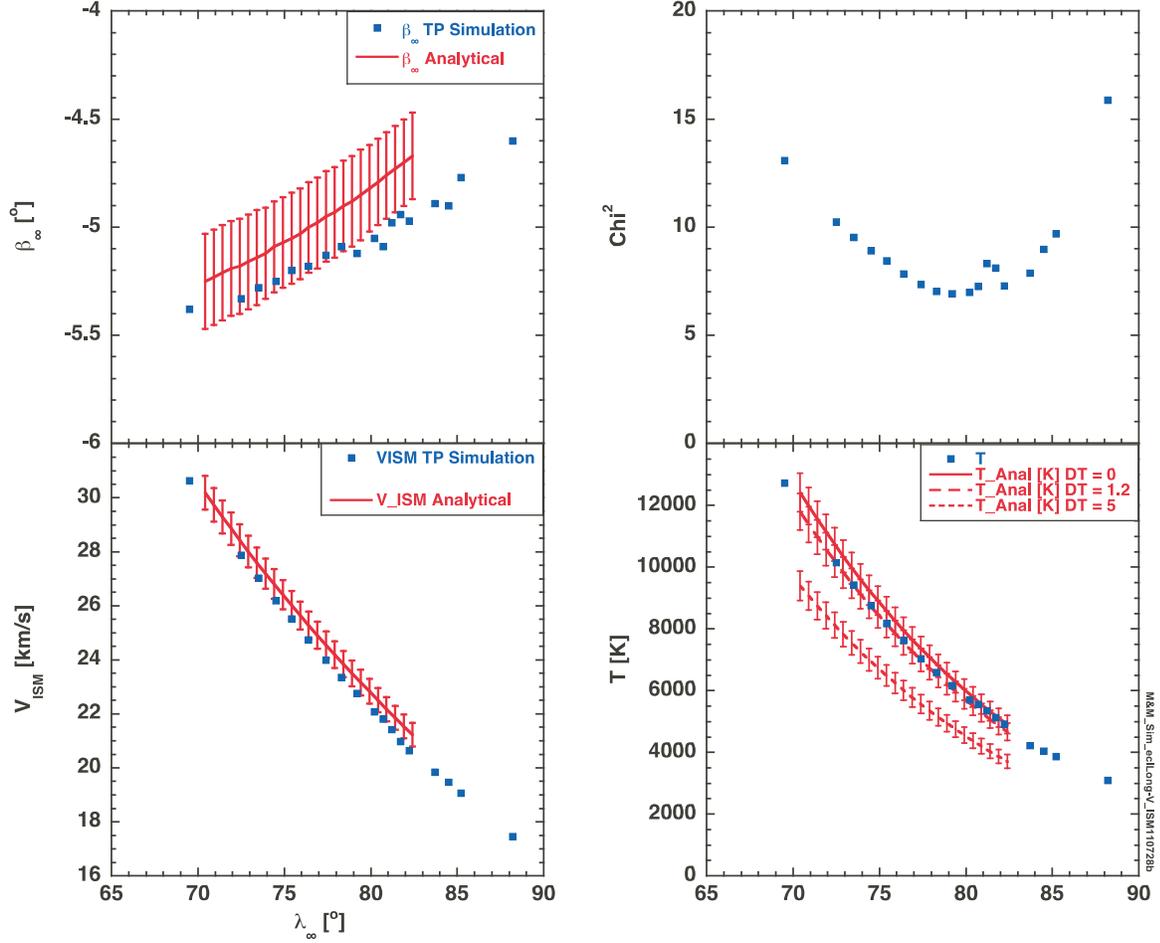}
		\caption{Comparison of analytic results obtained by \citet{mobius_etal:12a} with the results from this paper. Shown are relations between the flow direction and flow latitude (upper left panel), inflow velocity (lower left panel), and gas temperature (lower right panel). Upper right panel presents $\chi^2$ values as function of the flow direction that corresponds to the simulation points used in the remaining panels; basically it is a repetition of Fig. \ref{figChi2tD00}. The temperature panel shows a relation which assumes that all events are transmitted by the data system (upper line), along with two relations that assume a rate-dependent data loss in the transfer to the data system, whose magnitude is currently not well known yet. The system loading-related loss is modeled like a characteristic dead time of 1.2~ms (middle line) and 5~ms (lower line) \citep{mobius_etal:12a}.}
		\label{figNumVsAnalyt}
		\end{figure*}
		\clearpage
	
Another test was to compare the $v\left(\lambda\right)$, $T\left(\lambda\right)$ and $\beta\left(\lambda\right)$ relationships for the parameter sets forming the line in Fig.~\ref{figChi2tD00} with the approximate analytical relation specified by \citet{mobius_etal:12a}. Fig.~\ref{figNumVsAnalyt} shows that there is a very good agreement between the numerical and analytical relationships. 
	
From this analysis we conclude that the parameters of the pristine population of the Local Interstellar Cloud gas are those as found as the $\vec{p}_{\mathrm best}$ solution from the fitting of the numerical model of the gas flow to the data from the 2009 and 2010 observation seasons: $\lambda = 79.2\degr$, $\beta = -5.1\degr$, $v = 22.8 \, \mathrm{km}\,\mathrm{s}^{-1}$, $T = 6200\, \mathrm{K}$. The region in the 4D parameter space where the solutions are acceptable represent a relatively narrow range of tightly related values, whose 2-parameter cuts are shown in Fig. \ref{figParamCuts}. The inflow parameters fitted separately to the observations from the two seasons agree very well with the value obtained from the combined two seasons. There is not any reasonable trace of year-to-year change in the parameters within statistical uncertainties. 

\citet{mobius_etal:12a} discuss the effect of count losses due to the loading of the IBEX-Lo to data system interface and the data system at high count rates. This data loss increases with count rate and thus effectively broadens the angular flow distributions, which described like an effective dead time of the data transfer system, has been estimated from the data to range between 1.2 and at most 5~ms. To assess the influence of this effect on the solution, we introduced approximate corrections to our orbit-averaged data sets for dead times of 1.2~ms and 3~ms and repeated the minimizations. The results called for lower temperatures of $\sim4300$ -- 4700~K, flow longitude of $\sim 82.2\degr$, latitude $\sim -5.0\degr$~and speed $\sim 20.7\, \mathrm{km} \, \mathrm{s}^{-1}$. Even though the absolute values of minimized $\chi^2$ were greater than without the correction and were increasing with increasing dead time, the solutions were still within the acceptable region of solutions obtained for no dead time correction. We consider this result as a confirmation of robustness of the estimates of our acceptable range of NISHe flow parameters.

Apart from the Gaussian core that is produced by the pristine population of neutral interstellar helium, the observations suggest some additional populations of neutral gas are present in the earliest and latest orbits of the observation seasons. These additional populations are visible at ecliptic longitudes above $\sim 180\degr$ and below $\sim 95\degr$, as shown in the upper row of panels in Fig.~\ref{figESvsGT} and in the upper-right panel of Fig.~\ref{figSpinAxEffect}. Another evidence of additional populations may be the quasi-regular behavior of the contributions to $\chi^2$ from individual orbits, shown in Fig.~\ref{figChi2Contrib}. However, these might also be due to losses of counts in the sensor during high count rate intervals, as discussed by \citet{mobius_etal:12a}. 

At Earth longitudes above $\sim 180\degr$, IBEX observes the neutral interstellar hydrogen population as discussed in greater detail by \citet{saul_etal:12a}. An unexpected feature is the signal observed at ecliptic longitudes below $95\degr$ during the 2010 season, which we were not able to model with any reasonable parameters for the pristine NISHe gas. We interpret this signal as the discovery of an additional source of neutral helium in or near the heliosphere. This population is seen best in the 2010 observations (with some traces in 2009), as shown in the upper-right panel of Fig. \ref{figESvsGT}, because in 2009 science operations were not carried out when the Earth was in the longitude interval in question. We elaborate on this aspect in the discussion section. 

	\section{Discussion}
The results obtained here differ from the earlier consensus results which were based mostly on the observations from the GAS/Ulysses experiment \citep{witte:04}. The values about 7 for $\chi^2$ obtained from our fits must be contrasted with $\chi^2$ values for the earlier consensus parameter set suggested by \citet{witte:04, mobius_etal:04a}, which are on a level of ~139 for the 2009 observations series, $\sim164$ for the 2010 series, and $\sim144$ for both seasons together. 

We do not believe the difference between the consensus result \citep{mobius_etal:04a} and those found here is due to spatial/temporal variation in the interstellar gas ahead of the heliosphere. While we have analyzed just two years of data and are still unable to determine whether the small differences between the seasons are physical, we note that the results reported by Witte and his collaborators from the measurements by GAS/Ulysses were consistent within the stated uncertainties starting from the first report by \citet{witte_etal:93} and through the last ones by \citet{witte_etal:04a, witte:04}. In our opinion, it is highly unlikely that an abrupt change in the gas parameters at the entrance to the heliospheric interface occurred during the $\sim 6$~years between the end of GAS observations and beginning of the IBEX observations, with no changes during the earlier and later times. Additional evidence supporting this position is the fact that the signal we observe is compatible with a Maxwellian distribution with a small addition of the extra population we discovered, which we believe is due to processes operating within the heliospheric interface rather than within the interstellar gas. A near-Maxwellian distribution suggests a particle distribution in quasi-thermal equilibrium. The spatial scale of the near-equilibrium conditions must be comparable to at least the length of Sun's path through the interstellar gas since the beginning of Ulysses observations, i.e., since 1990. For a speed of the Sun relative to the gas 5 AU/y, this is only $\sim 100$~AU, i.e., shorter than the mean free path for charge exchange reaction in this environment. Charge exchange is the interaction mostly responsible for equilibrating the neutral component with the ionized component, which in turn is interrelated with disturbances in the local magnetic field. This suggests that the Sun is likely in a local region of space that is homogeneous on spatial scales at least on the order a few hundred of AU. Based on spectroscopic observations of interstellar matter lines along sightlines toward stars in the Hyades, \citet{redfield_linsky:01a} concluded that inhomogeneities in the LIC are on a spatial scale of $10^5$~AU.  

Information on interstellar gas in the Galactic neighborhood of the Sun is available mostly from observations of interstellar absorption lines in the spectra of nearby stars, reviewed very recently by \citet{frisch_etal:09a, frisch_etal:11a}. This information is derived from interpretation of absorption profiles collected from the lines of sight to nearby and more distant stars, i.e., at spatial scales from a few to a hundred parsecs from the Sun. The interstellar absorption scale is thus longer from the distance scale of our measurements by more than 3 orders of magnitude, which makes direct comparisons challenging. Indeed, the classical interpretation of interstellar lines of sight calls for an assumption that the gas is composed of separate clouds with some temperature and a certain level of turbulence \citep[the $\xi$ factor]{redfield_linsky:04a}. The lines are typically fitted with a number of components featuring Voigt profiles with different parameters, until a satisfactory agreement of the observed and fitted shape is obtained \citep[see, e.g.,][]{lallement_etal:95a, linsky_etal:00a, redfield_linsky:04a}. 

The Sun seems to be within an old remnant of a series of supernova explosions \citep[see discussion in ][]{redfield_linsky:00,fuchs_etal:09a, frisch_etal:11a} and modeling of such remnants suggests the material is expected to be turbulent and fragmented at many spatial scales \citep{breitschwerdt_etal:09a}. Spectroscopic measurements analyzed by \cite{frisch_etal:02a} suggest that the Sun might be still in one of the complex of local interstellar clouds called the Local Cloud, but quite close to its boundary, which might be between the Sun and the nearest star $\alpha$~Cen. This cloud is expected to have a velocity relative to the Sun about 25 km~s$^{-1}$, compatible with the results from GAS/Ulysses. The adjacent cloud, the so-called G-cloud \citep{lallement_etal:95a} is expected to be a few km/s faster. \citet{frisch_etal:02a} do not rule out that the Sun might be within a gradient of the gas velocity between the two clouds. More recent analysis by \citet{redfield_linsky:08a}, based on a more extensive observations material, suggests that the flow vector of the Local Cloud (converted to the J2000 ecliptic coordinates) is $ v = 23.84 \pm 0.90$~km~s$^{-1}$, $\lambda = 78.5\degr \pm 3\degr$, $\beta = -7.8\degr \pm 3\degr$, which (within the error bars) is in a very good agreement with our findings. Also the temperatures agree well: their $7500 \pm 1300$~K (plus the microturbulence parameter $\xi = 1.62 \pm 0.75$~km~s$^{-1}$) with our 6200~K.  

Another source of information on the kinematics of the interstellar material just outside the heliosphere might be interstellar dust. Interstellar dust grains were unambiguously identified on a number of spacecraft in the inner and outer heliosphere \cite[for review see][]{krueger_gruen:09a} and the direction of inflow seemed to be in a very good agreement with the helium inflow direction we have obtained: downwind ecliptic longitude of $79\degr \pm 20\degr$ and latitude $-8 \degr \pm 3\degr$ \citep{frisch_etal:99a}. This is in better agreement with our result than with the consensus result from \citet{mobius_etal:04a}. However, after 2005 the direction of inflow changed by $\sim 30 \degr$ southward \citep{krueger_etal:07a} due to a still unexplained phenomenon, which questions the direct usability of the inflow direction of interstellar dust for the studies of kinematics of the very local interstellar material.

The fact that our observations suggest that multiple solutions for the flow vector of the NISHe gas are almost equally possible is not surprising. Already very early determinations of the NISHe flow vector based on EUV observations of the neutral helium glow \citep{chassefiere_etal:88a, chassefiere_etal:88b} suggested many possible solutions, with either a slower speed and lower temperature, or a higher speed and higher temperature, just as our results do. The range of velocities they reported is compatible with ours, but their temperatures are higher by $\sim 2000$~K than ours. Also \cite{witte_etal:93} seem to suggest the existence of an ``alley'' in $\chi^2$ space, but never expanded on this in their subsequent papers. Hence our result is not in conflict with the bulk of the prior knowledge, even though it is statistically significantly different from the solution obtained from GAS/Ulysses; in contrast, our results are in very good agreement with the results of a recent and highly sophisticated study of the local gas kinematics by \citet{redfield_linsky:08a}.

The most pronounced and possibly far-reaching result is, in our opinion, the new flow velocity: $\sim 22.8 \, \mathrm{km}\, \mathrm{s}^{-1}$ as compared with the earlier $26.4 \, \mathrm{km}\, \mathrm{s}^{-1}$. This is a reduction of $\sim 15\%$, which results in a decrease in the ram pressure the interstellar gas exerts on the heliosphere by $\sim 25\%$. 

The size of the heliosphere is a result of the pressure balance between the outward pressure of the solar wind, and of the inward pressure from the Local Interstellar Cloud material. The components of the inward and outward pressures have been extensively discussed in the literature (see, e.g., \citet{fahr_etal:00, baranov:09a}) and include ram and thermal pressure of the solar wind thermal core and pickup ions, magnetic field pressure and pressure of the anomalous and Galactic components of cosmic rays. While the main pressure components from the core of the solar wind at the termination shock are known from in situ measurements of the solar wind \citep{richardson_etal:08a, burlaga_etal:08a}, the pressure from pickup ions and the components of the inward pressure from the LIC are known only indirectly, mostly from modeling based on the limited observations available. The LIC pressure components include mostly the ram pressure of the ionized component, mediated by the interaction with various neutral components, and supposedly the pressure from the external magnetic field. Thermal pressure plays a minor role. The change in the ram pressure at the LIC side must result in a change in the magnitudes of the other pressure components, to the first approximation without a change in the net pressure. Since the distance to the termination shock, known from the distances of the Voyagers crossings \citep{stone_etal:05a, burlaga_etal:08a} and modeling of the heliospheric size in the presence of an external magnetic field \citep{pogorelov_zank:06a,  pogorelov_etal:09a, grygorczuk_etal:11a}, is specifically driven by the ram pressure and external magnetic field, the reduction in the LIC velocity will require recalculation of the present heliospheric models and is likely to change the estimates of the external field as well as the proportion between the primary and secondary populations of neutral interstellar hydrogen in the heliosphere and their bulk velocity and temperature inside the termination shock. These changes in turn will affect the pickup ion production in the solar wind and the pickup ion supply to the inner heliosheath, again changing the pressure balance in the heliosphere, because of the very important role of pickup ions for the thermal pressure in this region. 

The change in the longitude of the flow direction by $3.8\degr$, from $75.4\degr$ to $79.2\degr$, is seemingly small, but it has a notable effect on the orientation of the Hydrogen Deflection Plane (HDP) suggested by \citet{lallement_etal:05a}. By definition, the HDP is the plane that contains the inflow vectors of neutral interstellar hydrogen and helium as observed in the inner heliosphere. It is believed that the flow vector of helium is practically unaffected by the interactions going on at the heliospheric interface. However, the flow of hydrogen should be strongly disturbed.  Since an external magnetic field is supposed to introduce a distortion of the heliosphere from axial symmetry, the direction of the H flow should be different from the direction of He. The discovery of a difference between the flow directions of He and H suggested that the heliosphere is indeed distorted and the most obvious cause is action of the external magnetic field. \citet{lallement_etal:05a} surmised that the external magnetic field vector may be in the HDP, a suggestion that obtained mixed support from the heliospheric modeling community \citep{pogorelov_etal:09a, izmodenov_etal:05a, izmodenov_alexashov:06a, zank_etal:09a}. But if indeed the external $\vec{B}$ vector is located in the HDP, as the more recent simulations suggest, then the new inflow direction of the NISHe gas reported in this paper, together with a refined direction of the H flow reported by \citet{lallement_etal:10a}, suggest a new geometry of the magnetic field in the LIC near the heliosphere. The magnetic field vector should be located in the HDP determined by the normal direction $\left(\lambda=357.51\degr ,\, \beta=58.51\degr\right)$, while the normal obtained from the previous estimates of the flow vectors by \citet{witte:04} and \citet{lallement_etal:05a, lallement_etal:10a} is $\lambda = 349.52\degr$, $\beta=32.29\degr$. Even though the error bars for both these determinations are big, the difference is significant and equal to $\sim21\degr$.  

Comparison of our extensive simulations with measurements suggests that a secondary population of neutral helium must be present at Earth orbit because, despite the fact that the simulations covered a very wide range of interstellar gas parameters values, we were unable to reproduce count rate profiles observed at orbits before 13 during the 2009 campaign and before 60 during the 2010 campaign, as illustrated in the upper row of Figs~\ref{figESvsGT} and \ref{figSpinAxEffect}. In particular it is clear that peak heights observed during the 2010 season at ecliptic longitudes lower than $\sim 95\degr$ are not reproduced well by the simulations performed with parameters best fit to Orbits 61 to 68. In fact, we were unable to fit these observations with any parameter set from the $\sim 4000$ tried. A similar situation happens for the orbits corresponding to Earth longitudes greater than $\sim 180\degr$. In this case we interpret the excess signal as due to interstellar hydrogen, as suggested by the simulations shown in Fig.~\ref{figHvsHeFlux}. The H signal is discussed in greater detail by \citet{saul_etal:12a}.   

Earlier studies \citep{mueller_zank:03a, mueller_zank:04a} suggested a possible secondary population of neutral helium from the charge exchange between the He$^+$ ions and H atoms in the outer heliosheath at a level of about 1\% of the primary. In these studies the only source of the secondary He population was the charge exchange reaction between the interstellar He$^+$ ions and neutral H in the outer heliosheath. In contrast, we believe that such a population could come from charge exchange between neutral interstellar He atoms and interstellar He$^+$ ions within the piled up and heated plasma in the outer heliosheath. In the following we will qualitatively assess whether such a hypothesis is justified. 

The ionic state of the interstellar He gas in the LIC is thought to be 0.611 He, 0.385 He$^+$, and 0.00436 He$^{++}$, as obtained by \citet{slavin_frisch:08a} as one of results of a research program \emph{Diagnostic of interstellar hydrogen} by an ISSI Working Group ``Interstellar Hydrogen in the Heliosphere'' (see \citet{richardson_etal:08a, bzowski_etal:08a, pryor_etal:08a} for other results from this campaign). The plasma in the outer heliosheath is compressed, slowed down and heated, as all modern heliospheric models suggest \citep{mueller_etal:08a}. An illustration of the gas parameters along the upwind direction can be found, e.g., in \citet{izmodenov_etal:05b}. The plasma density increases from the interstellar value of 0.06 cm$^{-3}$ to $\sim 0.14\, \mathrm{cm}^{-3}$ and the temperature from $\sim6000$ K to $\sim 35 000$~K. The typical plasma bulk speed in the outer heliosheath along the upwind direction is just $\sim 4 \, \mathrm{km}\, \mathrm{s}^{-1}$ sunward, while the primary components of both H and He maintain their interstellar speed. The plasma pile up results in a net difference in bulk velocities between the two interacting components, which adds to the typical relative speed of He atoms with respect to the ions.

Assuming the ionization state of the gas does not change in the outer heliosheath relative to the unperturbed LIC, we can calculate the density of the He$^+$ and He$^{++}$ ions in the heliosheath by multiplying the densities from the LIC by the typical plasma compression factor $0.14/0.06 = 2.33$. The base number for this calculation is the density of neutral interstellar He in the LIC equal to 0.015~cm$^{-3}$ \citep{witte:04}. The reaction rates defined as:
		\begin{equation}
		\beta=n_{\mathrm{target}}v_{\mathrm{rel}}\sigma_{\mathrm{cx}}\left(v_{\mathrm{rel}}\right)
		\label{eqCXRate}
		\end{equation}
will critically depend on the relative speed between the colliding partners. For the temperature 35000~K, the mean speed of He atoms and ions will be
		\begin{equation}
		u_{\mathrm{T},\mathrm{He}}=\sqrt{\frac{8kT}{\pi m_{\mathrm{He}}}}=13.6 \, \mathrm{km}\, \mathrm{s}^{-1}
		\label{eqUTHe}
		\end{equation}
in the reference frame co-moving with the gas. Simultaneously, the most probable speed of neutral He at 6300~K will be $5.8\, \mathrm{km}\, \mathrm{s}^{-1}$ and most probable speed of protons at 35000~K will be $27.1\,\mathrm{km}\, \mathrm{s}^{-1}$.
	
\begin{figure*}[t]
		\centering

		\plotone{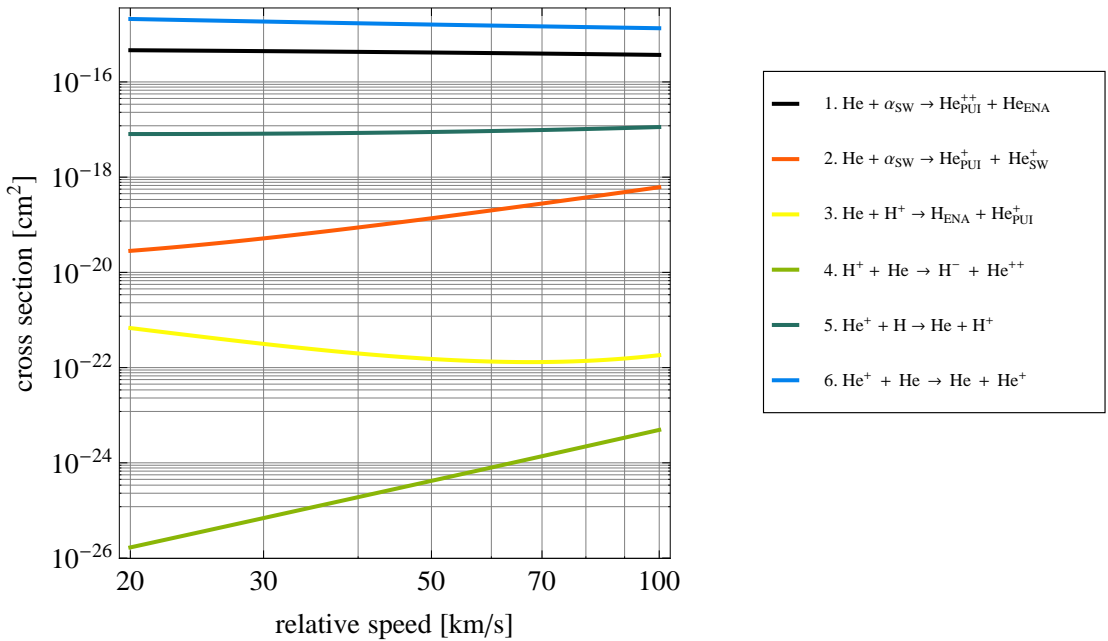}
		\caption{Cross sections for charge exchange reactions between hydrogen and helium atoms and ions as function of collision speed \citep{phaneuf_etal:87}. PUI denotes pickup ions, ENA energetic neutral atoms (in contrast to the atoms with energies close to typical energies of the neutral interstellar gas), SW is short for solar wind.}
		\label{figHeCrossSections}
		\end{figure*}

The process of filtration of the primary population and simultaneous production of the secondary has a kinetic character and thus here we are only able to crudely assess reaction rates in order to determine which potential processes must be taken into account and which can be neglected. To that end, we approximate the relative velocity of the colliding partners by means of harmonic sum of their bulk and thermal velocities:
		\begin{equation}
		v_{\mathrm{rel},\,12}=\sqrt{\left|v_{\mathrm{B1}}-v_{\mathrm{B2}}\right|^2 + u_{\mathrm{T1}}^2+u_{\mathrm{T2}}^2}
		\label{eqRelVelOutHS}
		\end{equation}
where $v_{\mathrm{B}}$ is the bulk velocity of collision partners 1, 2 and $u_{\mathrm{T}}$ is the most probable speed given by Eq.~(\ref{eqUTHe}). We assume here that the primary components of both He and H flow along the upwind line in the outer heliosheath maintaining their original unperturbed temperature and velocity of 6165~K and $22.756 \, \mathrm{km}\, \mathrm{s}^{-1}$, while the secondary components have the temperature and bulk velocity of the ambient plasma within the outer heliosheath, adopted here as 35000~K and $4 \, \mathrm{km}\, \mathrm{s}^{-1}$, respectively. Based on these numbers and on the calculations of the thermal velocities presented above it is clear that the relative velocity between the collision partners in the outer heliosheath will be less than $100 \, \mathrm{km}\, \mathrm{s}^{-1}$. Accordingly, in Fig. \ref{figHeCrossSections} we show the cross sections for potentially relevant charge exchange reactions for the relative velocities below $100 \, \mathrm{km}\, \mathrm{s}^{-1}$ \citep{phaneuf_etal:87}.

The loss rates of the primary population from various reactions in the upwind direction are listed in Table \ref{tabPriLoss}. Losses of the primary He atoms in the outer heliosheath due to the charge exchange reaction between He and He$^+$ are by far the strongest, larger by 2 orders of magnitude than the losses from solar photoionization. 
	
	\begin{table*}
	\begin{center}
	\caption{Losses rates of the primary population of He in the outer heliosheath and other relevant parameters.\label{tabPriLoss}}
	\begin{tabular}{l c c c c}
	\tableline\tableline
	reaction & rel.speed (km s$^{-1}$) & $\sigma \, \left(\mathrm{cm}^2\right)$ & density $\left(\mathrm{cm}^{-3}\right)$ & rate (s$^{-1}$)\\ \tableline
	$\mathrm{He}+\alpha_{\mathrm{SW}}\rightarrow \mathrm{He}_{\mathrm{PUI}}^{+}+\mathrm{He}_{\mathrm{SW}}^{+}$ & 23.9 & $3.6 \times 10^{-20}$ & 0.00025 & $2.1 \times 10^{-17}$\\
	$\mathrm{He}+\alpha_{\mathrm{SW}}\rightarrow \mathrm{He}_{\mathrm{PUI}}^{++}+\mathrm{He}_{\mathrm{ENA}}$ & 23.9 & $4.6 \times 10^{-16}$ & 0.00025 & $2.7 \times 10^{-13}$\\
	$\mathrm{He}^{+} + \mathrm{He}\rightarrow \mathrm{He}+\mathrm{He}^{+}$ & 23.9 & $2.0 \times 10^{-15}$ & 0.035 & $1.7 \times 10^{-10}$\\
	$\mathrm{He}+\mathrm{He}^{+}\rightarrow \mathrm{He}_{\mathrm{ENA}}+\mathrm{He}_{\mathrm{PUI}}^{+}$ & 33.5 & $2.6 \times 10^{-22}$ & 0.14 & $1.2 \times 10^{-16}$\\
	$\mathrm{He}^{+}+\mathrm{He}\rightarrow \mathrm{He}^{-}+\mathrm{He}^{++}$ & 33.5 & $1.0 \times 10^{-25}$ & 0.14 & $4.8 \times 10^{-20}$\\
	photoion.@ 150 AU & & & & $4.4 \times 10^{-12}$\\ \tableline
	\end{tabular}
 	\end{center}
	\end{table*}
	
The gain reactions for the secondary component of neutral He in the outer heliosheath are detailed in Table \ref{tabSecGain}. Also in this case the $\mathrm{He}^{+} + \mathrm{He}$ charge exchange reaction dominates.

	\begin{table*}
	\begin{center}
	\caption{Gain rates for the secondary neutral He population in the outer heliosheath and other relevant parameters.	\label{tabSecGain}}
	\begin{tabular}{l c c c c}
	\tableline\tableline
	reaction & rel.speed (km s$^{-1}$) & $\sigma \, \left(\mathrm{cm}^2\right)$ & density $\left(\mathrm{cm}^{-3}\right)$ & rate (s$^{-1}$)\\ \tableline
	$\mathrm{He}+\alpha_{\mathrm{SW}}\rightarrow \mathrm{He}_{\mathrm{PUI}}^{++}+\mathrm{He}_{\mathrm{ENA}}$ & 23.9 & $4.6 \times 10^{-16}$ & 0.00025 & $2.7 \times 10^{-13}$\\
	$\mathrm{He}^{+}+\mathrm{H}\rightarrow \mathrm{He}+\mathrm{He}^{+}$ & 25.8 & $8.1 \times 10^{-18}$ & 0.035 & $7.3 \times 10^{-13}$\\
	$\mathrm{He}^{+} + \mathrm{He}\rightarrow \mathrm{He}+\mathrm{He}^{+}$ & 23.9 & $2.0 \times 10^{-15}$ & 0.035 & $1.7 \times 10^{-10}$\\ \tableline
	\end{tabular}
 	\end{center}
	\end{table*}

Thus from this simplified, qualitative analysis it follows that the main source of losses of the secondary population of interstellar helium in the outer heliosheath is charge exchange between the He$^+$ and neutral interstellar He. It is also the main source of the secondary population of neutral He produced in the outer heliosheath. The secondary loss mechanism is solar photoionization; the remaining reactions are unimportant. Details, however, strongly depend on particulars such as the exact ionization state of helium in the interstellar gas, the temperature and density of the material in the outer heliosheath, the bulk speed and temperature of interstellar gas etc. Discussion of these aspects is outside the scope of this paper, we only mention that the parameter values we used in this estimate were consequently adopted in agreement with the results from the coordinated \emph{Diagnostic of interstellar hydrogen} ISSI program and thus we consider them as realistic. 

With the estimates on the typical reaction rates on hand, we can make an order-of-magnitude estimate of the percentage losses of the primary population along the upwind direction. We stress that this is just an order of magnitude estimate that we make to check if the unexpected helium population we have observed can potentially be explained as the secondary population of interstellar He; comprehensive modeling is needed in order to obtain estimates suitable for comparison with our observations. 

The order-of-magnitude estimate of the production of the secondary population of He along the stagnation line in the outer heliosheath can be obtained from the ``optical density'' against losses, calculated as:
		\begin{equation}
		\tau_{\mathrm{He},\, \mathrm{gain}}=1-\mathrm{exp}\left(-\frac{50 \mathrm{AU}}{v_{\mathrm{B}, \, \mathrm{He}}} \beta_{\mathrm{He}, \, \mathrm{loss}}\right)\simeq 0.1
		\label{eqPriLoss}
  	\end{equation}
Hence we conclude that the production of secondary neutral He component in the outer heliosphere may be much more intense than previously thought and the hypothesis that IBEX discovered the secondary population of neutral interstellar He that comes up in the outer heliosheath is plausible. 

	\section{Summary and conclusions}
In this study, we performed an extensive modeling campaign to identify the best observing conditions and features of the expected signal from the NISHe gas measured by the IBEX-Lo detector and to check which elements must be included in the simulations used to establish the parameters of the flow of the NISHe gas in the Local Interstellar Cloud. 

We showed that if the distribution function of the NISHe gas in the LIC is Maxwellian, then the count rates of the neutral He atoms observed by IBEX-Lo as function of spin angle of the IBEX spacecraft should feature Gaussian cores. 

We identified the range of Earth ecliptic longitudes where the helium signal is expected to dominate over the signal from neutral interstellar hydrogen: from $\sim 110\degr$ to $\sim 170\degr$, which corresponds to orbits 13 through 20 and 61 through 68.

In order to maintain maximum fidelity of the simulations, the exact solar distances and velocity vectors of both the Earth and IBEX spacecraft, the exact correspondence between the select ISM flow observation times and times for which the simulations are done, and the true shape of the collimator aperture and transmission function must all be included. Simplifications of these aspects cause inaccuracies that reduce the quality of the fits to unacceptable levels. \citet[this issue]{hlond_etal:12a} showed that since the boresight of the IBEX-Lo sensor matches the value provided by IBEX attitude control system to $\sim 0.1\degr$, no further corrections for the viewing geometry are needed in the modeling.

Based on these results, we analyzed direct measurements of the flow of neutral interstellar helium gas at Earth orbit obtained from the IBEX-Lo experiment onboard the Interstellar Boundary Explorer, performed during two observation campaigns at the beginning of 2009 and 2010. By numerical fitting of a model of the gas flow and measurement process to the data, we determined the flow vector and temperature of the neutral helium gas in the Local Interstellar Cloud immediately in front of the heliosphere. The flow vector differs from the previously measured, being $\sim 3.8 \, \mathrm{km}\, \mathrm{s}^{-1}$ slower and $3.8\degr$ greater in ecliptic longitude; the gas inflow direction is $79.2\degr$, latitude $-5.1\degr$, velocity $22.8 \, \mathrm{km}\, \mathrm{s}^{-1}$ and temperature 6200~K. The uncertainties of the parameters are correlated with each other and the acceptable ranges are shown in Fig. \ref{figParamCuts}. We estimate that the normal to the Hydrogen Deflection Plane differs by $\sim21\degr$ from the previous determination and points toward ecliptic (longitude, latitude) $\lambda=357.5\degr$, $\beta=58.5\degr$. These new findings are in a very good agreement with the conclusions from a recent sophisticated study of gas kinematics in the Local Interstellar Medium and hence should drive major revisions in the state-of-the-art models used to represent our heliosphere.

A comparison of the best model with the measurements indicates that IBEX also observed a new source of neutral helium in or near the heliosphere. A preliminary and rough estimate based on the prior knowledge of the interstellar conditions and on the plasma parameters in the outer heliosheath suggests that much more of the primary interstellar He may be transformed into the neutral secondary population than previously thought, mostly due to the charge exchange between the neutral He atoms and interstellar He$^+$ ions, and that the secondary population of He may be appreciably more abundant. We hypothesize that IBEX discovered this population.

	\acknowledgments
Acknowledgments: M.B. and M.A.K. wish to thank Mr J.~Kurek and Dr. M. Denis for their technical and engineering assistance with the SRC PAS computer cluster used for the simulations. The use of the solar EUV data from CELIAS/SEM and TIMED/SEE and of the solar wind data from the OMNI-2 series is gratefully acknowledged. The authors from the SRC PAS were supported by the Polish Ministry for Science and Higher Education grants NS-1260-11-09 and N-N203-513-038. This work was supported by the Interstellar Boundary Explorer mission as a part of NASA's Explorer Program.

\bibliographystyle{apj}
\bibliography{iplbib}{}

\end{document}